\documentclass[numberedappendix,apj]{emulateapj}

\shorttitle{}
\shortauthors{Graves et al.}

\begin{document}

\title{Ages and Abundances of Red Sequence Galaxies as a Function of
LINER Emission Line Strength}

\author{Genevieve J. Graves\altaffilmark{1}, S. M. Faber\altaffilmark{1}, Ricardo P. Schiavon\altaffilmark{2},
\& Renbin Yan\altaffilmark{3}}

\altaffiltext{1}{UCO/Lick Observatory, University of California, Santa
Cruz}
\altaffiltext{2}{University of Virigina}
\altaffiltext{3}{University of California, Berkeley}


\begin{abstract}

Although the spectrum of a prototypical early-type galaxy is assumed
to lack emission lines, a substantial fraction (likely as high as
30\%) of nearby red sequence galaxy spectra contain emission lines
with line ratios characteristic of low ionization nuclear
emission-line regions (LINERs).  We use spectra of $\sim 6000$
galaxies from the Sloan Digital Sky Survey (SDSS) in a narrow redshift
slice ($0.06 < z < 0.08$) to compare the stellar populations of red
sequence galaxies with and without LINER-like emission.  The spectra
are binned by internal velocity dispersion ($\sigma$) and by emission
properties to produce high $S/N$ stacked spectra.  The recent stellar
population models of R. Schiavon (2006) make it possible to measure
ages, [Fe/H], and individual elemental abundance ratios [Mg/Fe],
[C/Fe], [N/Fe], and [Ca/Fe] for each of the stacked spectra.  We find
that red sequence galaxies with strong LINER-like emission are
systematically 2--3.5 Gyr (10--40\%) younger than their emission-free
counterparts at the same $\sigma$.  This suggests a connection between
the mechanism powering the emission (whether AGN, post-AGB stars,
shocks, or cooling flows) and more recent star formation in the
galaxy.  We find that mean stellar age and [Fe/H] increase with
$\sigma$ for all galaxies.  Elemental abundance [Mg/Fe] increases
modestly with $\sigma$ in agreement with previous results, and [C/Fe]
and [N/Fe] increase more strongly with $\sigma$ than does [Mg/Fe].
[Ca/Fe] appears to be roughly solar for all galaxies.  At fixed
$\sigma$ galaxies with fainter $r$-band luminosities have lower [Fe/H]
and older ages but similar abundance ratios compared to brighter
galaxies.

\end{abstract}

\section{Introduction}

An important recent result in the field of galaxy evolution is the
build-up of the red sequence between $z=1$ and the present.  The red
sequence luminosity function determined by the COMBO-17 survey
\citep{bel04} shows an accumulation of massive red galaxies since
$z=1$. This result was confirmed by \citet{fab06} using combined data
from COMBO-17 and the DEEP2 redshift survey.  \citet{fab06} contrasted
the evolving luminosity function of red galaxies with the luminosity
function of massive blue galaxies, which remains roughly constant over
the same epoch.  With so much transition in the relatively recent
past, there should be visible evidence of recent red-galaxy creation
in the local universe; that is, nearby galaxy samples might contain
galaxies that are currently evolving onto the red sequence, as well as
recent red sequence arrivals.

Stellar population models predict that once a galaxy has ceased to
form stars it will move to the red sequence rapidly.  A galaxy whose
star formation is quenched suddenly after a history of continuous
constant star formation will arrive on the red sequence within $\sim
400$ Myr, while a galaxy that experiences a substantial burst of star
formation before quenching will take only $\sim 1$ Gyr to become red
\citep{har06}.  The red sequence may therefore include galaxies which
contain a fractional population of stars with young ages.  Under the
possibly more reasonable assumption of star formation that declines
with time, the fraction of young stars will be smaller, will
contribute less to the total galaxy light, and presumably the galaxy
will move to the red sequence more rapidly.

Recently, \citet{yi05}, \citet{kav06}, and \citet{schaw06} have used
near-ultraviolet (NUV) photometry from the GALEX Medium Imaging Survey
in conjunction with Sloan Digital Sky Survey (SDSS) optical photometry
to examine the colors of early-type galaxies.  They show that
morphologically selected early-type galaxies have uniformly red
optical colors but exhibit a wide range of NUV-optical
colors. According to their analysis, the observed blue NUV-optical
colors in some early-type galaxies are indicative of low levels of
residual star formation; they use \citet{bru03} stellar population
models to demonstrate that blue {\it NUV}-optical colors are sensitive
to very low levels of recent star formation and determine that at
least 30\% of red sequence galaxies have 1--3\% of their mass in young
(age $< 1$ Gyr) stars, with typical ages for the young component of
300--500 Myr.

Although the NUV-optical colors of these morphologically selected
galaxies show signs of recent star formation, the galaxies in their
sample lack optical emission lines, almost without exception.
\citet{kav06} show that their morphological selection criterion
removes virtually all galaxies with emission lines characteristic of
H\textsc{ii} regions.  In addition, they exclude all galaxies with
emission lines characteristic of active galactic nuclei (AGN) and
low ionization nuclear emission regions (LINERs), using the criterion
established by \citet{kau03}.  This is to remove from the sample
galaxies that might have substantial ultraviolet (UV) light generated
by a source other than the stars.  However, by excluding
all galaxies with AGN-like emission lines, they may have removed many
red sequence galaxies from their sample which are not forming stars.
These galaxies could include many of the most recent arrivals on the red
sequence.  

\citet{kau03} show that the host galaxies of luminous
([O\textsc{iii}]$\lambda5007$ emission line luminosities in excess of
$10^7 L_{\odot}$) AGN appear to be young or post-starburst systems,
while those of low-luminosity active nuclei (a category dominated by
LINERs in their sample) have morphologies and stellar populations that
are comparable to those of early-type galaxies with no
emission. Recently, \citet{yan06} have shown that up to 30\% of red
sequence galaxies in the SDSS have emission lines, with line ratios
characteristic of LINERs.  Another $\sim 50$\% of the red sequence
consists of ``quiescent'' galaxies with no detectable emission, while
the remaining $\sim 20$\% are divided among Seyferts, dusty
star-forming galaxies, and Transition Objects, which likely contain
both Seyfert and H\textsc{ii} region emission \citep{yan06}.  A
significant fraction of the red sequence population therefore consists
of galaxies with optical line emission, although these galaxies are
not necessarily forming stars.  These galaxies are often
systematically excluded from stellar population analysis of red
sequence or early-type galaxies, even though they are a large class of
objects and may harbor clues to the evolution of galaxies onto the red
sequence.

This paper presents a stellar population analysis of red sequence
galaxies, including both quiescent galaxies with no emission and those
with LINER-like emission.  The fact that galaxies with LINER-like
emission lines make up 30\% of the red sequence, in addition to having
morphologies and stellar populations that appear to be consistent with
the larger population of early-type galaxies generally, justifies a
stellar population analysis performed in parallel with quiescent red
sequence galaxies.  Also, if LINERs are genuinely active nuclei and if
their host galaxies form the low luminosity end of the AGN host galaxy
sequence described in \citet{kau03}, these galaxies with LINER-like
emission may be candidates for ``post post-starburst'' systems that
have recently stopped forming stars and moved onto the red sequence.
Signs of such recent quenching could then be detected through a
careful differential stellar population analysis of LINERs versus
non-LINERs.

There is still no consensus on the physical mechanism that powers
LINER emission.  Possibilities include photoionization by a central
AGN accretion disk as in higher ionization Seyferts (\citealt{fer83},
\citealt{hal83}), excitation through shocks (as in the models of
\citealt{dop95}), accretion of warm gas in cooling flows
(e.g. \citealt{hec81}), or photoionization by post asymptotic giant
branch (AGB) stars in either old \citep{bin94} or young \citep{tan00}
stellar populations.  This topic is covered by the \citet{fil03}
review, which includes a nice summary of these mechanisms.

There are clearly LINERs whose emission is strongly dominated by low
luminosity cental AGN, identified by either a central non-stellar
source (an X-ray or UV point source, or a compact radio core) or by a
signature of AGN accretion (broad line emission or even double-peaked
broad H$\alpha$ emission).  Examples of each of these are referenced
in \citet{fil03}.  However, there is also substantial evidence for
ionized gas in early type galaxies that extends {\it multiple
kiloparsecs} from the galaxy center, as shown in narrow-band imaging
of early-type galaxies (e.g., \citealt{gou94}, \citealt{mac96}) and
more recently in the SAURON integral field unit spectroscopy of
elliptical and lenticular galaxies \citep{sar06}.  These studies imply
that a large fraction of luminous elliptical galaxies contain {\it
distributed} ionized gas with optical emission line ratios that are
characteristic of LINERs.  We must be careful with terminology, since
the original definition of the term LINER \citep{hec80} required that
the emission be nuclear (``low ionization {\it nuclear} emission-line
region'').  These galaxies are therefore not nuclear LINERs, but they
do exhibit emission with LINER-like ratios.  We shall therefore refer
to them as ``LINER-like'' galaxies, without any assumption about the
spatial distribution of the ionized gas within the galaxy or the
ionization mechanism.  From SDSS data alone, whose spectral fibers
cover a substantial portion of the galaxy, it is not possible to
decide whether any given galaxy is a classic AGN LINER, an extended
LINER-like galaxy, or some combination of the two.

The fact that LINER-like emission is common in red sequence galaxies
prompts many questions: is it an evolutionary phase that all
early-type galaxies pass through?  Do galaxies pass through this stage
more than once?  Perhaps many times?  Is it a cyclical process?  Is
the gas external in origin, as might be suggested by the kinematic
decoupling of gas and stars seen by \citet{cao00} and \citet{sar06},
or is it produced by mass-loss from stars within the galaxy?  What is
the relation, if any, between the central AGN LINER emission and the
more extended LINER-like emission?  In this paper, we seek to address
some of these questions by examining the stellar populations of the
host galaxies to determine whether or not the existence of LINER-like
emission in a red sequence galaxy is related to its star formation
history.

Past stellar population analysis of early-type or red sequence
galaxies has tended to either exclude all galaxies with emission
(e.g., \citealt{nel05} and \citealt{ber03b}) or to include galaxies
with weak LINER-like emission in the main sample.  In the latter case,
authors have made a correction for Balmer line emission infill using
an empirical correction based on the strength of the
[O\textsc{iii}]$\lambda5007$ (e.g., \citealt{gon93}, \citealt{tra00a},
\citealt{ram05}) or [O\textsc{ii}]$\lambda3727$ (e.g.,
\citealt{sch_letter06}) emission lines, or else restrict themselves to
using higher-order Balmer lines such as H$\gamma$ and H$\delta$, where
contamination from emission is less.

In this work, we analyze galaxies with and without LINER-like emission
in parallel, in order to contrast their stellar population properties.
LINER-like nebulae produce only modest quantities of Balmer emission,
and that emission is well-characterized (due to the selection process
described in \S\ref{infill_correction}).  This allows us to make
accurate Balmer infill corrections for multiple Balmer lines and to
compare results.  In order to obtain very high signal to noise ($S/N$)
spectra for the stellar population analysis, we stack several hundred
similar galaxies to produce highly accurate determinations of the mean
stellar population properties of the galaxies that go into each
composite.  We use the new single stellar population (SSP) models of
\citet{sch07}, which allow the determination of not only mean ages and
total metallicities, but of several individual elemental abundances as
well; iron, magnesium, carbon, nitrogen, and calcium are all
separately determined.  Section \ref{data} presents the data sample,
the criteria for grouping similar galaxies used to construct the
composite spectra, and the procedure used to correct for Balmer
emission infill in the LINER-like galaxies.  Section
\ref{linestrengths} gives results of our line strength measurements as
a function of galaxies properties.  Section \ref{modeling} describes
the stellar population modeling process and important caveats, then
\S\ref{results} uses SSP models to interpret the line strength
measurements in terms of stellar population properties.  In
\S\ref{discussion}, we discuss the implications of our results on the
possible source of the LINER-like emission in red sequence galaxies,
and \S\ref{conclusions} summarizes our conclusions.

\section{Data}\label{data}

This paper presents an analysis of red sequence galaxy spectra from
the Sloan Digital Sky Survey (SDSS), a massive imaging and
spectroscopic survey conducted at the Apache Point Observatory with a
dedicated 2.5 m telescope.  The SDSS has imaged more than one quarter
of the sky in five filter bands: {\it u, g, r, i,} and {\it z}
(\citealt{fuk96}, \citealt{sto02}).  The survey has also obtained
spectroscopic observations of over one million galaxies and quasars
using a purpose-built dual fiber spectrograph, with resolution $R =
\lambda/\Delta\lambda \approx 1800$, wavelength coverage $\lambda =$
3800--9200{\AA}, and fiber diameter $d = 3''$.  The data presented in
this paper are taken as part of the SDSS main galaxy sample, which
includes over 500,000 galaxies with redshifts $0.00 \leq z \leq 0.30$.
The main galaxy sample is selected to include galaxies brighter than
$r < 17.77$.

The following imaging and spectroscopic parameters for each galaxy
were taken from the NYU Value Added Catalog (\citealt{bla05},
\citealt{ade05}), Data Release 4 (DR4): redshift $z$, velocity
dispersion $\sigma_{\rm{fib}}$ as measured in the SDSS fiber spectrum,
$ugriz$ fluxes measured in the SDSS flux units ``maggies'' (see
\citealt{bla05}), and median $S/N$ per {\AA} for each spectrum.  The
$ugriz$ fluxes were converted to absolute magnitudes assuming standard
$\Lambda$CDM cosmology ($\Omega_{\Lambda} = 0.7$, $\Omega_m = 0.3$),
with a Hubble constant of $H_0 = 70$ km s$^{-1}$ Mpc$^{-1}$.  The
magnitudes were then K-corrected to $z=0.1$ using Mike Blanton's {\it
kcorrect} code, v.  3\_2 \citep{bla03a}.  We chose to K-correct to
$z=0.1$ both to enable easy comparison with colors and magnitudes for
SDSS galaxies in \citet{bla03b} and because we limited our galaxy
sample to galaxies with $0.06 < z < 0.08$ (see \S\ref{select}); the
K-correction to $z=0.1$ is small and therefore any errors due to
uncertainty in the K-correction are minimized.  Throughout this paper,
all magnitudes and colors are quoted at $z=0.1$, unless otherwise
noted.

These values were supplemented with the following measurements from
the SDSS DR4 Catalog Archive Server: the best-fit deVaucouleurs radius
($r_{e}$) and axis ratio ($b/a$).  Our analysis also uses emission
line equivalent widths (EWs) for H$\alpha$, H$\beta$,
[O\textsc{ii}]$\lambda$3727, [O\textsc{iii}]$\lambda$5007,
[O\textsc{i}]$\lambda$6300, and [N\textsc{ii}]$\lambda$6583 measured
in SDSS DR4 galaxies by \citet{yan06}, with the zeropoint corrections
defined in their Table B3.  Before making their emission line
measurements, Yan et al. fit and subtract off a stellar absorption
line template spectrum to obtain good measurements of even those
emission lines that fall on top of stellar absorption features.  The
interested reader should consult their paper for further details on
this process.

\subsection{Sample Selection}\label{select}

The sample is chosen to match the selection criteria of the
\citet{yan06} magnitude limited sample; objects are from the SDSS main
galaxy sample and are spectroscopically confirmed to be galaxies.  We
then limit the sample to a narrow range in redshift, $0.06 < z <
0.08$, making this magnitude-limited sample roughly volume-limited and
minimizing the impact of evolution on this analysis.  The $1.5''$ SDSS
spectral fiber radius corresponds to 1.7--2.3 kpc in this redshift
range.

The aim of this work is to study red sequence galaxies with no active
star formation.  We use two selection criteria to select galaxies
which are not currently forming stars: a color cut (described below)
and the emission line properties of the galaxies.  \citet{yan06}
demonstrated that the distribution of [O\textsc{ii}] to H$\alpha$
emission line EW ratios is bimodal such that some emission line
galaxies have strong H$\alpha$ emission but relatively weak
[O\textsc{ii}] emission while others have strong [O\textsc{ii}] and
weak H$\alpha$.  This is best seen in a plot of log EW(H$\alpha$)
vs. log EW([O\textsc{ii}]), as in Figure \ref{sample_cuts}a (see also
\citealt{yan06}, Figure 2).  

They showed that the left peak with high EW([O\textsc{ii}]) and low
EW(H$\alpha$) (they call these ``High-[O\textsc{ii}]/H$\alpha$''
galaxies) is dominated by red sequence galaxies, whereas the right
peak with larger EW(H$\alpha$) (``Low-[O\textsc{ii}]/H$\alpha$''
galaxies) is dominated by blue galaxies.  This led them to explore the
possibility that the High-[O\textsc{ii}]/H$\alpha$ galaxies are
star-forming systems while the Low-[O\textsc{ii}]/H$\alpha$ galaxies
contain emission that is not powered by star formation. They confirm
this hypothesis through the use of line ratio classification diagrams
first proposed by \citet{bal81} and commonly known as ``BPT
diagrams''.

Following \citet{yan06}, we define red sequence galaxies as those with
\begin{equation}\label{redcut}
^{0.1}(g-r) > -0.025 (\mbox{ }^{0.1}M_{r} - 5 \log h) + 0.42, 
\end{equation}
here with $h = 0.70$. This includes 22,501 galaxies in the chosen
redshift range.  This color cut is shown in Figure \ref{sample_cuts}c
as the dashed line.  The color cut has been chosen to be conservative
and excludes galaxies that lie in the ``valley'' between the red
sequence and the bulk of blue galaxies.  On the red sequence, we use
H$\alpha$ and [O\textsc{ii}] EWs to characterize galaxies according to
their emission line properties.  We use the \citet{yan06} relation
\begin{equation}\label{o2hacut}
\mbox{EW([O\textsc{ii}])} > 5\mbox{EW(H}\alpha\mbox{)}-7,
\end{equation}
to separate High-[O\textsc{ii}]/H$\alpha$ from
Low-[O\textsc{ii}]/H$\alpha$ galaxies.  This dividing line is shown as
the solid line in Figures \ref{sample_cuts}a and \ref{sample_cuts}b.
Figure \ref{sample_cuts}a shows the same information as Figure
\ref{sample_cuts}b, but with linear axes.  The latter format does not
highlight the bimodality as clearly but is a more natural way to
examine the cut used to define the High-[O\textsc{ii}]/H$\alpha$
sample.  

To confirm that the High-[O\textsc{ii}]/H$\alpha$ objects are not star
forming galaxies, Figure \ref{bpt} shows BPT diagrams of emission line
galaxies in the redshift range $0.06 < z < 0.08$.  Red points show the
red sequence High-[O\textsc{ii}]/H$\alpha$ objects as defined above;
black points show all other emission-line galaxies.  Only galaxies
with a $\geq 3\sigma$ detection in all lines are shown on each plot.
Panel (b) contains fewer objects than (a) because
[O\textsc{i}]$\lambda$6300 is a weak line and undetected in 2/3 of the
High-[O\textsc{ii}]/H$\alpha$ and 1/2 of all other galaxies shown in
(a).  The solid lines in each panel show the standard classification
scheme dividing Seyferts, LINERs, and star forming galaxies.  The
dotted lines show the \citet{kew01} division between star forming
galaxies and active nuclei; the dashed line in (a) shows the division
used by \citet{kau03}.  The dash-dot line in (b) shows the
\citet{ho97} division between LINERs and transition objects (TOs).
Based on the \citet{kau03} division between star forming galaxies and
active nuclei, the standard division between LINERs and Seyferts, and
the \citet{ho97} division between LINERs and TOs, the red sequence
High-[O\textsc{ii}]/H$\alpha$ galaxies consist of 88.4\% LINERs, 8.6\%
TOs, 1.9\% Seyferts, and 1.1\% star forming galaxies.  Most of the
High-[O\textsc{ii}]/H$\alpha$ objects classified as TOs lie very near
the division between LINERs and TOs, well within the typical errors,
and are not convincingly a separate class of objects.  There are
High-[O\textsc{ii}]/H$\alpha$ galaxies that are not shown here because
their H$\beta$, [O\textsc{iii}]$\lambda$5007,
[O\textsc{i}]$\lambda$6300, and/or [N\textsc{ii}]$\lambda$6583 emission
lines are weaker than $3\sigma$ detections, but they are assumed to
possess the same underlying emission line properties as those shown.

The analysis in this paper uses two samples of galaxies classified by
their [O\textsc{ii}] and H$\alpha$ emission lines: galaxies with
neither H$\alpha$ nor [O\textsc{ii}] detected (at the 2$\sigma$ level)
and High-[O\textsc{ii}]/H$\alpha$ galaxies with both lines detected.
We shall hereafter refer to these as ``quiescent'' and ``LINER-like''
galaxies respectively.  The color distributions of these samples are
similar and concentrated on the red sequence.  Figure
\ref{sample_cuts}d shows CMDs of the quiescent and LINER-like
galaxies, with the color cut defining the red sequence overplotted as
the dashed line (as in Figure \ref{sample_cuts}c).  While the colors
of the LINER-like and quiescent galaxies are similar, the luminosity
distribution is clearly different.  This is partially a selection
effect; the LINER-like galaxies are required to have both
[O\textsc{ii}] and H$\alpha$ detected at the $2\sigma$ level, thus
fainter galaxies with lower $S/N$ spectra are less likely to meet the
selection criteria.  Taking this effect into account, there still
appears to be a decline in the fraction of red sequence LINER-like
galaxies at faint magnitudes.  This will be explored in more detail in
future work.  Only galaxies redder than the dashed line of Figures
\ref{sample_cuts}c-d are included in our sample.  Spectra of the
quiescent and LINER-like red sequence galaxies were downloaded from
the SDSS Data Archive Server (http://das.sdss.org).

As defined, our sample of galaxies contains overlap with the sample of
AGNs and their hosts described by \citet{kau03}.  They select galaxies
on the basis of their log([N\textsc{ii}]$\lambda$6583/H$\alpha$) and
log([O\textsc{iii}]$\lambda3727$/H$\beta$) ratios, including galaxies
that are classified as Seyferts by the criteria of \citet{ho97}, as
well as LINERs and some Transition Objects.  They claim that all of
these objects harbor AGN of some kind, and divide their sample into
low-luminosity and high-luminosity AGN, with the dividing line at log
$L_{\rm{[O\textsc{iii}]}}/L_{\odot} = 7.0$.  Careful study of their
Figure 3 reveals that at least 85\% of the objects that they classify
as LINERs fall into their low-luminosity AGN sample, while only $\sim
20$\% of their Seyferts fall into this sample.  Comparisons to the
results of \citet{kau03} are therefore most appropriate between our
sample and their ``low-luminosity AGN'' sample.  However, they apply
no color criterion to their selection of AGN hosts.  Of the galaxies
in our Figure \ref{bpt} which meet the \citet{kau03} definition of
AGN, 55\% are bluer than our red sequence cut-off and 37\% meet the
more conservative color criterion for blue galaxies in \citet{yan06}.
The \citet{kau03} sample clearly includes many more young (and
possibly actively star forming) galaxies than our sample.  Their
analysis of other properties of the host galaxies suggest that the
hosts of strong AGN are young galaxies but that the hosts of their
low-luminosity AGN sample are similar to early-type galaxies, so many
or most of these blue galaxies may be in their high-luminosity AGN
sample.  However, their low-luminosity AGN sample is still likely to
contain a significant fraction of galaxies bluer than any present in
our analysis.

\subsection{Emission Infill Correction}\label{infill_correction}

Our analysis uses the strength of the stellar H$\beta$, H$\gamma$, and
H$\delta$ absorption features, in conjunction with metal lines, to
measure SSP ages for the galaxies in our sample.  In the LINER-like
galaxies, stellar Balmer line absorption will be partially or
completely filled in by emission from the ionized gas within the
galaxy.  This infill severely impacts stellar population analysis, and
a correction must be made.  If we use the H$\alpha$ and H$\beta$
emission line strengths measured by \citet{yan06} to make an emission
infill correction, we merely recover the Balmer absorption line
strengths of their subtracted absorption line spectral template, and
further stellar population modeling will be meaningless.
\citet{gon93} defined an H$\beta$ infill correction based on the EW of
[O\textsc{iii}]$\lambda 5007$ emission, which has been used (sometimes
in modified form) by many other groups.  \citet{tra00a} use
$\Delta$H$\beta$ = 0.7 EW([O\textsc{iii}]) which, assuming the
continuum level is similar at H$\beta$ and [O\textsc{iii}], is a good
match to the observed [O\textsc{iii}]/H$\beta$ ratios for LINERs shown
in Figure \ref{bpt}.  However, [O\textsc{iii}]$\lambda 5007$ is weaker
in LINERs than [O\textsc{ii}]$\lambda 3727$ due to their lower
ionization.  For the SDSS spectra used here, with typical
$S/N_{\rm{med}} \sim 22$ {\AA}$^{-1}$, measurements of [O\textsc{iii}]
are less reliable than those of [O\textsc{ii}].  In the LINER-like
galaxy sample, fewer than one third of all galaxies have
[O\textsc{iii}] EW detections at or above a 3$\sigma$ confidence level
whereas more than 75\% have $> 3\sigma$ detections in [O\textsc{ii}].
Because the LINER-like galaxies lie on a narrow sequence in
EW([O\textsc{ii}])-EW(H$\alpha$) space (see Figure \ref{sample_cuts}),
[O\textsc{ii}] can provide an estimate of H$\alpha$ emission which is
less dependent upon the Balmer line strengths of the stellar template
fit used to measure emission lines in \citet{yan06}.  Having estimated
the H$\alpha$ EW from [O\textsc{ii}], we can use a theoretically
predicted Balmer decrement to derive the expected emission line
strengths of H$\beta$, H$\gamma$, and H$\delta$.

The existence of a tight roughly linear scaling relation between
EW([O\textsc{ii}]) and EW(H$\alpha$) suggests that both emission lines
are produced by the same mechanism.  The classification of quiescent
vs. LINER-like galaxies was based in part on a $S/N$ criterion that
both emission lines be detected at a 2$\sigma$ confidence level.  It
is reasonable to assume that galaxies exist with very faint emission
that is undetected above the stellar continuum in the $S/N_{\rm{med}}
\approx 22$ {\AA}$^{-1}$ SDSS spectra.  These galaxies would most
likely be intrinsically LINER-like galaxies with emission line
equivalent widths approaching zero and would be classed by us in the
quiescent sample.  Using a linear least-squares fit of EW(H$\alpha$)
as a function of EW([O\textsc{ii}]) and forcing the fit to go through
the origin, we find that EW(H$\alpha$) $\approx$ 0.197
EW([O\textsc{ii}]).  This fit is shown as the dashed line in Figures
\ref{sample_cuts}a-b.  Using this fit, the measured EW([O\textsc{ii}])
for each galaxy is converted into an estimated EW(H$\alpha$) and then
into an estimated H$\alpha$ flux, using the median spectral continuum
level around H$\alpha$ in each spectrum.  The wavelength range used
for computing the H$\alpha$ continuum level is the same as \citet[Table
3]{yan06}.

The theoretically predicted Balmer decrement H$\alpha$/H$\beta$ for
LINER- and AGN-like emission line ratios is 3.1 \citep{ost05}, with
further decrements H$\gamma$/H$\beta = 0.46$ and H$\delta$/H$\beta$ =
0.26 for the higher order Balmer lines. These values are fairly
independent of the exact excitation mechanism.  However, if there is
dust within the galaxy in or around the emission line region,
H$\beta$, H$\gamma$ and H$\delta$ emission will be dimmed with respect
to H$\alpha$.  The Balmer decrement actually measured for LINER-like
galaxies in the \citet{yan06} sample is H$\alpha$/H$\beta =
4.1\pm0.3$, corresponding to reddening with $\tau = 0.28$ between
H$\alpha$ and H$\beta$, or $E(B-V) = 0.47$.

As already noted, the Balmer emission measurements in the Yan et
al. analysis are highly dependent on the choice of stellar template,
and Balmer line emission is weak in LINER-like objects; H$\beta$ is
detected at the 3$\sigma$ level in less than 23\% of the galaxies in
the LINER-like sample.  In light of this, we also examined the sample
of bright galaxies with $B_T \le 12.5$ mag compiled by Ho, Filippenko,
\& Sargent a decade ago.  (\citealt{ho95}, \citealt{ho97} and other
papers in the series).  In \citet{ho97}, they classify emission line
galaxies based upon their emission line ratios (specifically
log([O\textsc{iii}]$\lambda5007$/H$\beta$),
log([N\textsc{ii}]$\lambda6583$/H$\alpha$), and
log([O\textsc{i}]$\lambda6300$/H$\alpha$)) as Seyfert, LINER,
H\textsc{ii}, or Transition Objects.  For all objects in their catalog
which are classified as LINERs and which satisfy our color cut, the
median H$\alpha$/H$\beta$ Balmer decrement is 3.27, only slightly
higher than that predicted by theory.  This suggests that there is
very little dust absorption within these systems, corresponding to
a reddening of $\tau = 0.053$ between H$\alpha$ and H$\beta$, equivalent
to $E(B-V) \approx 0.050$.  The distribution of H$\alpha$/H$\beta$ in
their sample of LINERs is independent of galaxy luminosity and
emission line strength.  However, their spectra cover the nuclear
region (the central $\sim 200$ pc) only, and therefore may not be an
appropriate counterpart to the SDSS fiber spectra used in our
analysis, which typically cover 0.35--1.1 galaxy effective radii
corresponding to several kpc.

In view of the uncertainty in the value of the proper Balmer
decrement, the analysis below was repeated for both
H$\alpha$/H$\beta = 3.27$ (as in \citealt{ho97}) and H$\alpha$/H$\beta
= 4.1$ (as in \citealt{yan06}).  To obtain an emission infill
correction for H$\gamma$ and H$\delta$, we assumed intrinsic Balmer
decrements of H$\gamma$/H$\beta = 0.46$ and H$\delta$/H$\beta = 0.26$,
with a reddening law that scales as $\tau \propto \lambda^{-0.7}$
(following \citealt{kau03} and \citealt{car98}).  In
\S\ref{ssp_models} we will show that the choice of Balmer decrement
affects the stellar population ages measured from the Balmer lines,
particularly in H$\beta$.  The higher order Balmer lines (and
H$\delta$ in particular) are less affected, and the evidence is
strong that while uncertainties in the Balmer decrement introduce
uncertainties into our age measurements, they do not qualitatively
change our main results.

\subsection{Composite Spectra}\label{composite_spectra}

Detailed SSP modeling requires high $S/N$ spectra, ideally $S/N \ge
100$ {\AA}$^{-1}$.  Individual SDSS spectra in this sample typically
have $S/N_{\rm{med}} \sim 22$ {\AA}$^{-1}$, making it necessary to
stack dozens of similar spectra to achieve adequate $S/N$ for SSP
modeling.  In this paper, we explore the dependence of stellar
population age, [Fe/H], and abundance ratios on aperture-corrected
galaxy velocity dispersion\footnotemark ($\sigma$), and the observed
LINER-like emission line strength as measured by the strength of the
[O\textsc{ii}]$\lambda3727$ emission line.\footnotetext{Following
\citet{ber03a} and \citet{jor95}, velocity dispersions are aperture
corrected to a standard one-eighth of the effective radius using
$\sigma = \sigma_{\rm{fib}} (r_{\rm{fib}} / \frac{1}{8} r_o)^{0.04}$,
where $\sigma_{\rm{fib}}$ is the velocity dispersion measured in the
fiber spectrum, $r_o$ is the circular galaxy radius in arc seconds and
$r_{\rm{fib}}$ is the spectral fiber radius, $1.5''$ for SDSS.  This
correction is small, of order $\sim$6\%.}  To do this, we divide the
LINER-like galaxy sample into six bins in $\sigma$, with $\sigma =
70$--$120$ km s$^{-1}$, $\sigma = 120$--$145$ km s$^{-1}$, $\sigma =
145$--$165$ km s$^{-1}$, $\sigma = 165$--$190$ km s$^{-1}$, $\sigma =
190$--$220$ km s$^{-1}$, and $\sigma = 220$--$300$ km s$^{-1}$.  Bins
in $\sigma$ are chosen to contain roughly equal numbers of galaxies.
Within each $\sigma$-bin, the LINER-like galaxies are further divided
into bins of weak (EW([O\textsc{ii}]) $<$ 5 \AA) and strong
(EW([O\textsc{ii}]) $>$ 5 \AA) emission strength.  Each bin contains
over 100 galaxies and has a $S/N_{med} \geq 230$ \AA$^{-1}$.  In
total, there are 2141 galaxies with weak emission and 1775 with strong
emission.

Once the LINER-like galaxies have been divided into weak and strong
subsamples, there are almost five times the number of quiescent
galaxies as there are either weak or strong LINER-like galaxies.  For
comparison with the weak and strong LINER-like galaxies, we construct
a sample of 2000 quiescent galaxies.  Figure \ref{o2_hist} shows the
distribution of EW([O\textsc{ii}]) for the 10,284 in the quiescent
sample as the open histogram along with a gaussian fit to the
distribution. The distribution peaks $\sim0.3$ {\AA} below zero.  This
may mean that the zeropoint correction determined in \citet{yan06} is
uncertain at this level.  For LINER-like emission with
EW([O\textsc{ii}]) $\sim 5 \times$ EW(H$\alpha$) and H$\alpha$/H$\beta
\sim 3$, this produces $\Delta$H$\beta \sim 0.02$ {\AA}.  An
uncertainty at this level is insignificant in the stellar population
modeling process, thus we ignore the possible small discrepancy in
zeropoint.  

To create a comparison sample of quiescent galaxies, we construct a
gaussian distribution centered about EW([O\textsc{ii}]) = 0 {\AA} with
$\sigma_{\rm{[O\textsc{ii}]}} = 1.56$ {\AA} to match the width of the
distribution in Figure \ref{o2_hist}.  This distribution is truncated
at $\pm 2 \sigma_{\rm{[O\textsc{ii}]}}$ in order to exclude outliers
which may contain low-level undetected emission.  From this truncated
gaussian, we randomly select 2000 galaxies for our analysis.  The
EW([O\textsc{ii}]) distribution of the final quiescent sample is shown
as the shaded histogram in Figure \ref{o2_hist}.

The 2000 quiescent galaxies are then divided into six bins in
$\sigma$ as were the weak and strong LINER-like galaxies.  The
sample therefore includes 18 bins in total: six bins in $\sigma$ by
three bins in emission line strength, labelled as ``quiescent'',
``weak'', and ``strong''.  The parameters of the various galaxy bins
are summarized in Table \ref{bin_table}.

In each individual spectrum, absorption lines are broadened by the
internal velocity dispersion of the stars in the galaxy, which has the
effect of reducing the depths of the lines in the higher-$\sigma$
galaxies.  To make a consistent comparison between separate bins in
$\sigma$, all the spectra are smoothed to an effective $\sigma = 300$
km s$^{-1}$ to match the highest-$\sigma$ galaxies in the
sample.\footnotemark \footnotetext{Note: Galaxies are binned by
aperture-corrected $\sigma$, as this is the best estimate of an {\it
intrinsic} property of the galaxy.  For the purpose of measuring
absorption line indices, all galaxies are smoothed based on their {\it
observed} $\sigma_{\rm{fib}}$ to simulate the effect of uniformly
observing all galaxies with broadening of 300 km s$^{-1}$.} The
spectra are coadded within each bin, using an algorithm which averages
the unsmoothed spectra within small sub-bins in $\sigma_{\rm{fib}}$
before smoothing to $\sigma = 300$ km s$^{-1}$ and coadding the
sub-bins.  This minimizes the effect of far outlier pixels by
averaging out noise spikes in a single pixel before smoothing the
spectra.  The regions around bright skylines at 5577{\AA}, 6300{\AA},
and 6363{\AA} (observed frame) are masked in the individual galaxies
to produce clean stacked spectra.  In addition to an average spectrum,
the total $S/N$ at each resolution element is computed to produce an
error spectrum for each stacked spectrum.

It is possible, even likely, that the quiescent galaxy bins in fact
contain galaxies which do have faint emission lines but the galaxy
spectra are too low $S/N$ for the emission lines to be detected at the
$> 2\sigma$ level that would qualify them for the LINER-like
classification.  We have attempted to remove some of these cases by
truncating the EW([O\textsc{ii}]) distribution at $\pm 2
\sigma_{\rm{[O\textsc{ii}]}}$, as described above.  If a substantial
number of weak emission line objects remain, these could produce
non-negligible emission infill in the stacked spectra of quiescent
galaxies.  A check against this is to measure the [O\textsc{ii}]
emission in the stacked quiescent galaxy spectra, which have extremely
high $S/N$.  This is done in the same way as for the individual
spectra, subtrating off a stellar absorption line template before
making an emission line measurement, using the procedure described in
the appendix of \citet{yan06}.  The EW([O\textsc{ii}]) values measured
for the quiescent stacked spectra are -0.01 {\AA}, 0.04 {\AA}, 0.41
{\AA}, -0.27 {\AA}, 0.28 {\AA}, and -0.44 {\AA} for the $\sigma = $
70--120 km s$^{-1}$, 120--145 km s$^{-1}$, 145--165 km s$^{-1}$,
165--190 km s$^{-1}$, 190--220 km s$^{-1}$, and 220--300 km s$^{-1}$
stacked spectra, respectively.  The largest of these [O\textsc{ii}]
EWs, 0.41 {\AA}, corresponds to an H$\alpha$ EW of 0.082 {\AA} and an
H$\beta$ EW of 0.027 {\AA}, using the scaling of equation
\ref{o2hacut} and the theoretical Balmer decrement.  This level of
contamination in H$\beta$ is within the measurement errors of the
H$\beta$ Lick index for these high $S/N$ spectra (see
\S\ref{measure_lines}).

The stacked spectra with $70$ km s$^{-1} < \sigma < 120$ km s$^{-1}$
are shown in Figure \ref{em_spectra} as examples of the quality and
resolution of the stacked spectra.  The black line shows the quiescent
galaxy bin, the green line shows the galaxies with weak LINER-like
emission, and the red line shows the galaxies with strong LINER-like
emission.  The weak and strong LINER-like bins have been corrected for
emission infill.  The weak and strong bins {\it without} the emission
infill correction are over plotted as dotted lines; the only
differences appear in the Balmer absorption lines where the emission
infill correction has been made.  Prominant emission features are
indicated by vertical dashed lines and are labelled.  Other than the
emission features, the spectra are remarkably similar, which shows
both the high $S/N$ of the stacked data and the subtleness of the
spectral differences that are studied in this paper.

\subsection{Sample Completeness}\label{completeness}

The galaxy sample presented here is a magnitude-limited sample, as
described in \S\ref{select}, chosen within a narrow range in redshift
($0.06 < z < 0.08$).  The galaxy bins, however, are defined in terms
of $\sigma$.  This means that the galaxy bins will be incomplete at
faint magnitudes, where galaxies of a given $\sigma$ may exist but be
too faint to make it into the sample.  Because galaxy magnitude and
velocity dispersion are related through the Faber-Jackson relation
\citep{fab76}, the lowest-$\sigma$ bins will be most affected by this
incompleteness.  If the stellar populations of fainter galaxies differ
systematically from those of brighter galaxies at the same $\sigma$,
the results obtained from the stacked spectra will be biased at low
$\sigma$.  There is every reason to expect that the stellar
populations will indeed differ, since fainter magnitudes at the same
$\sigma$ likely imply larger mass-to-light ratios.  The incomplete
low-$\sigma$ bins will be biased toward brighter, and therefore
possibly younger and/or more metal-poor stellar populations.

The top panel of Figure \ref{cmd_sig} shows a color-magnitude diagram
(CMD) of the quiescent and LINER-like galaxy samples.  Galaxies have
been selected by the color cut described in \S\ref{select}, so only
red sequence galaxies are included (compare to Figure
\ref{sample_cuts}c).  The data points are color-coded by $\sigma$ as
explained in the caption.  The dashed line at $\mbox{ }^{0.1}M_{r} =
-19.89$ shows the completeness limit at $z=0.08$.  The solid line at
$\mbox{ }^{0.1}M_{r} = -19.63$ shows the magnitude at which the sample
is incomplete at the 50\% level (i.e. missing half the galaxies at that
magnitude).  Thus the dashed line shows where the sample begins to
miss galaxies and the solid line shows a point at which the sample has
become severely incomplete.

In the lower panels of Figure \ref{cmd_sig}, CMDs for each of the six
ranges in $\sigma$ are shown separately.  In each CMD, the gray points
show the entire sample of quiescent and LINER-like galaxies.
Overplotted are colored contours showing the distribution of galaxies
within a given range of $\sigma$.  Solid and dashed lines indicate
the 50\% and 100\% completeness limits respectively, as in the top
panel.  Below each CMD is a histogram of the $\mbox{ }^{0.1}M_{r}$
values for the galaxies in each $\sigma$ range.

Two well-known trends are immediately visible: higher $\sigma$
galaxies are brighter (Faber-Jackson), and higher $\sigma$ galaxies
are redder, producing the slope of the red sequence.  What is worthy
of note is that at fixed $\sigma$, there is little if any slope to the
color-magnitude relation.  This is consistent with the claim of
\citet{ber05} that the color-magnitude relation is purely a result of
color-$\sigma$ and magnitude-$\sigma$ relations which are more
fundamental, and Figure \ref{cmd_sig} gives a straight-forward visual
confirmation of their result.

By comparing the distribution of the galaxies in each $\sigma$ range
(both the 2-dimensional distribution in the CMD as shown in the
contour plots, and the 1-dimensional histogram in $\mbox{
}^{0.1}M_{r}$), we can determine the level of incompleteness in each
$\sigma$ range.  The lowest-$\sigma$ range, shown in purple, is
significantly incomplete; the distribution of $\mbox{ }^{0.1}M_{r}$
has not even reached a peak at the completeness limit (dashed line)
and the turn-over in the histogram may be entirely due to
incompleteness.  In this $\sigma$ range, the stacked spectra will
be missing the contribution of the faintest galaxies and will thus
give an estimate of mean stellar population properties which is biased
toward brighter galaxies at fixed $\sigma$.  

To estimate the incompleteness of each $\sigma$ range, we assume that
the highest $\sigma$ range ($\sigma = 220$--300 km s$^{-1}$) is
complete and that its distribution in $\mbox{ }^{0.1}M_{r}$
represents the true distribution in $\mbox{ }^{0.1}M_{r}$ for each
range.  The histogram of $\mbox{ }^{0.1}M_{r}$ for this range (shown in
red in the lower right corner of Figure \ref{cmd_sig}) shows an
extended tail to fainter magnitudes.  This tail is also visible in the
$\sigma = 190$--200 km s$^{-1}$ range and is suggested in the
$\sigma = 165$--190 km s$^{-1}$ range, thus lending support to the
assumption of similar $\mbox{ }^{0.1}M_{r}$ distributions in all
$\sigma$ ranges.  In each $\sigma$ range, the complete (bright) end of
the normalized $\mbox{ }^{0.1}M_{r}$ distribution is fit to the
normalized $\sigma = 220$--300 km s$^{-1}$ distribution and the
discrepancy between the current range and the $\sigma = 220$--300 km
s$^{-1}$ range at the faint end is used to estimate the
fractional incompleteness of the lower $\sigma$ range.  We find that
the missing galaxies constitute approximately 51\%, 28\%, 19\%, 11\%,
and 4\% of the $\sigma = 70$--120, 120--145, 145--165, 165--190, and
190--220 km s$^{-1}$ ranges, respectively.  The estimated fractional
incompleteness for each $\sigma$ range is shown in Table \ref{bin_table}.  

This indicates that the lower $\sigma$ ranges are significantly
incomplete.  If there are systematic differences between the stellar
populations of bright and faint galaxies at fixed $\sigma$, the
average stellar population parameters derived from the stacked spectra
of {\it decreasing} $\sigma$ ranges will be {\it increasingly} biased
toward the bright galaxies at fixed $\sigma$.  In appendix
\ref{lowsig_corr} we show that in fact the fainter galaxies at fixed
$\sigma$ have systematically weaker metal absorption lines than the
bright galaxies, but that their Balmer absorption lines are the same.
This suggests that the fainter galaxies are more metal poor and
somewhat older than the brighter galaxies at fixed $\sigma$.  We
will correct for the effect of this trend in our incomplete data in
\S\ref{measure_lines}.

\section{Lick Index Line Strengths}\label{linestrengths}

\subsection{Line Strength Measurements}\label{measure_lines}

In each stacked spectrum, we measure absorption line strengths of the
Lick indices, as defined in \citet{wor94} and \citet{wor97}.  These
include the Balmer lines H$\beta$, H$\gamma_F$, and H$\delta_F$, as
well as numerous ``metal lines'': five lines dominated by iron
absorption (Fe4383, Fe4531, Fe5270, Fe5335, and Fe5406), and lines
which are sensitive to the abundances of other elements (Mg$_1$,
Mg$_2$, Mg {\it b}, CN$_1$, CN$_2$, G4300, C$_2$4668, Ca4227, Ca4455,
NaD, TiO$_1$, and TiO$_2$).  The Lick indices are measured using the
IDL code {\bf Lick\_EW}, part of the {\bf EZ\_Ages} code package
described in Graves \& Schiavon (in preparation).  {\bf Lick\_EW} uses
the error spectra described in \S\ref{composite_spectra} to compute
the statistical errors in the index measurements due to the finite
$S/N$ of the stacked spectra.  The error calculation is based on the
formulae of \citet{car98} as given in their equations (33) and (37).
In addition to the Lick indices, the 4000 {\AA} break has been
measured using the narrow definition of D$_n$4000 given in
\citet{bal99}, with errors calculated from \citet{car98} equation
(38).  Lick index line strengths for each stacked spectrum are given
in Tables \ref{indices_table1}--\ref{indices_table3}.

Five Lick indices are excluded from this analysis.  These include the
wide Balmer indices H$\gamma_A$ and H$\delta_A$, which are designed to
measure the broad Balmer wings of A stars and are not relevant for the
analysis of old ($>1$ Gyr) stellar populations.  Also excluded are the
reddest iron lines Fe5709 and Fe5782 because they are weaker features
than the bluer iron lines and therefore represent noisier
measurements.  Finally, we exclude the iron line Fe5015 because it is
strongly contaminated by [O\textsc{iii}]$\lambda$5007 emission in the
LINER-like galaxies.

In order to have a uniform sample, the stacked spectra are all
smoothed to $\sigma = 300$ km s$^{-1}$ to match the $\sigma$ of the
most massive galaxies (see \S\ref{composite_spectra}).  The combined
resolution of $\sigma = 300$ km s$^{-1}$ and the native SDSS
resolution is at {\it lower} resolution than the Lick/IDS instrument
resolution at which the Lick indices are defined.  The index
measurements must therefore be corrected back to effective Lick/IDS
resolution.  This correction is performed by {\bf Lick\_EW} using
velocity dispersion corrections determined from smoothed SSP spectra
from \citet[Table A2a]{sch07}.  The indices reported here are
therefore effectively at Lick/IDS resolution.  However, no zeropoint
correction is performed to place them on the original Lick/IDS system.
This is for two reasons.  Firstly, the Lick standard stars are not
observed by the SDSS---they are too bright---thus the data needed to
make this correction do not exist.  More importantly, the SSP models
which we use to interpret the index measurements in \S\ref{ssp_models}
are defined using flux-calibrated spectra and are themselves not
zeropoint-shifted to the Lick/IDS system.  Because the SDSS spectra
are also flux-calibrated, any zeropoint offset between the indices
measured in SDSS spectra and the models should be small (see
discussion in \citealt{sch07}).

Figure \ref{indices} shows the indices measured in the stacked spectra
as a function of the mean $\sigma$ in each bin.  Colors correspond to
the emission line properties of the galaxies as follows: quiescent
galaxy bins are shown in black, those with weak LINER-like emission in
green, and those with strong LINER-like emission in red.  In each of
the panels a--d, the left column shows the Lick index line strengths
measured in the stacked spectra.  The error bars include statistical
errors due to finite $S/N$ only---no systematic errors are included.
The lines show least squares fits of index strength onto $\sigma$ for
each bin in emission line strength.

In \S\ref{completeness}, we demonstrated that the lower $\sigma$ bins
are significantly incomplete at faint magnitudes.  A correction for
this effect, described in detail in Appendix \ref{lowsig_corr}, is
applied to the measured Lick index line strengths; these corrected
line strengths are shown in the right columns of panels a--d in Figure
\ref{indices}.  For the moment, we note merely that the correction is
a small effect in all indices and has no qualitative effect on the
following discussion.

No attempt has been made to correct the line strengths for variations
in aperture.  The range of redshifts in this sample spans a factor of
30\% in angular diameter distance, but because the redshift
distributions of quiescent, weak, and strong LINER-like galaxies are
similar, the differences in mean physical aperture radius between the
samples under comparison are negligible.  

\subsection{Line Strength Trends with $\sigma$ and Emission
  Properties}\label{linestrength_trends} 

In SSP models with ages over 1 Gyr, Balmer line strengths {\it
decrease} with increasing SSP age.  In the same models, the 4000 {\AA}
break (D$_n$4000), the CH band (G4300), and all of the metal line
strengths {\it increase} with increasing SSP age (\citealt{wor94},
\citealt{wor97}, \citealt{bru03}, \citealt{tho03}, \citealt{sch07}).
However, line strengths are also affected by the metallicity of the
SSP.  In the same models, increasing the metallicity changes the line
strengths in the same direction as increasing the SSP age: Balmer line
strengths {\it decrease} while D$_n$4000, G4300, and metal line
strengths {\it increase} with increasing SSP metallicity.  These
effects are summarized in Table \ref{stellarpop_indices} for those
unfamiliar with SSP model results.

Figure \ref{indices}a shows the Balmer lines H$\beta$, H$\gamma_F$,
and H$\delta_F$, as well as D$_n$4000 and G4300.  All of these indices
vary systematically with $\sigma$ such that, as $\sigma$ increases,
the Balmer lines get weaker and D$_n$4000 and G4300 get stronger.
Panels b--d in Figure \ref{indices} show line strengths for the metal
lines.  These indices show the same trends as D$_n$4000, and G4300:
composite galaxy spectra with higher $\sigma$ have {\it stronger}
metal absorption lines than those with low $\sigma$.  The stellar
populations of the galaxies presented here vary systematically with
$\sigma$, as has been seen previously by many authors.

It is also clear from Figure \ref{indices} that, at fixed $\sigma$,
galaxies with strong LINER-like emission (red points) typically have
stronger Balmer line absorption and weaker D$_n$4000, G4300, and metal
lines than galaxies with weak LINER-like emission (green), which in
turn have stronger Balmer lines and weaker D$_n$4000, G4300, and metal
lines than the quiescent galaxies (black).  This suggests that, in
addition to a relationship between the stellar populations and
$\sigma$, stellar populations also vary with the emission line
strength of the galaxy.  This trend exists not only in the Balmer
lines, which are sensitive to errors in the emission infill correction
made in \S\ref{infill_correction}, but also in D$_n$4000, G4300, and
the metal lines which are not affected by contamination from either
the emission or the chosen emission infill correction.

The sodium line NaD is the one exception to the trend that metal lines
weaken as LINER-like emission strength increases; NaD absorption
strengths {\it increase} with increasing LINER-like emission and
typically vary by several tenths of an {\AA} between quiescent and
strong emission galaxies at fixed $\sigma$.  NaD is a resonance line
and is known to be significantly affected by interstellar absorption.
In their sample of Galactic globular clusters, \citet{gor93} assume
that up to 0.8 {\AA} of the observed NaD line strengths is due to
interstellar rather than stellar absorption.  If the intrinsic {\it
stellar} NaD line strength in galaxies actually follows the behavior
of all the other metal lines and {\it decreases} with increasing
emission, then the observed {\it increase} of NaD absorption with
emission suggests that LINER-like galaxies contain more cold
interstellar material than their quiescent counterparts, which serves
to enhance the observed NaD absorption.  This is discussed in greater
detail in \S\ref{ages}.

The trends visible in Figure \ref{indices} (with the exception of NaD,
as discussed above) are consistent with the hypothesis that stars in
galaxies with higher $\sigma$ are typically {\it older} than those in
lower-$\sigma$ galaxies, and that at fixed $\sigma$ galaxies with
LINER-like emission have {\it younger} stars on average than their
quiescent counterparts.  However, as shown in Table
\ref{stellarpop_indices}, increasing SSP age changes all index
strengths in the same direction as increasing SSP metallicity.  This
means that the observed trends with $\sigma$ and emission strength may
be trends in age, trends in metallicity, or some combination of the
two.  Fortunately, although the response of the various lines to
changes in age is qualitatively similar to the response to changes in
metallicity, there are quantitative differences: the Balmer lines are
more sensitive to changes in age and less sensitive to changes in
metallicity than are the metal lines.  By comparing SSP model
predictions to pairs of measured line strengths that differ
quantitatively in their sensitivity to age and metallicity, we can
break the ``age-metallicity degeneracy.''  This is done in
\S\ref{ssp_models}.

\subsection{Effect of the Infill Correction on Balmer Line
  Strengths}\label{infill_effect} 

Figure \ref{indices} shows that galaxies with LINER-like emission
typically have stronger Balmer absorption lines than quiescent
galaxies at the same $\sigma$.  However, the strength of this effect
depends critically upon the assumptions made in
\S\ref{infill_correction} in correcting for emission infill in these
galaxies.  It is instructive to examine the total effect of the
emission infill corrections to understand their significance for Balmer
absorption line strengths.

Figure \ref{balmer_comp}a-c plots the Balmer lines as measured in
the composite spectra without any emission infill correction.  As in
Figure \ref{indices}, black, green, and red data points show
quiescent, weak, and strong LINER-like emission galaxy bins,
respectively.  The emission infill is a large effect in H$\beta$,
showing significant infill from emission in both weak and strong
LINER-like galaxies.  However, the infill in H$\gamma_F$ is
substantially less, and is least significant in H$\delta_F$ due to
the Balmer decrement.  Even without any emission infill correction at
all, H$\delta_F$ still shows a trend of stronger Balmer absorption in
stronger LINER-like galaxies.  Moreover, making no infill correction
at all is obviously wrong, as can be seen by the substantial infill in
the uncorrected H$\beta$.  Any infill correction will only make the
trend in H$\delta_F$ stronger.

Figures \ref{balmer_comp}d-f and \ref{balmer_comp}g-i show the
Balmer lines measured with the two different assumptions for the
Balmer decrement that were described in \S\ref{infill_correction}:
H$\alpha$/H$\beta = 4.1$, as measured by \citet{yan06}, and
H$\alpha$/H$\beta = 3.27$, as seen in the \citet{ho97} red sequence
LINERs.  The choice of Balmer decrement makes some difference in the
H$\beta$ measurement (as can be seen comparing Figure
\ref{balmer_comp}f and \ref{balmer_comp}i), but it is a small effect for
H$\gamma_F$ (Figure \ref{balmer_comp}e and \ref{balmer_comp}h) and is
practically negligible for H$\delta_F$ (Figure \ref{balmer_comp}d and
\ref{balmer_comp}g).

The infill correction H$\alpha$/H$\beta = 4.1$ gives H$\beta$
absorption line strength values that cross at high $\sigma$; at
$\sigma < 190$ km s$^{-1}$ galaxies with LINER-like emission have
stronger H$\beta$ absorption than their quiescent counterparts, but
above $\sigma = 220$ km s$^{-1}$, the H$\beta$ absorption in the
LINER-like galaxies is weaker than in the quiescent galaxies.  In
\S\ref{ssp_models} we will show that this results in age trends that
reverse for galaxies with $\sigma < 190$ (where LINER-like galaxies
are younger than the quiescent galaxies) versus those with $\sigma >
220$ (where LINER-like galaxies are older than the quiescent
galaxies).  It is difficult to find a physical justification for this
reversal of the trend in H$\beta$ (and therefore age) with emission
strength, and furthermore this reversal is not corroborated by
H$\gamma_F$ or H$\delta_F$; the simpler explanation is that assuming a
Balmer decrement of H$\alpha$/H$\beta = 4.1$ {\it underestimates} the
appropriate emission infill correction.  Figure \ref{balmer_comp}f
shows that the reversal described here is hardly a significant effect
but would be larger for even larger values of the Balmer decrement;
therefore this argument suggests that H$\alpha$/H$\beta = 4.1$ is an
upper limit on the Balmer decrement.

Using H$\alpha$/H$\beta = 3.27$ gives consistent trends with
LINER-like emission strength versus $\sigma$ for all Balmer lines,
with H$\beta$ line strengths that converge to a single value for all
emission strengths at high $\sigma$.  If the intrinsic Balmer
decrement for these LINER-like galaxies is similar to that of AGN
(which have H$\alpha$/H$\beta = 3.1$, \citealt{ost05}), then this is
very close to the theoretical (unreddened) lower limit on the Balmer
decrement.  It is therefore likely that these two values enclose the
true mean value, and $3.27 <$ H$\alpha$/H$\beta < 4.1$ for these
systems.

The assumption that LINER-like emission is similar to AGN emission is
not necessarily a legitimate one.  Emission from stellar-ionized
H\textsc{ii} regions has H$\alpha$/H$\beta = 2.86$ in the absence of
dust absorption \citep{ost05}.  For the sake of completeness, Figure
\ref{balmer_comp} also shows the Balmer lines with an infill
correction assuming this theoretical H\textsc{ii} region value. The
strength of H$\beta$ changes noticeably, but the effect on H$\delta_F$
is within the measurement errors.  In summary, H$\alpha$/H$\beta =
3.27$ appears to be a reasonable value for the mean reddened Balmer
decrement and it is almost certain that $2.86 <$ H$\alpha$/H$\beta <
4.1$.  From Figure \ref{balmer_comp}, it is clear that the exact value
of the mean H$\alpha$/H$\beta$ has some effect on H$\beta$ line
strengths but does not qualitatively changing the trends described in
\S\ref{linestrength_trends} above, and that H$\gamma_F$ and
H$\delta_F$ are only marginally affected by the exact value of the
decrement.

\section{Stellar Population Modeling: Procedures and Caveats}\label{modeling}

\subsection{SSP Models}\label{ssp_models}

Stellar population models make it possible to convert measured index
strengths into fundamental stellar population parameters (i.e., age,
[Fe/H], and elemental abundance ratios).  In this paper, we have
chosen to use the models of Schiavon (2006; hereafter S06), which are
based on fitting functions.  Although the models produce SSPs rather
than more complicated star formation histories, they include the
effects of multiple individual elemental abundances in addition to
mean age and [Fe/H].

The S06 models calculate predicted Lick index values for a model SSP
with a specified abundance pattern.  Non-solar abundance patterns
change the SSP model through two different channels.  The first is the
effect on the isochrone of a non-solar abundance pattern, which
changes the temperature and luminosity of the main sequence turn-off
and red giant branch.  The S06 models incorporate this effect
(although in a limited way) by providing the option to construct the
SSP model based on one of two isochrones: the solar-scaled isochrone
of \citet{gir00} or an $\alpha$-enhanced isochrone from \citet{sal00}
with average [$\alpha$/Fe] = +0.42.\footnotemark \footnotetext{The
\citet{sal00} $\alpha$-enhanced isochrone was computed using the
following abundances: [O/Fe] = +0.50, [Ne/Fe] = +0.29, [Mg/Fe] =
+0.40, [Si/Fe] = +0.30, [S/Fe] = +0.33, [Ca/Fe] = +0.50, [Ti/Fe] =
+0.63, [Ni/Fe] = +0.02.  All other abundances are solar.}  The second
way in which non-solar abundances enter the SSP models is through
their effect on the various line opacities of stellar atmospheres.  In
modeling line opacities, the S06 models take into account the effects
of the elements C, N, O, Mg, Ca, Na, Si, Cr, and Ti.  Stellar line
opacities due to various elements and molecules may lie within the
index passband of a Lick index, enhancing the strength of the measured
index.  They may also lie in the blue and red continuum regions on
either side of the index against with the index absorption strength is
measured.  In this latter case, increased absorption {\it weakens} the
measured Lick index strength.  The \citet{kor05} index response
functions are used to calculate the effect on the model Lick indices
due to each of the elements listed above.

It should be emphasized that the S06 models are computed {\it at fixed
[Fe/H]}, rather than at fixed total metallicity ([Z/H]), as is the
case in many other stellar population models.  This has several
benefits.  The first is one of simplicity and conceptual transparency.
For models computed at fixed [Z/H], changing the $\alpha$-enhancement
of the model also changes [Fe/H].  Total metallicity is in fact
dominated by oxygen (an $\alpha$-element), so that changing
[$\alpha$/Fe] a small amount results in a large change in [Fe/H] in
order to keep [Z/H] constant.  It is therefore conceptually more
complicated to keep track of how enhancing [$\alpha$/Fe] affects
stellar populations because one must disentangle the effects of
increasing $\alpha$-elements from the effects of decreasing [Fe/H].
One case where this is particularly important is that of the higher
order Balmer lines, which are strongly affected by Fe \textsc{i}
absorption lines in their continuum \citep{tho04}.  In fixed
[Z/H] models, changing [$\alpha$/Fe] dramatically changes [Fe/H],
which in turn has a strong effect on H$\gamma_F$ and H$\delta_F$.
Because the S06 models are computed at fixed [Fe/H], the model
predictions for H$\gamma_F$ and H$\delta_F$ are relatively insensitive
to [$\alpha$/Fe].  This is discussed in detail in \S\ref{hd_ages}.

A second benefit of computing models at fixed [Fe/H] rather than fixed
[Z/H] is that oxygen, which dominates total metallicity, is
notoriously difficult to measure in the integrated light of stellar
populations.  While [O/H] remains essentially unmeasured, it is safer
not to claim knowledge of total [Z/H].  This will be discussed in more
detail in \S\ref{o_fe} and \S\ref{abuns}.  A related point is that
[Fe/H] is a parameter that is more directly related to measurable
quantities (i.e., iron absorption lines such as Fe5270 or Fe5335) than
is [Z/H], which depends on the abundances of many different elements.
Some of these elements are not measured by any stellar absorption
line, including the dominant element oxygen.

The S06 models take as input a set of abundance ratios and calculate
Lick index values for the given abundances at a range of ages and
[Fe/H]. The model produces grids of Lick index values for ages between
1.2 and 17.7 Gyr, and $-1.3 < $[Fe/H]$ < +0.2$ for the solar-scale
isochrone and $-0.8 < $[Fe/H]$ < +0.3$ for the $\alpha$-enhanced
isochrone.  The model grids can be used with index-index plots in the
style pioneered by \citet{wor94} to determine the age and [Fe/H] of an
observed stellar population.  Because the various Lick indices have
different sensitivities to each elemental abundance, a best-fitting
set of abundances can be found by tweaking the input abundances in the
S06 models until all of the index-index plots give consistent values
for the age and [Fe/H] of the stellar population in question.

The S06 models are limited to a subset of the Lick indices due to the
restricted wavelength coverage of the stellar spectra used to
construct the fitting functions.  Only the Balmer indices (H$\beta$,
along with both the broad and narrow H$\gamma$ and H$\delta$ indices),
Fe4383, Fe5015, Fe5270, Fe5335, Mg$_2$, Mg {\it b}, G4300, C$_2$4668,
CN$_1$, CN$_2$, and Ca4227 are modeled.  This set of indices makes it
possible to constrain the abundances [Mg/Fe], [C/Fe], [N/Fe], and
[Ca/Fe], but [O/Fe], [Na/Fe], [Si/Fe], [Cr/Fe], and [Ti/Fe] are
relatively unconstrained.  In our analysis, we predominantly use the
solar-scaled isochrone and set [O/Fe] to match the isochrone
(i.e. [O/Fe] = 0.0).  We also show results computed using the
$\alpha$-enhanced isochrone and with [O/Fe]=+0.3 for comparison.
Because Na, Si, and Ti are $\alpha$-elements, we force their
abundances to follow [Mg/Fe], while the iron-peak element Cr tracks Fe
([Cr/Fe] = 0.0).

\subsection{Fitting for Ages and Abundances}\label{fitting}

Fitting for the best set of input abundance ratios is done using the
IDL code package {\bf EZ\_Ages} (Graves \& Schiavon, in preparation).
Briefly, {\bf EZ\_Ages} implements the method developed by
\citet{sch07} to determine abundances and ages of stellar populations.
It uses an index-index plot of H$\beta$ and $\langle$Fe$\rangle$
(defined by $\langle$Fe$\rangle$ = $0.5$(Fe5270 + Fe5335)) to measure
a ``fiducial'' age and [Fe/H].  These indices are chosen because they
are relatively insensitive to other, non-solar abundance ratios.  {\bf
EZ\_Ages} then uses an index-index plot of H$\beta$ and Mg {\it b} to
constrain [Mg/Fe].  This is done by varying [Mg/Fe] as input to the
S06 model until the age and [Fe/H] measured in the H$\beta$-Mg {\it b}
plot match the fiducial values from the H$\beta$-$\langle$Fe$\rangle$
plot.  In similar fashion, C$_2$4668 is used to determine [C/Fe],
CN$_2$ is used to determine [N/Fe] once [C/Fe] is set, and Ca4227 is
used to get [Ca/Fe].  The entire process is then iterated in order to
produce a self-consistent set of abundances.  {\bf EZ\_Ages} then
measures separately an age from H$\gamma_F$ and an age from
H$\delta_F$ using the best-fitting abundances.  The result is a
measurement of [Fe/H], [Mg/Fe], [C/Fe], [N/Fe], and [Ca/Fe], as well
as ages measured from each of the three Balmer lines, for each set of
measured Lick indices.

This abundance fitting must be done in a proscribed order.  This takes
into account the complicated effects of the various element abundances
on the Lick indices used in the fitting procedure.  The goal is to
only introduce one new abundance ratio at a time.  The Mg {\it b}
index is sensitive to age, [Fe/H], and [Mg/Fe], but is relatively
insensitive to all other abundance ratios (\citealt{kor05},
\citealt{ser05}).  Therefore, once the fiducial age and [Fe/H] have
been determined, Mg {\it b} can be used to fit for [Mg/Fe].  The
abundance ratios [C/Fe], [N/Fe], and [Ca/Fe] are more difficult to
isolate.  The [N/Fe] abundance ratio is difficult because its effect
is seen most strongly in the CN$_1$ and CN$_2$ indices where [C/Fe]
and [O/Fe] are also important factors; the Ca4227 index is highly
sensitive to CN, as well as being sensitive to [Ca/Fe] because of a
strong CN band in the blue continuum region around the index
(described in detail in \citealt{pro05}); and [C/Fe] can best be
isolated in the C$_2$4668 index, however this index is also sensitive
to [O/Fe] (\citealt{kor05}, \citealt{ser05}).  As mentioned above,
oxygen is very difficult to measure in unresolved stellar populations.

In the {\bf EZ\_Ages} abudance fitting process, [C/Fe] is set first by
fitting the C$_2$4668 index, because C$_2$4668 depends only on age,
[Fe/H], [C/Fe], and [O/Fe].  Then CN$_2$ can be used to fit [N/Fe]
because [C/Fe] is already determined.  Only once [C/Fe] and [N/Fe] are
determined can Ca4227 be used to fit for [Ca/Fe], in order to account
for the strong dependence of Ca4227 on CN.  The fact that [O/Fe]
remains unknown contributes a significant uncertainty to all of these
abundance determinations, but is unavoidable.  When presenting
abundance measurements for our data in \S\ref{abuns}, we will
illustrate the effect of changing [O/Fe] by 0.3 dex.  In practice,
changing [O/Fe] by this amount appears to have no effect on [Fe/H] or
[Mg/Fe], some effect on [C/Fe], a substantial effect on [N/Fe], and
almost no effect on [Ca/Fe].

In order to illustrate the effect of elemental abundances on different
line indices and our abundance-fitting process, we show in Figures
\ref{grid1_a} and \ref{grid1_b} a set of index-index plots for all of the Lick
indices included in the S06 model.  In each plot, solid lines connect
models at fixed [Fe/H], while the dotted lines connect models of
constant age.  The three data points are from the stacked spectra of
quiescent galaxies with $\sigma = 70$--120 km s$^{-1}$, $\sigma =
145$--165 km s$^{-1}$, and $\sigma = 220$--300 km s$^{-1}$ plotted as
the triangle, star, and square, respectively.  The index measurement
errors are smaller than the size of the plotting symbols.

The model shown here is computed with the {\bf EZ\_Ages} output
best-fitting abundances for the $\sigma = 70$--120 km s$^{-1}$ stacked
spectrum: [Mg/Fe] = 0.120, [C/Fe] = 0.100, [N/Fe] = -0.002, and
[Ca/Fe] = 0.010, calculated with a solar-scale isochrone, with solar
[O/Fe], with Na, Si, and Ti abundances set to equal [Mg/Fe], and with
solar [Cr/Fe].  Figure \ref{grid1_a} shows the Lick indices which are
used by {\bf EZ\_Ages} in the abundance fitting.  Note that the
$\sigma = 70$--120 km s$^{-1}$ data point (triangle) which was used in
the abundance fitting falls in the same place with respect to the grid
lines in each index-index plot.  This is the essence of the abundance
fitting process and signifies that a mutually self-consistent set of
stellar population parameters and abundance ratios has been found.

A check on the robustness of the abundance results is to compare these
index-index plots with those in Figure \ref{grid1_b}, which were {\it
not} used in the abundance fitting process.  The location of the
$\sigma = 70$--120 km s$^{-1}$ data point in each grid in Figure
\ref{grid1_b} is consistent with its position in the grids of Figure
\ref{grid1_a}, except for the G4300 index which is too strong in the
best-fitting model and H$\gamma_F$ which is too weak in the
best-fitting model.  A similar discrepancy in G4300 was noted in
\citet{sch07} in comparisons with Galactic clusters M67 and NGC 6528.
\citet{san06} also show anomalous behavior in the G4300 index with
respect to C$_2$4668 in their sample of early-type galaxies.  The
G4300 index in globular clusters appears to plateau for iron
abundances with [Fe/H] $\geq -0.7$; above this [Fe/H] the G band
feature continues to increase with metallicity but absorption lines in
the continuum against which the G4300 line strength is measured also
increase in strength in such a way as to maintain relatively constant
values of G4300 over a range of metallicity.  The C$_2$4668 index is
in a region of the spectrum where it is possible to define continua
which are less affected by absorption.  We thus consider C$_2$4668 to
be a more reliable indicator of carbon enhancement than G4300.  As for
H$\gamma_F$, we will show in \S\ref{hd_ages} that H$\gamma_F$ and
H$\delta_F$ generally give {\it younger} age estimates than H$\beta$
based on a SSP model and discuss the implications of this effect.

Unlike the $\sigma = 70$--120 km s$^{-1}$ data, the $\sigma = 145$--
165 km s$^{-1}$ and $\sigma = 220$--300 km s$^{-1}$ data points (star
and square) do not fall in the same location on each grid.  The Mg,
CN, and C$_2$4668 lines are particularly discrepant.  This is because
the grids were constructed to match the $\sigma = 70$--120 km s$^{-1}$
data.  In order to produce model grids that match these higher
$\sigma$ data points, [Mg/Fe], [C/Fe], and [N/Fe] must be increased
(which shifts the model grids for the Mg, CN, and C$_2$4668 indices in
Figures \ref{grid1_a} and \ref{grid1_b} to the right, toward the data).
This suggests that higher $\sigma$ galaxies have larger [Mg/Fe],
[C/Fe], and [N/Fe].  In \S\ref{abuns} we show that this is indeed in
the case.

For comparison, Figures \ref{grid6_a} and \ref{grid6_b} show
index-index plots computed with a different set of abundance ratios.
Here we have used the best-fitting abundances for the $220 < \sigma <
300$ km s$^{-1}$ data, which have [Mg/Fe] = 0.260, [C/Fe] = 0.270,
[N/Fe] = 0.220, and [Ca/Fe] = 0.040.  As before, the other
$\alpha$-elements (Na, Si, and Ti) follow Mg, while O and Cr have
solar abundances.  In Figure \ref{grid6_a}, the $220 < \sigma < 300$
km s$^{-1}$ data (squares) give the same age and [Fe/H] values in each
index-index plot.  Because this model has enhanced [Mg/Fe], [C/Fe],
and [N/Fe], the plots of H$\beta$ against Mg $b$, C$_2$4668, and
CN$_2$ are now consistent with the H$\beta$-$\langle$Fe$\rangle$ plot
for the $220 < \sigma < 300$ km s$^{-1}$ data.  In contrast, the lower
$\sigma$ data points in these plots lie at values of [Fe/H] that are
too low to match those from the H$\beta$-$\langle$Fe$\rangle$ plot.
Figure \ref{grid6_b} shows the indices that are not used in the
abundance fitting.  As before, the plots of H$\beta$ against Fe4383,
Mg$_2$, and CN$_1$ are consistent for the $220 < \sigma < 300$ km
s$^{-1}$ data, while the G4300 index produces ages that are too young.
Here the ages measured from H$\gamma_F$ and H$\delta_F$ are somewhat
younger than those measured from H$\beta$.

An important caveat applies to the ages and abundances presented in the
following sections.  The S06 model fitting functions are constructed
from flux-calibrated stellar spectra with no zeropoint correction to
bring them onto the original Lick/IDS system.  The SDSS spectra used
in this analysis are also flux-calibrated, so zeropoint offsets
between the model index system and the data should be small.  However,
as with most stellar population models based upon index measurements,
{\it relative} values of stellar population parameters are more
reliable than absolute measurements due to the zeropoint
uncertainties.

\subsection{Oxygen Abundances in Old Stellar Populations}\label{o_fe}

It is unfortunate that it has not been possible thus far to measure
oxygen abundances in unresolved stellar populations.  Not only is
oxygen the single most abundant metal in the Universe, but main
sequence turn-off temperatures are substantially affected by oxygen
abundances \citep{van01}.  An inability to determine the oxygen
abundance therefore means that the total metallicity is unknown and
the true isochrone shape is not well-known, raising uncertainties in
stellar population analysis.

Theoretical models of nucleosynthesis predict that Mg and O both
originate in Type II SNe (e.g. \citealt{whe89}, \citealt{woo95}) and
should track one another closely, but there is some evidence that this
is not always the case.  In a survey of 27 Milky Way bulge giants,
\citet{ful06} find that, although Mg appears to be enhanced at all
[Fe/H], [O/Fe] declines with increasing [Fe/H] and is solar or mildly
sub-solar at [Fe/H] $\ge 0.0$.  This indicates that, at least in some
stellar populations, the assumption that O varies in lockstep with Mg
is not valid.  In addition, there is growing evidence from X-ray
observations that the hot gas in elliptical galaxies has subsolar
values of [O/Fe] (\citealt{hum06}, \citealt{hum04} and references
therein) and also subsolar [O/Mg] \citep{hum06}, further undermining
the assumption that Mg and O vary together.

In light of these uncertainties, we treat [O/Fe] as undetermined.  The
S06 SSP models are computed for fixed [Fe/H] rather than fixed [Z/H]
so the uncertainty in total metallicity should not have a large effect
on the results of the stellar population modeling.  In the following
sections, we present results based on both the solar abundance and
$\alpha$-enhanced isochrones available with {\bf EZ\_Ages}, and at each
stage of the analysis illustrate the effect of changing the oxygen
abundance by exploring a range of values for [O/Fe].

\section{Stellar Population Modeling Results: Ages and
  Abundances}\label{results} 

\subsection{SSP Ages as a Function of $\sigma$ and Emission Line
  Strength}\label{ages}
 
The {\bf EZ\_Ages} code was used to estimate mean luminosity-weighted
ages and abundances from Lick index measurements for each of the
stacked spectra.  Using the best-fitting abundance models, we obtained
ages for each stacked spectrum, measured separately from the
H$\beta$-$\langle$Fe$\rangle$, H$\gamma_F$-$\langle$Fe$\rangle$, and
H$\delta_F$-$\langle$Fe$\rangle$ grids.  We shall hereafter refer to
these separate age measurements as ``H$\beta$ ages'', ``H$\gamma_F$
ages'', and ``H$\delta_F$ ages'', with the understanding that these
indicate the ages as measured from index-index plots using the named
Balmer line.  Figure \ref{ages_plot} shows these age measurements as a
function of the mean $\sigma$ in each galaxy bin and emission line
strength, for each of the stacked spectra.  The left panel shows the
measured ages when a Balmer decrement of H$\alpha$/H$\beta$ = 3.27 is
assumed, while the center panel assumes H$\alpha$/H$\beta$ = 4.1.  In
both cases, the {\bf EZ\_Ages} fitting has been computed with a
solar-scaled abundance isochrone.  The right panel shows the effect of
using the $\alpha$-enhanced isochrone and [O/Fe] = +0.3 in the {\bf
EZ\_Ages} fitting process.  The H$\beta$ ages, H$\gamma_F$ ages, and
H$\delta_F$ ages are plotted separately in each panel, from top to
bottom.  The error bars show the uncertainty in the ages due to
measurement errors in the Balmer lines and $\langle$Fe$\rangle$ only.
The solid lines are linear least squares fits of age onto $\sigma$.

The most striking result in this figure is that the galaxies with
strong LINER-like emission are systematically {\it younger} than their
quiescent counterparts at the same $\sigma$, with weak LINER-like
galaxies falling in between.  This trend exists in all of the age
measurements, regardless of which Balmer line is used, and for either
Balmer decrement.  As expected from the line strength variations shown
in Figure \ref{balmer_comp}, the H$\beta$ ages are most affected by
the chosen decrement, while the H$\delta_F$ ages are essentially
unaffected.  In \S\ref{infill_effect}, we argued that the two values
of the Balmer decrement used here likely represent lower and upper
boundaries for reasonable estimates of H$\alpha$/H$\beta$.  Thus the
panels of Figure \ref{ages_plot} show the range of possible SSP age
differences between quiescent and LINER-like galaxies.  The use of an
$\alpha$-enhanced isochrone with [O/Fe] = +0.3 (right panel) results
in H$\beta$ ages that are consistently younger than those measured
using a solar abundance isochrone but does not change the trend in age
with emission strength.

The difference in measured age between quiescent and strongly emitting
galaxies is substantial---ranging from around 3.5 Gyr in the lowest
$\sigma$ bin to 2 Gyr in the highest $\sigma$ bin with
H$\alpha$/H$\beta$ = 3.27 and a solar-scaled isochrone, as measured in
all Balmer lines.  Galaxies with LINER-like emission contain stars
which are on average 10--40\% {\it younger} than those in quiescent
galaxies of the same $\sigma$ (using H$\beta$ ages).

A second effect that is apparent in Figure \ref{ages_plot} is that
galaxies with high $\sigma$ are on average {\it older} than those with
low $\sigma$.  This trend is apparent in ages measured by all Balmer
lines, independent of the chosen Balmer decrement, and for quiescent
as well as emission line galaxies (with the exception of H$\delta_F$
ages in quiescent galaxies which are $\sim 7$ Gyr, regardless of
$\sigma$).  This suggests that more massive galaxies formed a larger
percentage of their stars in the distant past than did less massive
galaxies, which seems to be consistent with the so-called
``downsizing'' phenomenon (\citealt{cow96}, see also \citealt{ber03b},
\citealt{nel05}, \citealt{tho05}, \citealt{tre05}, and many others).

Interestingly, the \citet{sch07} analysis of SDSS stacked spectra from
\citet{eis03} shows no increase in H$\beta$ age with galaxy absolute
magnitude but a definite increase in H$\delta$ age, using the same
models as in this work.  That sample is selected predominantly by
morphology and color, with no selection applied based on emission line
properties.  Because of this, the \citet{eis03} sample likely contains
not only quiescent and LINER-like galaxies, but a small contaminating
fraction of galaxies with Seyfert and H\textsc{ii} region emission as
well.  The emission infill correction applied by \citet{sch07} is
inaccurate for these galaxies, and appears to cause the discrepancy in
the trends seen in \citet{sch07} with the results presented here.
Appendix \ref{eis_data} presents a detailed comparison between
\citet{sch07} and this work.

The stellar population modeling described above shows that the
LINER-like galaxies have younger ages than their quiescent
counterparts at fixed $\sigma$; it is natural to assume that they
should therefore have slightly {\it bluer} colors.  In fact, the
opposite is true.  As shown in Figure \ref{g_r_sig}, the LINER-like
galaxies typically have slightly {\it redder} colors than the
quiescent galaxies at the same $\sigma$.  The effect is small, ranging
up to $0.037$ mag in {\it $^{0.1}$(g-r)} but is consistent across all
galaxy bins.  If the LINER-like galaxies are indeed younger than the
quiescent galaxies at the same $\sigma$, they must also be subject to
a larger degree of intrinsic interstellar reddening, on average.  This
hypothesis is corroborated by the observation that LINER-like galaxies
appear to have stronger NaD absorption than their quiescent
counterparts, suggesting that they are subject to a greater amount of
internal interstellar absorption (see the discussion in
\S\ref{linestrength_trends}).  

More quantitatively, the younger ages measured for the LINER-like
galaxies in the spectral line fitting (Figure \ref{ages_plot}) should
make the LINER-like galaxies $\sim 0.06$ mag {\it bluer} in $B-V$,
based on the stellar population models of \citet{bru03}.  Instead,
they are observed to be $\sim 0.04$ mag {\it redder} in
$B-V$.\footnotemark \footnotetext{The $^{0.1}(g-r)$ color has been
converted to rest frame $B-V$ using Mike Blanton's {\it kcorrect}
code, v.  3\_2 \citep{bla03a} along with the empirical transformations
from the SDSS $ugriz$ magnitude system to Johnson $B$ and $V$ bands
given by \citet{jor06}.}  This implies that the typical LINER-like
galaxy is subject to internal reddening of $E(B-V) \approx 0.10$
beyond the (typically small) internal reddening of quiescent red
sequence galaxies.  This modest amount of reddening is reasonably
consistent with the value $E(B-V) = 0.05$ that was used in our
preferred computation of the Balmer decrement in
\S\ref{infill_correction}, and is substantially smaller than the value
$E(B-V) = 0.47$ that gives an upper boundary to reasonable values of
H$\alpha$/H$\beta$ above.

In the Milky Way, $E(B-V) = 0.10$ in the Galactic plane roughly
corresponds to interstellar NaD absorption of $\sim 0.35$ {\AA}
\citep{gor93}.  This is generally consistent with the excess of NaD
absorption in the LINER-like galaxies needed to bring the NaD stellar
absorption strengths into line with the rest of the metal lines shown
in Figure \ref{indices}.  The relation between $E(B-V)$ and NaD
interstellar absorption observed in the Milky Way varies substantially
outside the Galactic plane, making a calculation of $E(B-V)$ from
observed NaD absorption necessarily inaccurate.  However, it is clear
that there is at least reasonable agreement between the internal
reddening suggested by the median galaxy colors and the observed
interstellar NaD absorption in the LINER-like galaxies.  Thus,
LINER-like galaxies typically suffer from a greater degree of internal
reddening than their quiescent counterparts at fixed $\sigma$ but this
excess reddening is small, around $E(B-V) = 0.1$ on average.

\subsection{Ages Measured by H$\gamma_F$ and H$\delta_F$}\label{hd_ages}

A third effect that is visible in Figure \ref{ages_plot} is that ages
measured with different Balmer lines vary systematically, such that
bluer Balmer lines produce younger age measurements.  This is true for
the quiescent galaxies as well as the LINER-like galaxies, so it
cannot be an artifact of the emission infill correction.

The Balmer lines H$\gamma_F$ and H$\delta_F$ have been shown to be
sensitive to non-solar abundance ratios in models computed at fixed
[Z/H].  The effect of non-solar abundance ratios is strongest at high
metallicity \citep{tho04}, and is thus particularly relevant for early
type galaxies, whose metallicities are typically nearly solar or
super-solar, unlike globular clusters, which are typically
$\alpha$-enhanced but are comparitively metal-poor.  \citet{tho04}
have demonstrated that, in models computed at fixed [Z/H], including
the effects of $\alpha$-enhancement raises the age estimates from
H$\gamma_F$ and H$\delta_F$, and could thus in principal solve a
discrepancy in which the bluer Balmer lines are producing younger age
estimates than H$\beta$.

However, the sensitivity of H$\gamma_F$ and H$\delta_F$ is primarily a
sensitivity to [Fe/H], rather than [$\alpha$/Fe], and is an example of
the complication of computing models at fixed [Z/H].  The abundance
ratio dependence of H$\gamma_F$ and H$\delta_F$ is primarily due to
the large number of iron absorption lines in the regions surrounding
the H$\gamma_F$ and H$\delta_F$ indices \citep{tho04}.  For models at
fixed [Z/H], choosing a model with too low a value of [$\alpha$/Fe]
(i.e., a solar-scaled model when the galaxy is $\alpha$-enhanced)
means that the model has too {\it high} a value of [Fe/H] at fixed
[Z/H].  This makes the Fe \textsc{i} absorption lines in the continuum
regions on either side of the index passband too strong, lowering the
continuum measurement and consequently lowering the measured value of
the Balmer absorption index in the models.  This makes the observed
Balmer index look stronger in comparison to the models, and results in
a {\it younger} age measurement.  Using the correct [$\alpha$/Fe]
model fixes this problem, which has led \citet{tho04} to claim that
$\alpha$-enhancement effects cause younger ages to be measured in
higher order Balmer absorption lines.  However, the effect is
dominated by Fe \textsc{i} lines.  For models computed at fixed
[Fe/H], the effect of the iron abundance on H$\gamma_F$ and
H$\delta_F$ is accounted for, and does not depend on the
$\alpha$-enhancement of the model.  Thus the claimed
[$\alpha$/Fe]-dependence of ages measured from H$\gamma_F$ and
H$\delta_F$ is in fact a dependence on [Fe/H] that is translated into
an $\alpha$-enhancement effect in models at constant [Z/H]. The S06
models, which are computed at fixed [Fe/H] rather than fixed [Z/H] do
not have this entangled dependency.

The S06 models produce consistent age measurements from H$\beta$,
H$\gamma_F$, and H$\delta_F$ when they are used to analyze known SSPs
such as the Galactic globular clusters 47 Tuc, M5, and NGC 6528, and
the open cluster M67 \citep{sch07} which span a wide range of SSP
parameters.  Both 47 Tuc and M5 are metal-poor systems for which the
$\alpha$-enhancement effect on H$\gamma_F$ and H$\delta_F$ should be
weak \citep{tho04}, and M67 is not significantly $\alpha$-enhanced,
thus these clusters are less relevant comparisons for high metallicity
$\alpha$-enhanced galaxies.  However, NGC 6528 is relatively old, has
roughly solar [Fe/H] and is mildly $\alpha$-enhanced, with $age =
11\pm2$ Gyr, [Fe/H]$ = 0.0\pm0.15$, [Mg/Fe]$ = +0.10\pm0.04$ and
[O/Fe]$ = +0.1\pm0.05$ (\citealt{zoc04}, \citealt{bar04},
\citealt{car01}, summarized in Table 10 of \citealt{sch07}) and is
thus an excellent SSP analog for early type galaxy data.  The
S06 best-fitting model for NGC 6528 has [Fe/H]$ = -0.2$, [Mg/Fe]$ =
+0.1$, [O/Fe]$ = +0.15$, and produces H$\beta$, H$\gamma_F$, and
H$\delta_F$ ages of 13, 10, and 12 Gyr, respectively, which are
relatively consistent within the uncertainties in the models.  

In the red sequence galaxies analyzed in \citet{sch07}, however, the
models produce age measurements that decrease as bluer Balmer lines
are used.  \citet{sch07} has shown that the most likely explanation
for this phenomenon is that galaxies are not SSPs but contain stars
with a range of ages.  This effect has also been reported by
\citet{san06}; they also interpret the differing age measurements from
different Balmer lines as due to extended star formation histories.

In \citet{sch07}, four possible explanations for this phenomenon are
discussed: the presence of a sub-population of young to
intermediate-age stars, a significant number of blue stragglers, the
presence of a metal-poor old stellar population with a blue horizontal
branch, or the effects of strongly non-solar abundance ratios which
are not constrained by the available Lick indices.  Among these
scenarios, \citet{sch07} concludes that a young-to-intermediate
subpopulation is the most likely; the blue horizontal branch scenario
cannot reproduce all three Balmer lines simultaneously, the fraction
of blue stragglers required to reproduce the observed line strengths
is 10--100 times larger than blue straggler populations observed in
Galactic globular clusters, and the abundance ratios that reproduce
the observed line strengths seem extreme: [O/Fe] = +1.0, or [Ti/Fe] =
$-0.7$.

In contrast, a small population of young-to-intermediate stars is
sufficient to produce differences in Balmer line strengths similar to
those observed.  \citet{sch07} shows that a 11.2 Gyr SSP combined with
a superimposed young population with age = 0.8 Gyr and comprising
0.5--1.0\% of the total stellar mass produces measured H$\beta$ age
$\sim 5$ Gyr and H$\delta_F$ age $\sim 3.5$ Gyr.  These ages
correspond to the measured H$\beta$ and H$\delta_F$ ages of the
$\sigma = 70$--120 km s$^{-1}$ strong LINER-like emission stacked
spectrum shown in Figure \ref{ages_plot}, with H$\alpha$/H$\beta$ =
3.27, which can therefore be modeled as such a two-burst system.
This simple example serves to illustrate that measurement and
modeling of multiple Balmer lines in a single system has the
potential to reveal extended star formation histories.  More
sophisticated analysis of these galaxy spectra using composite stellar
populations is beyond the scope of this work but is clearly worthy of
future investigation.  

It is noteworthy that almost all of the stacked spectra in our
sample---quiescent and LINER-like, low $\sigma$ and high
$\sigma$---display this discrepancy in ages as measured by different
Balmer lines.  The only exception is the lowest $\sigma$ quiescent
galaxy bin, for which H$\beta$, H$\gamma_F$, and H$\delta_F$ ages are
relatively consistent at 8.4, 5.9, and 7.0 Gyr, respectively.  

One could postulate that the younger ages in bluer Balmer lines,
instead of indicating a composite stellar population, were in fact
caused by an inadequate treatment of $\alpha$-enhancement in the S06
models, although we have argued above that H$\gamma_F$ and H$\delta_F$
are strongly affected by $\alpha$-enhancement only in stellar
population models computed at fixed [Z/H].  A systematic variation of
H$\gamma_F$ and H$\delta_F$ with [$\alpha$/Fe] could produce the trend
observed in the quiescent galaxies: that the higher $\sigma$ galaxies
(which have higher [Mg/Fe], as shown below in \S\ref{abuns}) show
younger ages in H$\gamma_F$ and H$\delta_F$ than in H$\beta$, but that
all ages are nearly consistent in the lowest $\sigma$ galaxy bins
which have nearly solar abundance ratios.  However, it would be
difficult to then explain the fact that the weak and strong LINER-like
galaxies {\it do} show younger H$\gamma_F$ and H$\delta_F$ ages at
all $\sigma$.  In the next section, we will show that the weak and
strong LINER-like galaxies have virtually identical abundance ratios
to the quiescent galaxies at the same $\sigma$, thus it is difficult
to invoke abundance ratio effects to fully explain the trends in
Figure \ref{ages_plot}.

If we were to take the age trends in Figure \ref{ages_plot} at face
value, the implication is that for quiescent galaxies, those with
large $\sigma$ have a spread of stellar ages, while those a lower
$\sigma$ are relatively close to SSPs.  This would be in disagreement
with compelling work by \citet{tho05} who find more extended star
formation histories in lower $\sigma$ early type galaxies than in
higher $\sigma$ galaxies.  It would also disagree with recent evidence
from star forming galaxies at redshift $z \sim 1$ that lower mass
galaxies have more extended star formation episodes \citep{noe07}.
However, younger ages in H$\gamma_F$ and H$\delta_F$ {\it are} seen in
Figure \ref{ages_plot} for the galaxy bins with LINER-like emission,
at all $\sigma$.  If the galaxies which have experienced relatively
recent star formation episodes are predominantly found in the weak and
strong LINER-like classes, the classification by emission line
strength may have isolated the low $\sigma$ galaxies whose star
formation ceased long ago and who may therefore be more likely to
resemble SSPs.  It is not immediately clear, however, why this would
not also be the case for more massive galaxies.  A more detailed
investigation of complex model star formation histories would be
necessary to further interpret the trends seen in Figure
\ref{ages_plot}, which we defer to future work.

Detailed analysis of the age differences between Balmer lines may in
fact be pushing the data farther than is reasonable, given the
uncertainties in the stellar population modeling process and the
unknown contribution of oxygen.  In addition, any conclusion that is
strongly dependent on the $\sigma = 70$--120 km s$^{-1}$ and $\sigma =
120$--145 km s$^{-1}$ bins should be regarded with caution, as those
bins are known to be substantially incomplete (see
\S\ref{completeness}).

An important caveat that pertains to both this work and the
\citet{sch07} study is that each of these stacked spectra are
composites of the spectra of several hundred galaxies.  The
differences in ages measured from various Balmer lines may therefore
merely indicate that the galaxies included in each bin have a range of
stellar population ages, rather than that each individual galaxy
contains a composite stellar population.

\subsection{Abundances as a Function of $\sigma$}\label{abuns}

Iron abundances and elemental abundance ratios measured with {\bf
EZ\_Ages} are shown in Figure \ref{abuns_plot} as a function of
$\sigma$.  Plotted abundances are computed using a solar abundance
isochrone with H$\alpha$/H$\beta=3.27$; the abundances computed with
H$\alpha$/H$\beta=4.1$ or the $\alpha$-enhanced isochrone (not shown)
are very similar.  As before, black, green, and red data points show
results for quiescent, weak, and strong LINER-like galaxies
respectively.  The dotted line in each panel shows the solar value for
reference.  Error bars are 1$\sigma$ errors computed in {\bf EZ\_Ages}
by summing in quadrature the errors in each abundance due to
individual Lick index measurement errors.  The effect of errors in the
fiducial ages and [Fe/H] are propagated through the abundance ratio
estimates.  Graves \& Schiavon (in preparation) will discuss the error
calculations in greater detail.

Unlike the mean, lightweighted SSP ages that were shown in Figure
\ref{ages_plot}, the [Fe/H] and abundance ratios of the stacked galaxy
spectra do not appear to vary with emission line strength; at fixed
$\sigma$, quiescent and LINER-like galaxies appear to have the same
average abundance pattern within the errors.  This means that the
metal absorption line variations with emission line strength that are
apparent in Figure \ref{indices} are caused by variations in mean
galaxy age, rather than by abundance differences.

The chemical abundances do however depend on $\sigma$.  The solid
black lines in Figure \ref{abuns_plot}a--e are linear least
squares fits of the abundances onto $\sigma$.  A single line is used
to fit quiescent, weak, and strong LINER-like galaxies simultaneously.
For ease of comparison, the scale of the y-axis is the same in panels
a--e.  All abundances and abundance ratios increase with increasing
$\sigma$ such that high $\sigma$ galaxies are both more metal-rich
and more enhanced in light elements with respect to Fe than are the
low $\sigma$ galaxies.  

[Fe/H] increases with increasing $\sigma$ as [Fe/H] $\propto
\sigma^{0.45}$, and is observed to range from about 1/2 solar at
$\sigma \sim 100$ km s$^{-1}$ to slightly sub-solar at $\sigma \sim
250$ km s$^{-1}$.  Because early-type galaxies are known to have
substantial metallicity gradients (\citealt{car93}, \citealt{oga06}),
the zero-point of the [Fe/H] relation is dependent upon the aperture
used to extract the spectral data. The SDSS 1.5$''$ spectral fibers
typically cover 0.35--1.1 effective radii and thus sample a large
fraction of the galaxy light.  Nuclear spectra of early-type galaxies
(e.g. those of \citealt{tra00b} and \citealt{tho05}) should appear
substantially more metal-rich due to the metallicity gradients of the
galaxies.

The galaxies in this sample are Mg-enhanced at all $\sigma$, with
the highest $\sigma$ galaxies being the most enhanced ([Mg/Fe]
$\propto \sigma^{0.36}$).  Both [C/Fe] and [N/Fe] increase strongly
with $\sigma$, from solar values in the low $\sigma$ galaxies to
substantial C and N enhancement at high $\sigma$.  Finally, Ca
appears to follow Fe in all galaxies.  Table \ref{abun_table} gives
the slope of the solid line fit for each elemental abundance or
abundance ratio, along with the 1$\sigma$ uncertainty in the slope.

The tendency of Ca to remain relatively constant with $\sigma$, rather
than to follow Mg and other $\alpha$-elements as expected if Ca
production is dominated by Type II supernovae \citep{woo95}, has been
known for some time.  It has been seen both in the Ca4227 index, which
appears to be constant over a range of $\sigma$ (e.g.,
\citealt{vaz97}, \citealt{tho03}) and in the near-IR Ca \textsc{ii}
triplet, which may even {\it anti}-correlate with $\sigma$ (e.g.,
\citealt{sag02}, \citealt{cen03}, and references therein).  Recently,
\citet{pro05} have demonstrated that the Ca4227 index is stronly
affected by CN absorption in its blue continuum region and have
suggested that redefining the Ca4227 index to avoid this region
results in a Ca4227 index that is less sensitive to other element
abundances.  Unfortunately, the \citet{kor05} index sensitivities and
the S06 models do not include the redefined Ca4227r index.  In
principal, the abundance fitting process described in \S\ref{fitting},
which fits for [C/Fe] and [N/Fe] before using Ca4227 to fit for
[Ca/Fe], should account for the CN effect on Ca4227.  Indeed,
\citet{sch07} uses the same abundance fitting method as in this paper
and finds that [Ca/Fe] does increase with galaxy luminosity in a set
of stacked SDSS spectra.  However, the discrepancy between that result
and our Figure \ref{abuns_plot} appears to be an artifact due to
imperfect emission infill correction in the H$\beta$ feature in the
\citet{sch07} analysis.  Details of this are provided in appendix
\ref{eis_data}.

An interesting result from this Figure is that [C/Fe] and [N/Fe] show
the strongest variation with $\sigma$, with considerably steeper
slopes than [Mg/Fe].  The value of [Mg/Fe] is measured here using the
Mg {\it b} Lick index, and has been used in past works (e.g.,
\citealt{tra00a}) to trace the enhancement of $\alpha$-elements
([$\alpha$/Fe]).  If Mg is indeed a good proxy for the bulk of the
$\alpha$-elements, the relative strength of the abundance-$\sigma$
relations shows that [C/Fe] and [N/Fe] vary more strongly with
$\sigma$ than does [$\alpha$/Fe].

It has been suggested in previous work that [C/Fe] and [N/Fe] are
enhanced above solar ratios.  \citet{tra00a} prefer a model for
abundances in elliptical galaxies that has C and N enhanced along
with the $\alpha$-elements O, Ne, Na, Mg, Si, and S, but state that a
similar model with solar C is difficult to distinguish from their
preferred model based on the data.  \citet{san06} prefer an abundance
model in which the $\alpha$-elements Na, Si, Ca, O, and Ti are
enhanced at higher $\sigma$, with C enhanced but as a less steep
function of $\sigma$ than the $\alpha$-elements, with N enhanced more
steeply than the $\alpha$-elements, and with Mg enhanced most strongly
of all.  The results presented here differ from both \citet{tra00a}
and \citet{san06} in that we show [C/Fe] and [N/Fe] enhancement to
increase {\it more strongly} with $\sigma$ than does [Mg/Fe].  All
of these results are at odds with the C enhancement pattern found by
\citet{cle06}, who find that [C/H] {\it decreases} with increasing
$\sigma$.  It is clear that there is no consensus as to the
enhancement (or depletion) pattern of these light elements.

One difference in the modeling process is that the models used in our
analysis were computed at [O/Fe] = 0.0, whereas the other three works
mentioned above have oxygen enhanced along with the $\alpha$-elements.
The dashed line in Figure \ref{abuns_plot}a--e shows the effect of
fixing [O/Fe] at +0.3 during the abundance fitting process instead of
[O/Fe] = 0.0.  Both [Fe/H] and [Mg/Fe] appear unchanged and [Ca/Fe]
shows only a small difference, but [C/Fe] and [N/Fe] {\it increase} at
all $\sigma$ for the oxygen-enhanced models.  This is because oxygen
has a negative effect on the C$_2$4668 and CN index strengths used to
compute [C/Fe] and [N/Fe].  In the presence of increased oxygen, most
C will form CO rather than C$_2$ or CN, requiring larger C and N
abundances in order to produce the observed C$_2$ and CN line
strengths.  Note that although there is a vertical offset, the slopes
of the enhancement-$\sigma$ relations do not change between the
[O/Fe] = 0.0 and the [O/Fe] = +0.3 models.  If [O/Fe] increases with
$\sigma$ following [Mg/Fe] as expected if both Mg and O production
are dominated by Type II supernovae (SNe), it is clear that the high
$\sigma$ galaxies will be more enhanced above the [O/Fe] = 0.0 level
than the low $\sigma$ galaxies, causing the slopes of the
[C/Fe]-$\sigma$ and [N/Fe]-$\sigma$ relations to be even steeper.

Many stellar population models are computed at fixed [Z/H], whereas
the S06 models are computed at fixed [Fe/H]. To facilitate comparisons
between models, Figure \ref{abuns_plot}f shows [Z/H] as a function of
$\sigma$, for five different assumptions about [O/Fe] values.  [Z/H]
was computed using the solar elemental abundances of \citet{gre96}.  In
calculating [Z/H], Cr, Mn, Fe, Co, Ni, Cu, and Zn are treated as
Fe-peak elements and assumed to scale with the measured [Fe/H], while
Ne, Na, Mg, Si, S, and Ti are treated as $\alpha$-elements and assumed
to scale with the measured [Mg/Fe].  C, N, and Ca are allowed to vary
separately with their measured abundance ratios.  All other elements
are assumed to have solar abundances, excepting O.  The black lines in
Figure \ref{abuns_plot}f show [Z/H] computed assuming constant O
enhancement; [O/Fe] = 0.0, [O/Fe] = +0.3, and [O/Fe] = +0.5 are shown
as the solid, dashed, and dash-dot black lines, respectively.  The
blue lines show [Z/H] computed with [O/Fe] that varies with $\sigma$.
The solid blue line is computed assuming that O scales with the other
$\alpha$-elements ([O/Fe] = [Mg/Fe]), while the dashed blue line
has O scaling with Mg, but enhanced by a factor of two above Mg
([O/Fe] = [Mg/Fe] + 0.3).  The slopes of the relations are given
in Table \ref{abun_table}.  Unsurprisingly, a steeper relation results
when O scales with Mg because [O/Fe] increases with $\sigma$.  Figure
\ref{abuns_plot}f highlights a problem with using models computed at
fixed [Z/H]: total metallicity is dominated by O, which is highly
uncertain and not measured.

As discussed in Appendix A, Figure \ref{siglum_indices} shows that, at
fixed $\sigma$, galaxies with fainter magnitudes have weaker metal
line strengths than brighter galaxies but almost identical Balmer line
strengths, suggesting that fainter galaxies are more metal-poor than
their brighter counterparts at the same $\sigma$.  We have used the
S06 models to convert the line strengths from Figure
\ref{siglum_indices} into age and abundance estimates, following the
same procedure described above for quiescent versus LINER-like
galaxies.  This modeling (not shown) confirms that fainter galaxies
have lower values of [Fe/H] and older ages compared to brighter
galaxies at the same $\sigma$, while their abundance ratios are nearly
identical.

\subsection{Comparison with Previous Results}\label{abun_comp}

To put the results presented here into the context of existing similar
work, we have compiled results from four other SSP analyses of early
type galaxies.  Table \ref{abun_comp_table} gives the slopes of the
log $age$--$\log \sigma$, [Fe/H]--$\log \sigma$, [Z/H]--$\log \sigma$,
and [$\alpha$/Fe]--$\log \sigma$ relations reported here and in
previous work.  Interestingly, although models claiming to accurately
reproduce the effects of $\alpha$-enhancement are relatively recent,
the reported [$\alpha$/Fe] values agree better than the ages or
metallicities.  There appears to be a consensus value of
$d$[$\alpha$/Fe]$/ d \log \sigma$ = $0.32\pm0.04$.

All of the studies reviewed here find that typical galaxy ages
increase with $\sigma$, although there is substantial variation in the
reported values of $d \log age / d \log \sigma$, ranging from 0.24
\citep{tho05} to 1.15 \citep{ber03b}.  Our value of $d \log age / d
\log \sigma$ = 0.35 is toward the shallower end of the reported
trends.  It is noteworthy that the age trends shown in Figure
\ref{ages_plot} for LINER-like galaxies are steeper than those for
quiescent galaxies.  

As for $d$[Z/H]$ / d \log \sigma$, our value of $0.79\pm0.05$ computed
assuming [O/Fe]$=0.0$ (see Table \ref{abun_comp_table}) is in
excellent agreement with the \citet{tra00b} value of $0.76\pm0.13$,
and is a little steeper than the trends reported by \citet{tho05},
\citet{nel05}, \citet{smi07}, and \citet{ber03b}.  Because this work
analyzes stacked spectra, we can only quantify mean trends and thus
the age-metallicity hyperplane of \citet{tra00b} is not visible in
this work.

\section{Discussion}\label{discussion}

We have shown that, on average, red sequence galaxies with LINER-like
emission have {\it younger} SSP ages than quiescent galaxies at the
same $\sigma$; strong LINER-like galaxies are 2--3.5 Gyr younger than
their quiescent counterparts.  This suggests that there is a
connection between the star formation history of a galaxy and the
presence or absence of LINER-like emission within it.

Several caveats apply to this statement.  Most importantly, this study
is based on stacked spectra, and therefore analyzes only the mean
properties of galaxies in each bin of emission properties and
$\sigma$.  There may be a significant spread in the ages of galaxies
within each bin; not all LINER-like galaxies are necessarily younger
than all quiescent galaxies at the same $\sigma$.  \citet{tho05} show
a spread in the ages of early-type galaxies at fixed $\sigma$ which
they can match using Monte-Carlo simulations incorporating
measurements errors and an intrinsic age spread of 20--25\%.
\citet{tra00b} also show a large spread in ages at fixed $\sigma$.  We
have checked their reported ages as a function of [O\textsc{iii}]
emission line strength and confirm that the galaxies in their sample
with $> 2\sigma$ [O\textsc{iii}] emission line detections are
generally younger than their quiescent galaxies at similar $\sigma$.
It is possible that only a subset of LINER-like galaxies have younger
ages, but these must then make up a significant fraction of the
LINER-like galaxies in order to substantially lower the mean
LINER-like galaxy ages to match observations.

It is also clear from Figure \ref{ages_plot} that there cannot be a
one-to-one correlation between SSP age and LINER-like emission
strength because the mean H$\beta$ age of the highest-$\sigma$
LINER-like galaxies is {\it older} than the mean H$\beta$ age of
lower-$\sigma$ quiescent galaxies; specifically, if LINER-like
emission existed in all galaxies with ages below some threshold value,
all LINER-like galaxies would be younger than all quiescent galaxies,
regardless of $\sigma$, which is not seen.  However, this trend
disappears using H$\delta_F$, where all mean ages of LINER-like
galaxies are younger than all the H$\delta_F$ mean ages of quiescent
galaxies, even though this is not true of H$\beta$.  If, as suggested
in \citet{sch07}, higher order Balmer lines are more sensitive to the
presence of very young stars, this may imply that LINER-like galaxies
have a larger fraction of relatively younger stars than quiescent
galaxies, that is the LINER-like galaxies have had more recent star
formation than their quiescent counterparts.

This does not mean that the LINER-like galaxies are post-starburst
systems, galaxies which have undergone a recent epoch of powerful star
formation but are no longer forming stars.  Spectra of post-starburst
galaxies show strong higher-order Balmer absorption lines
characteristic of A stars.  They are often modeled by the linear
combination of an A star spectrum and a K star spectrum.  The ratio of
the linear coefficients of this two-component model ($A/K$) is used to
make the post-starburst classification.  Only 1.1\% of the LINER-like
galaxies have $A/K$ ratios from \citet{yan06} consistent with
identification as post-starburst galaxies ($A/K \geq 0.25$),
suggesting that the LINER-like galaxy population in general is more
than $\sim 0.5$--1 Gyr beyond any major recent episode of star
formation \citep{qui04}.  However, LINER-like galaxies might be ``post
post-starburst'' systems seen $\gtrsim 1$ Gyr after the post-starburst
phase, possibly a continuation of the post-starburst AGN population
seen in \citet{kau03} to older ages and lower ionization AGN.  It is
also possible that the LINER-like galaxies are undergoing low-level
star formation which is contributing to the optical emission lines.  

The interpretation of the age difference between quiescent and
LINER-like galaxies is not straightforward, due to the lack of
consensus as to the ionization mechanism driving LINER-like emission;
it is in fact likely that LINERs and LINER-like galaxies are a
heterogeneous population \citep{fil03}.  Possible mechanisms for
producing LINER-like emission are:
\begin{list}{}{}
\item[1.] Photoionization by an AGN continuum (e.g., \citealt{fer83}).
\item[2.] Photoionization by newly-formed O stars (e.g.,
  \citealt{shi92}).
\item[3.] Shock ionization (e.g., \citealt{dop95}).  The shocks could
  be driven by starburst winds, gas accretion from a satellite galaxy
  or the surrounding inter-galactic medium, episodic AGN feedback, or
  by galaxy interactions and/or mergers.
\item[4.] Cooling flows (e.g., \citealt{hec81}).
\item[5.] Photoionization at early times ($\lesssim500$ Myr, e.g.,
  \citealt{tan00}) or at late times (1--13 Gyr, e.g., \citealt{bin94})
  by post-AGB stars.
\end{list}

Of these possibilities, photoionization or shock ionization by young
stars and photoionization by post-AGB stars are the most directly
associated with the stellar population of the galaxy, and would thus
give a potentially straight-forward connection between the presence of
LINER-like emission and the star formation history of a galaxy.  In
particular, if the observed LINER-like emission in many of these
galaxies is powered by young stars, it would not be at all surprising
to discover that these galaxies have younger mean light-weighted ages.

However, there are several indications that this explanation is too
simple.  First, \citet{yan06} show that the LINER-like red sequence
galaxies in their analysis have higher observed
[O\textsc{ii}]/H$\beta$ ratios than either star-forming galaxies or
AGN (i.e. galaxies with Seyfert-like emission line ratios), and thus
LINER-like emission cannot be produced by a superposition of
H\textsc{ii} regions and AGN activity.  This does not preclude the
LINER-like emission being caused by star formation through a mechanism
other than H\textsc{ii} region emission, however H\textsc{ii} regions
produce strong H$\beta$ emission and their presence would tend to
weaken the [O\textsc{ii}]/H$\beta$ ratios of the galaxy.

There are galaxies with LINER-like emission that do show indications
of containing starbursts.  These are the IR-luminous galaxies
described in \citet{vei95} and \citet{stu06}.  These galaxies have
mid-IR SEDs and IR fine structure emission line ratios similar to
those observed in starburst galaxies \citep{stu06}.  In addition, they
have highly reddened but intrinsically blue optical continuum colors
and low Mg {\it b} EWs characteristic of relatively young stellar
populations in dusty galaxies \citep{vei95}.  We do not have mid-IR
data for the SDSS galaxies to determine whether the galaxies present
in our sample are indeed IR-luminous, but the optical continuum
colors, internal reddening (determined by observed H$\alpha$/H$\beta$
emission line ratios), and Mg {\it b} EWs of the IR-luminous
LINER-like galaxies are vastly different from those measured in the
galaxy sample presented here.  We therefore think it highly unlikely
that these IR-luminous LINER-like galaxies contribute significantly to
the LINER-like red sequence galaxies in this sample.

In the scenario where the bulk of the LINER-like emission on the red
sequence is produced by post-AGB star photoionization, a correlation
between LINER-like emission and stellar population age would imply
that either the amount of ionizing photons produced by the old stellar
population or the supply of gas (from stellar mass loss) would depend
upon the age of the stars in the galaxy, which is predicted by some
models (e.g., \citealt{mat89}). 

The other possible LINER mechanisms mentioned above (photoionization
by an AGN, shock ionization due to gas accretion, galaxy interactions,
or merger activity, and cooling flows) are not directly related to the
stellar populations of the galaxies.  To explain the correlation
between SSP age and emission properties with any of these mechanisms
requires a connection between the stellar population age of the galaxy
and some quantity external to the galaxy (accretion, interaction,
cooling flows) or internal to the galaxy but separate from the stellar
population (AGN activity).  This would also be true of a scenario in
which the post-AGB stars in the galaxy provide the ionizing photons
but where the emission strength is determined by the supply of gas to
be ionized, with the gas coming from an external source (rather than
from stellar mass loss).  

These ``external'' possibilities have interesting implications for
galaxy evolution, in particular the possibility of ionization by low
luminosity AGN.  This mechanism is favored by \citet{fil03} for
centrally-concentrated LINER emission, and has a tantalizing resonance
with recent work suggesting that AGN activity is involved in quenching
star formation in galaxies (e.g., \citealt{kau03}, \citealt{hop05},
\citealt{cro06}, and others).  However, there is some evidence that
the LINER-like emission in many if not most early-type galaxies is
extended throughout the galaxy, rather than being centrally
concentrated (e.g., \citealt{gou94}, \citealt{sar06}), suggesting that
photoionization by an AGN may not be the dominant mechanism driving
LINER-like emission in most early-type galaxies.

Unfortunately, it is outside the scope of this work to speculate as to
which mechanism (or mechanisms) dominate the observed LINER-like
emission in red sequence galaxies, making it difficult to assess the
physical cause of our observed result.  It would be particularly
interesting to perform an analysis similar to that presented here on a
sample of galaxies with spatially resolved spectra, so that true
nuclear LINERs could be analyzed separately from galaxies with
extended LINER-like emission.

\section{Conclusions}\label{conclusions}

This paper has presented an analysis of stacked spectra for a
well-defined set of SDSS red sequence galaxies.  Galaxies observed to
harbor LINER-like emission are analyzed in parallel with quiescent
galaxies, while galaxies with emission line ratios characteristic of
H\textsc{ii} regions, Seyferts, or Transition Objects have been
excluded.  Lick index absorption line strengths have been measured in
all the stacked spectra and have been used to study the stellar
population properties of red sequence galaxies as a function of galaxy
velocity dispersion and emission line properties. The main results of
this analysis can be summarized as follows:
\begin{list}{}{}
\item[1.] Red sequence galaxies with LINER-like emission (roughly
  one-third of the total red sequence) are systematically {\it
  younger} than their quiescent counterparts at similar $\sigma$ by
  several Gyr.  Despite this age difference they appear to have
  metallicities and abundance patterns similar to quiescent galaxies.
  This result is robust to the details of the emission infill
  correction, the enhancement pattern of the isochrone used to
  construct comparison models, and the (unknown) oxygen abundance of
  the galaxies.
\item[2.] Many properties of red sequence galaxies' stellar
  populations vary with galaxy velocity dispersion, including SSP age,
  [Fe/H], and abundance ratios.  Galaxies with higher $\sigma$ are
  typically older and more metal-rich than lower $\sigma$ galaxies,
  and are increasingly enhanced in Mg, C, and N.  Ca appears to scale
  with Fe at all $\sigma$.  
\item[3.] [C/Fe] and [N/Fe] vary strongly with $\sigma$; the slopes of
  the [C/Fe]--log $\sigma$ and [N/Fe]--log $\sigma$ relations are
  steeper than the [Mg/Fe]--log $\sigma$ relation.  
\item[4.] The SSP ages measured in stacked spectra from the bluer
  Balmer lines (H$\gamma_F$ and H$\delta_F$) are younger than those
  measured using H$\beta$.  This effect is not present when the S06
  SSP models are used to analyze known simple stellar populations.  If
  the interpretation presented in \citet{sch07} is correct, the
  discrepant age measurements observed in this analysis indicate the
  presence of composite stellar populations.  From the work presented
  here, it is not possible to tell whether the individual galaxies
  that go into each stacked spectrum are themselves composite stellar
  populations with similar star formation histories, or whether the
  individual galaxies are true SSPs of different ages which produce a
  composite population when they are combined together.
\end{list}
The correlation between the presence of LINER-like emission and
intrinsically younger stellar populations demonstrates that there is a
link between the mechanism driving the emission and the star formation
history of the galaxy.  If the LINER-like emission in red sequence
galaxies is primarily driven by AGN activity, this may be an
indication that AGN do indeed play a role in ending star formation
activity.  Alternatively, if the LINER-like emission in most of these
galaxies is controlled entirely by the properties of the stellar
population itself (e.g. gas lost from evolved stellar envelopes which
is ionized by UV flux from post-AGB stars), the correlation with
stellar population age may provide important constraints on models
for the UV flux from low mass stars or for the rate of mass loss in a
stellar population over time.

\acknowledgements

The authors wish to thank Michael R. Blanton, David W. Hogg, and
collaborators for making the NYU Value-Added Galaxy Catalog publicly
available.  They wish to thank William G. Mathews for valuable
discussion.  This work was made possible by support from National
Science Foundation grant AST 05-07483.

Funding for the creation and distribution of the SDSS Archive has been
provided by the Alred P. Sloan Foundation, the Participating
Institutions, the National Aeronautics and Space Administration, the
National Science Foundation, the U.S. Department of Energy, the
Japanese Monbukagakusho, and the Max Planck Society.  The SDSS Web
site is http://www.sdss.org/.

The SDSS is managed by the Astrophysical Research Consortium (ARC) for
the Participating Institutions.  The Participating Institutions are
the University of Chicago, Fermilab, the Institute for Advanced Study,
the Japan Participation Group, the Johns Hopkins University, the
Korean Scientist Group, Los Alamos National Laboratory, the
Max-Planck-Institute for Astronomy (MPIA), the Max-Planck-Institute
for Astrophysics (MPA), New Mexico State University, University of
Pittsburgh, University of Portsmouth, Princeton University, the United
States Naval Observatory, and the University of Washington.

\begin{appendix}

\section{Correcting for Incompleteness at Low $\sigma$}\label{lowsig_corr}

In \S\ref{completeness}, we showed that the magnitude limit of the
SDSS makes the galaxy sample presented in this paper significantly
incomplete in the lower $\sigma$ bins, with the lowest $\sigma$
bin missing as much as 51\% of the galaxies it should contain.
This causes the low $\sigma$ bins to be increasingly biased toward
the brightest galaxies in that bin.  

To understand the effect of this bias on the measured line strengths,
it is necessary to characterize the variation of mean absorption line
strengths as a function of galaxy magnitude at fixed $\sigma$.  We
therefore use the total sample of 10,284 galaxies with no H$\alpha$ or
[O\textsc{ii}] emission detected at the $2\sigma$ level to construct
high $S/N$ stacked spectra as a function of $\sigma$ and absolute
magnitude $\mbox{ }^{0.1}M_{r}$.  As in the selection of the sample of
2000 quiescent galaxies, we exclude objects with possible low-level
emission and maintain a roughly symmetric distribution about
EW([O\textsc{ii}]) = 0 by requiring -3.12 {\AA}$ <
\mbox{EW([O\textsc{ii}])} < 3.12 $ {\AA} (see
\S\ref{composite_spectra}).  This leaves us with a sample of roughly
9400 galaxies.

These galaxies are divided into six bins in $\sigma$, as before.  Each
of these six bins is then divided into three bins in absolute
magnitude, with equal numbers of galaxies in each magnitude bin.  The
median magnitudes of galaxies in the bright bins are $\sim 0.5$ mag
brighter than those in the middle bins at the same $\sigma$, which are
in turn $\sim 0.5$ mag brighter than those in the faint bins.
Galaxies are coadded within each bin (using the process described in
\S\ref{composite_spectra}), and the Lick indices and D$_n$4000 are
measured in each of the $\sigma$-magnitude binned spectra.

Measured line strengths are shown as a function of $\sigma$ in
Figure \ref{siglum_indices}.  In each $\sigma$ range, the brightest,
middle, and faintest magnitude bins are shown in red, green, and
blue, respectively.  As in Figure \ref{indices}, error bars include
statistical errors only, and the lines show a least squares fit of
index strength onto $\sigma$ separately for the
brightest, middle, and faintest galaxies at each
$\sigma$.  For ease of comparison, the axis scales in each
index plot are the same as in Figure \ref{indices}.

From Figure \ref{siglum_indices}, it is clear that the stellar
populations of red sequence galaxies {\it do} vary with absolute
magnitude at fixed $\sigma$.  The iron lines (Figure
\ref{siglum_indices}b) in particular are considerably stronger in
bright galaxies than in faint ones.  However, the Balmer lines and
G4300 (Figure \ref{siglum_indices}a) show little variation, while
other lines (Figure \ref{siglum_indices}cd) show some slight variation
with magnitude.  This suggests that at fixed $\sigma$, brighter
galaxies have higher metallicities and slightly younger ages compared
to fainter galaxies.

The low $\sigma$ bins of the quiescent and LINER-like galaxy sample
are biased toward brighter objects and therefore, because of the
variation in line strengths with magnitude, toward {\it stronger}
absorption lines for all lines shown in Figure
\ref{siglum_indices}b-d.  A correction for this effect is computed as
follows: we assume that the characteristics of the galaxies missing
from each $\sigma$ bin are similar to the faint galaxies that are
included in the bin, and that the mean measured properties of the bin
are similar to those of the middle magnitude bin.  For each measured
index, the difference between the middle magnitude bin and the faint
bin at fixed $\sigma$ is computed by subtracting the linear fit to the
middle magnitude bins (green lines in Figure \ref{siglum_indices})
from the faint bins (blue lines).  The line strength differential at
each $\sigma$ is multiplied by the corresponding incompleteness
fraction (see Table \ref{bin_table}), so that the $\sigma = 70$--120
km s$^{-1}$ bin has the largest correction because it is most
incomplete, while the $\sigma = 220$-300 bin has no correction because
it is complete by assumption (see \S\ref{completeness}).  This scaled
correction is then subtracted from the measured line strengths in the
quiescent, weak, and strong emission bins at each value of $\sigma$.
The scaled corrections for each index are given in Table
\ref{incomplete_table} for the five lowest $\sigma$ bins (the highest
$\sigma$ bin is assumed to be complete, so no correction is applied).
Implicit in this treatment is that the relative line strength
differences with magnitude are the same for LINER-like galaxies as for
quiescents.

The corrected line strengths are shown in Figure \ref{indices} in the
right-hand column of each panel.  Comparing the corrected with the
uncorrected (left-hand column) values, it is clear that the
corrections are small and are negligible in the Balmer lines and
G4300.  In the case of the metal lines (Figure \ref{indices}b-d), they
operate in the sense of slightly {\it enchancing} trends in line
strength with $\sigma$.  They have no effect on the relative line
strengths of galaxy bins with differing emission properties because
the same correction was applied to all bins at the same $\sigma$.

\section{Comparison with Result from Schiavon 2007}\label{eis_data}

Although the results presented in this work are generally consistent
with the analysis of stacked SDSS spectra in \citet{sch07}, there are
a number of differences worthy of deeper investigation.  \citet{sch07}
finds no trend in age with galaxy luminosity when using H$\beta$ to
compute mean ages, but finds that galaxy age increases with increasing
luminosity when ages are computed using H$\delta_F$.  In contrast, we
find age increasing with $\sigma$ regardless of which Balmer line is
used in the fitting process.  In addition, \citet{sch07} finds that
[Ca/Fe] increases with galaxy luminosity, whereas we find practically
no trend in [Ca/Fe] with $\sigma$.  It is difficult to directly
interpret these differences, as the \citet{sch07} results are
presented as a function of galaxy absolute magnitude, rather than
$\sigma$.

In order to make a more direct comparison between the results of
\citet{sch07} and those presented here, we have taken the same galaxy
sample used in our main analysis and divided the galaxies into bins
based on absolute magnitude, rather than $\sigma$.  As before,
galaxies with weak and strong LINER-like emission are stacked
separately from those with no detectable emission lines.  For this
process, we have used the galaxy absolute $r$-band magnitude,
K-corrected to z=0.0 in order to match the \citet{sch07} data which
made use of stacked spectra from \citet{eis03}.  We then measure Lick
indices and fit for stellar population mean ages, abundances, and
abundance ratios as described in \S\ref{linestrengths} and
\S\ref{modeling} above.  

Figure \ref{eis_abuns} shows the results of the stellar population
modeling for our luminosity-binned sample (black stars), as compared
to the results from \citet{sch07} (grey triangles).  Values of [Fe/H]
appear slightly higher in the \citet{sch07} results than in our
luminosity-binned sample, by a few hundredths of a dex.  The typical
values for the abundance ratios are comparable between the two data
sets, but the \citet{sch07} values show consistently stronger trends
with magnitude than do our data.  Again, the discrepancies are all
within a few hundredths of a dex.  The largest discrepancies between
the data sets are in the age measurements.  Here our data show
slightly older ages for more luminous galaxies, while the
\citet{sch07} results show slightly {\it younger} ages for more
luminous galaxies.

Because both the \citet{eis03} stacked spectra and the stacked spectra
in our sample were acquired as part of the same survey and have been
analyzed using the same method, any differences in the results should
be due to differences in the sample selection process.  The sample
selection criteria are in fact quite different between the two
samples.  Both samples are taken from the SDSS MAIN galaxy sample.  
The \citet{eis03} spectra used in \citet{sch07} are chosen using two
morphological criteria: that the galaxies have concentrated light
profiles (determined from the SDSS concentration parameter $c$, see
\citealt{eis03} for details) and by requiring that a de Vaucouleurs
fit to the light profile be 20 times more likely than an exponential
fit.  Most of these galaxies lie on the red sequence; an additional
color criterion is applied to exclude color outliers, removing about
5\% of the morphologically selected galaxies from the sample.  

It is important to highlight here that no spectral criteria are
applied.  Many of the galaxies in the \citet{eis03} stacked spectra
may therefore include emission lines from ionized gas.  In fact, the
principle component analysis in \citet{eis03} shows that there is a
significant component of emission in the galaxies in their sample.
\citet{yan06} have shown that red sequence galaixes (chosen by a color
criterion only, without any morphological selection) consist of about
48\% quiescent galaxies without emission lines, 29\% galaxies with
LINER-like emission, and 23\% galaxies with emission characteristic of
Seyferts or H\textsc{ii} regions.  The morphological selection applied
by \citet{eis03} may remove some fraction of the Seyferts and
star-forming galaxies from the red sequence, but a contaminating
population of galaxies with emission due to low-level star formation
or Seyfert activity likely exists at the roughly 10\% level.  

Before measured Lick indices and analyzing the stellar populations of
the stacked \citet{eis03} spectra, \citet{sch07} corrects for emission
infill in the stacked spectra by measuring the strength of the
[O\textsc{ii}]$\lambda$3727 line and applying the conversion from
[O\textsc{ii}] EW to H$\alpha$ EW determined in \citet{yan06}.  A
standard Balmer decrement is then used to correct H$\beta$,
H$\gamma_F$, and H$\delta_F$.  However, this conversion is appropriate
only for galaxies with LINER-like emission line ratios.  The
\citet{eis03} spectra may contain some fraction of Seyferts and
low-level star formation.  For these galaxies, the strength of
H$\alpha$ emission is much higher for a given [O\textsc{ii}] EW, thus
the emission infill correction performed by \citet{sch07} may {\it
undercorrect} for infill in the Balmer lines, to the extent that
galaxies in the sample have non-LINER-like emission line ratios.  As
discussed in \S\ref{infill_effect}, H$\beta$ will be much more
sensitive to the effects of infill than will the bluer Balmer lines.

Figure \ref{eis_indices} shows Lick index measurements as a function
of galaxy absolute magnitude for our sample (black stars) and for the
\citet{eis03} stacked spectra (grey triangles).  Only Lick indices
that are used in the abundance fitting process are shown.  The Balmer
lines (left panel) differ significantly between the two data sets,
while the metal lines Fe5270, Fe5335, Mg {\it b}, C$_2$4668, CN$_2$,
and Ca4227 are relatively consistent.  In our luminosity-binned data,
all three Balmer line strengths decrease as galaxy luminosity
increases.  The same trend is seen in H$\delta_F$ and H$\gamma_F$ for
the \citet{eis03} data, although the line strengths are offset toward
stronger absorption compared to our data.  This suggests that the
\citet{eis03} sample of galaxies includes a larger fraction of
intermediate-age galaxies than does our own.  

\begin{figure*}[b]
\epsscale{0.6}
\plotone{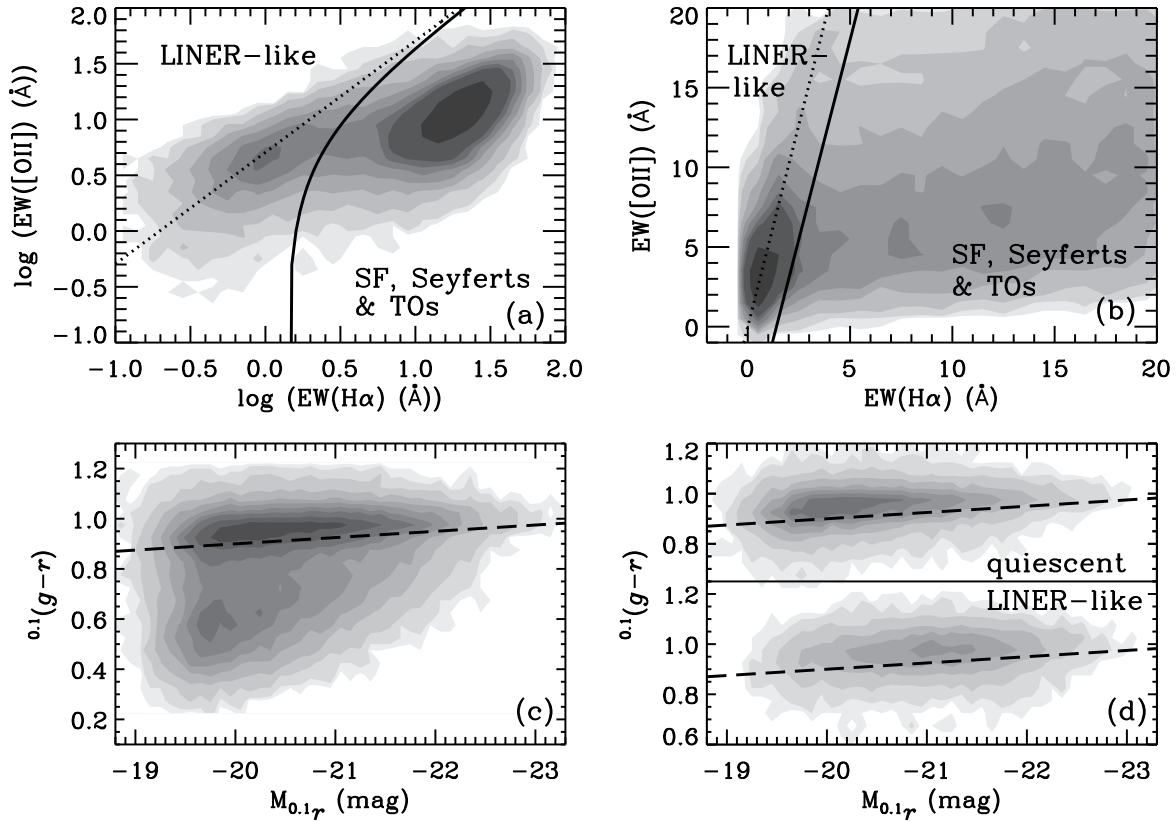}
\caption{(a) H$\alpha$ and [O\textsc{ii}] EWs for all galaxies in the
  redshift range $0.06 < z < 0.08$ and with both lines detected at $>
  2\sigma$.  The distribution is bimodal (see \citealt{yan06} for
  details).  The EW(H$\alpha$)/EW([O\textsc{ii}]) criterion of
  equation \ref{o2hacut} is used to define the LINER-like sample is
  shown as the solid line.  Galaxies to the right of the solid line
  were discarded as ``Low-[O\textsc{ii}]/H$\alpha$''contaminants with
  dusty star formation, Seyfert emission, or a mixture of Seyfert and
  star forming emission (Transition Objects).  The dashed line shows
  the linear least squares fit of EW(H$\alpha$) as a function of
  EW([O\textsc{ii}]) which was used to convert a measured
  [O\textsc{ii}] EW into an H$\alpha$ EW as part of the
  infill-correction process (see \S\ref{infill_correction}).  (b) The
  same as (a) but with linear-scaled axes.  (c) Color-magnitude
  diagram of all SDSS galaxies in the redshift range $0.06 < z <
  0.08$.  The dashed line shows eq. \ref{redcut} used to define the
  red sequence galaxies included in the sample.  (d) Color-magnitude
  diagram of LINER-like and quiescent galaxies selected for the sample
  presented here.  The dashed line shows the cut used to define the
  red sequence.  The LINER-like and quiescent samples have similar
  color distributions (mostly red sequence galaxies) but different
  magnitude distributions.  See text for details.  In all panels,
  contour levels in all panels indicate the density of objects in the
  plot corresponding to, from light to dark, 3, 10, 30, 60, 100, 150,
  200, 300, 400 galaxies per bin, where bin boundaries correspond to
  the tickmarks shown on the x and y axes.}\label{sample_cuts}
\end{figure*}

\begin{figure*}
\epsscale{1.0}
\plotone{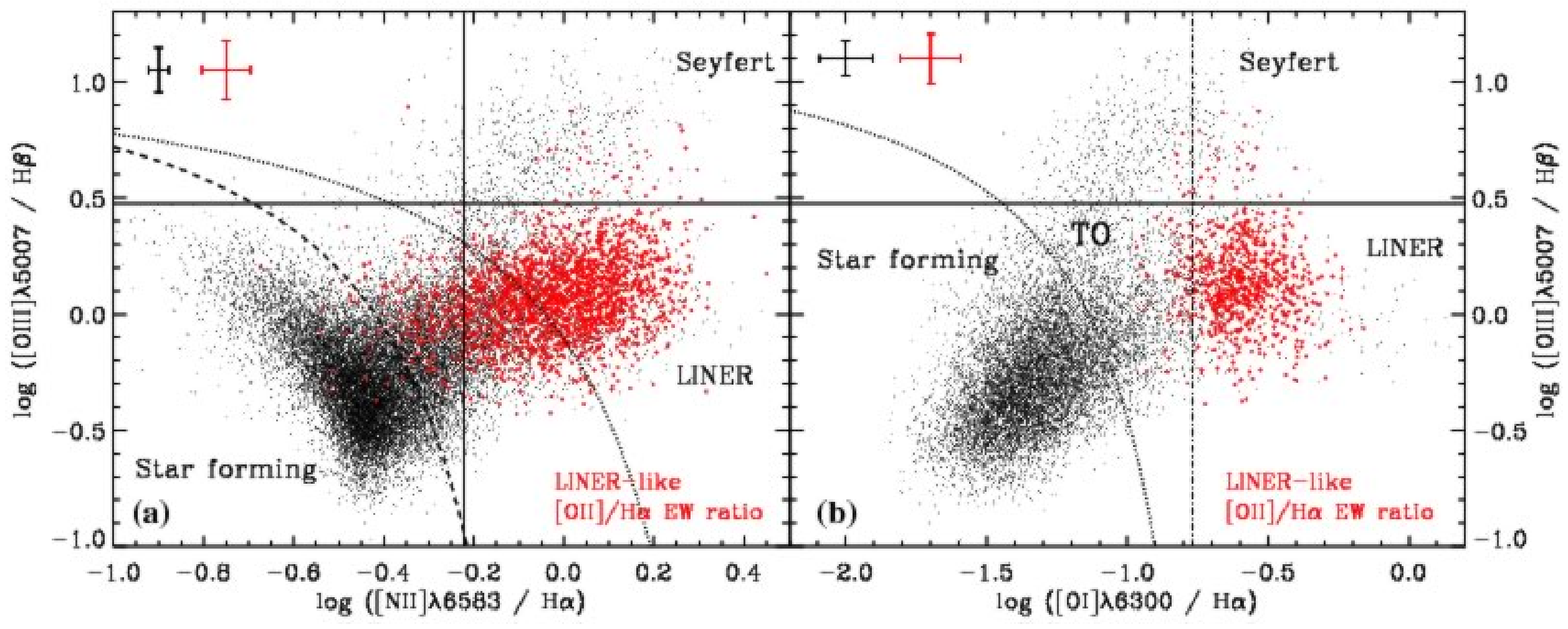}
\caption{BPT diagrams \citep{bal81} for galaxies with $0.06 < z <
  0.08$.  Red points show those galaxies classified as ``LINER-like''
  in our sample based on their EW([O\textsc{ii}])/EW(H$\alpha$)
  ratios; black points show all other galaxies. Only galaxies with all
  four emission lines detected at the 3$\sigma$ level are plotted in
  each panel.  The weak [O\textsc{i}] emission line is undetected in
  many galaxies, thus (b) contains fewer galaxies than (a).  Error
  bars in the upper left show the median error in line ratios
  separately for the LINER-like and other galaxies.  The solid lines
  show the standard classification scheme: Seyferts have
  [O\textsc{iii}]/H$\beta > 3$ and [N\textsc{ii}]/H$\alpha > 0.6$,
  LINERs have [O\textsc{iii}]/H$\beta < 3$ and [N\textsc{ii}]/H$\alpha
  > 0.6$, and star forming galaxies have [N\textsc{ii}]/H$\alpha <
  0.6$.  Dotted lines show the separation of \citet{kew01} between
  star forming galaxies and active nuclei; the dashed line in (a)
  shows the \citet{kau03} separation.  The dash-dot line in (b) shows
  the \citet{ho97} separation between LINERs and transition objects
  (TOs).  Using the \citet{kau03} separation between star forming
  galaxies and active nuclei, the standard separation between Seyferts
  and LINERs, and the \citet{ho97} separation between LINERs and TOs,
  our LINER-like sample defined using EW([O\textsc{ii}])/EW(H$\alpha$)
  contains 88.4\% LINERs, 8.6\% TOs, 1.9\% Seyferts, and 1.1\% star
  forming galaxies.}\label{bpt}
\end{figure*}

\begin{figure*}[b]
\epsscale{0.5}
\plotone{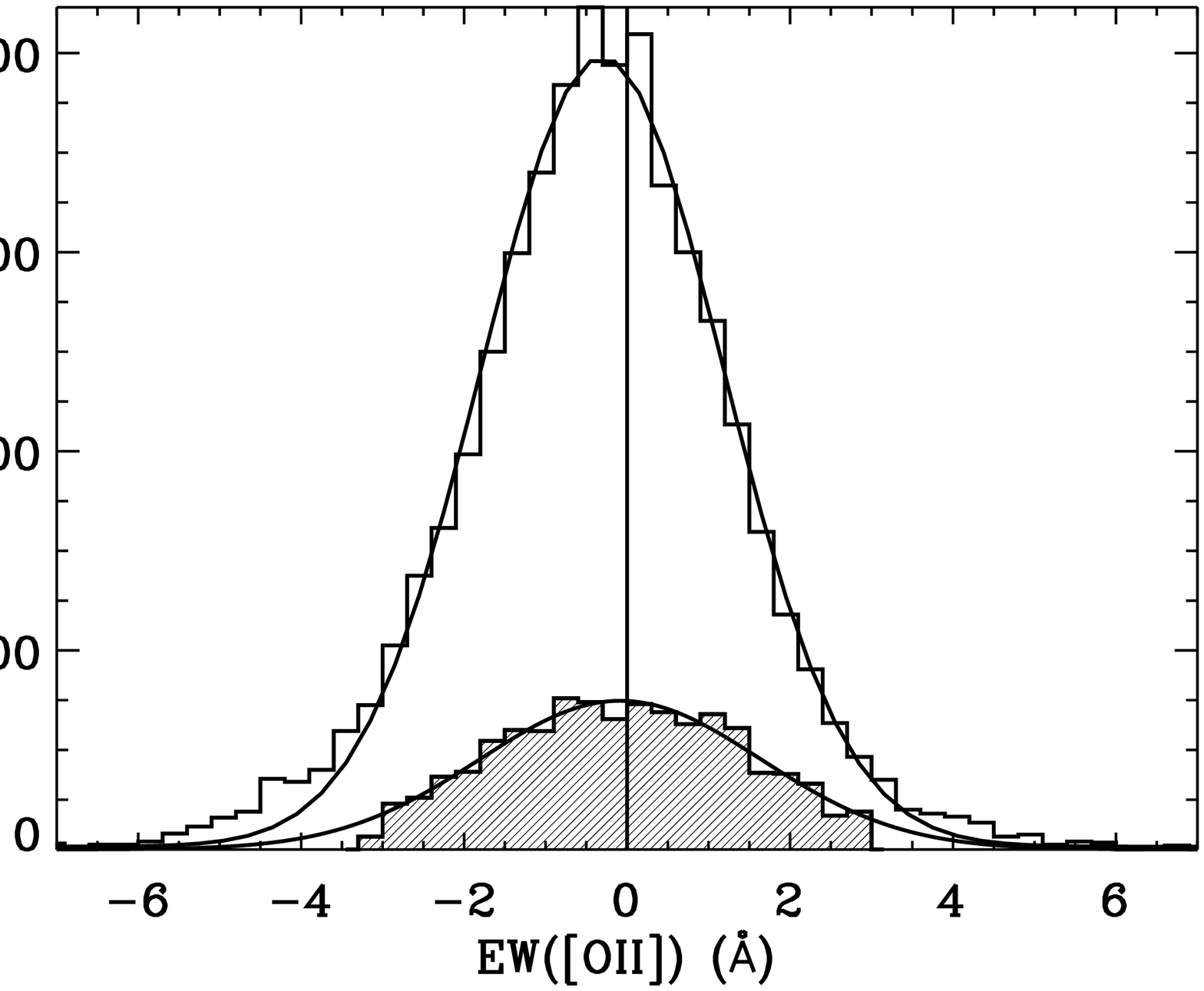}
\caption{Histogram of [O\textsc{ii}] EW in the quiescent galaxy
  sample.  The unfilled histogram shows all galaxies that meet the
  selection criteria of \S\ref{select}.  From these galaxies, a
  sub-sample of 2000 quiescent galaxies is chosen for comparison with
  the weak and strong LINER-like samples.  The sub-sample, shown as
  the filled histogram, is constructed to have a symmetric gaussian
  distribution around EW([O\textsc{ii}]) = 0, and is truncated at
  $\pm3.12$ \AA (the 2$\sigma$ spread of the total EW([O\textsc{ii}])
  distribution) to limit contamination from undetected faint emission
  lines. See \S\ref{composite_spectra} for details.}\label{o2_hist}
\end{figure*}

\begin{figure*}[b]
\epsscale{0.6}
\plotone{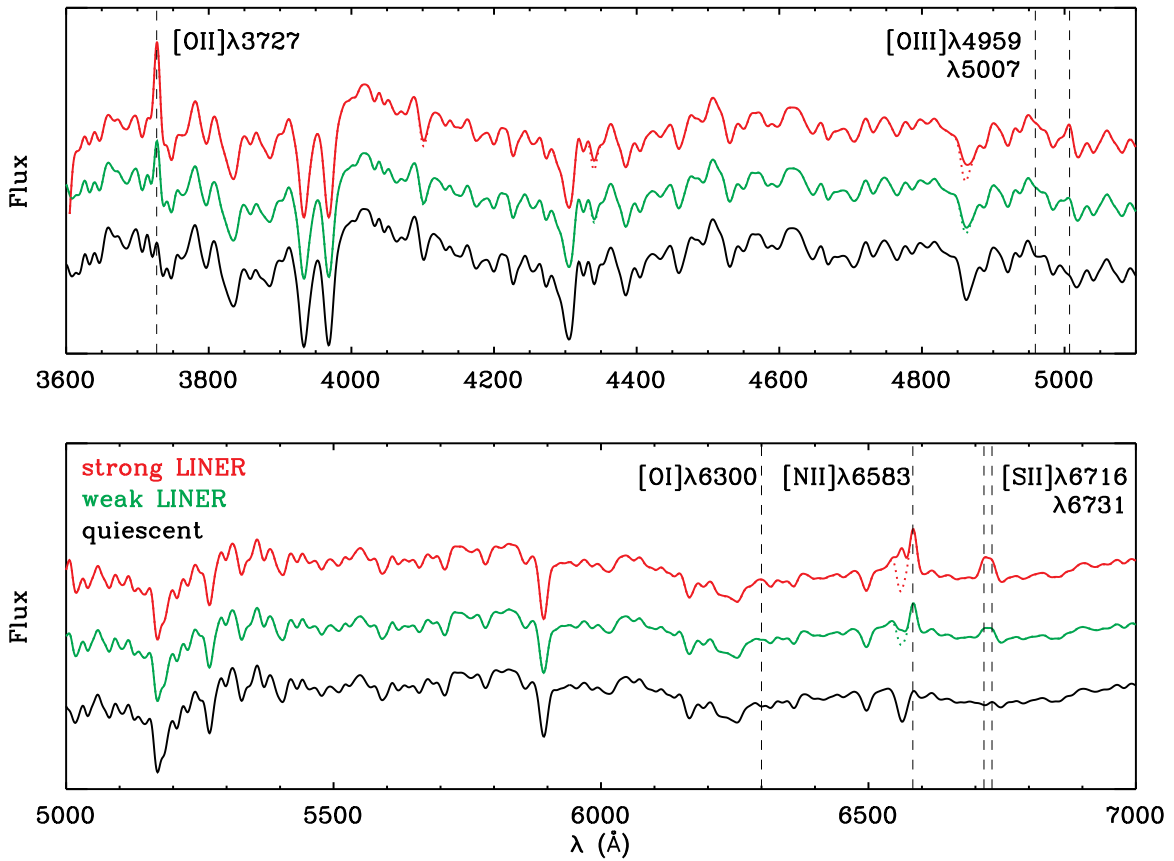}
\caption{Stacked spectra for the galaxy bins with $70 < \sigma < 120$
  km s$^{-1}$ (the lowest $\sigma$ bins), with and without emission
  infill corrections.  The black, green, and red lines show the
  quiescent, weak LINER-like, and strong LINER-like galaxy bins,
  respectively.  Solid lines show the uncorrected spectra, while
  dotted lines show the emission infill correction.  Differences are
  only visible in the Balmer lines where the corrections have been
  made.  The correction to H$\delta$ in particular is very small.
  Strong emission lines are indicated by vertical dashed lines and are
  labelled.  Except for the emission lines, the spectra are remarkably
  similar.  This illustrates the extremely high $S/N$ of the stacked
  spectra and shows the wealth of information in the absorption
  features, as well as the subtlety of the spectral variations that
  are quantified in this paper.}\label{em_spectra}
\end{figure*}

\begin{figure*}[b]
\epsscale{1.0}
\plotone{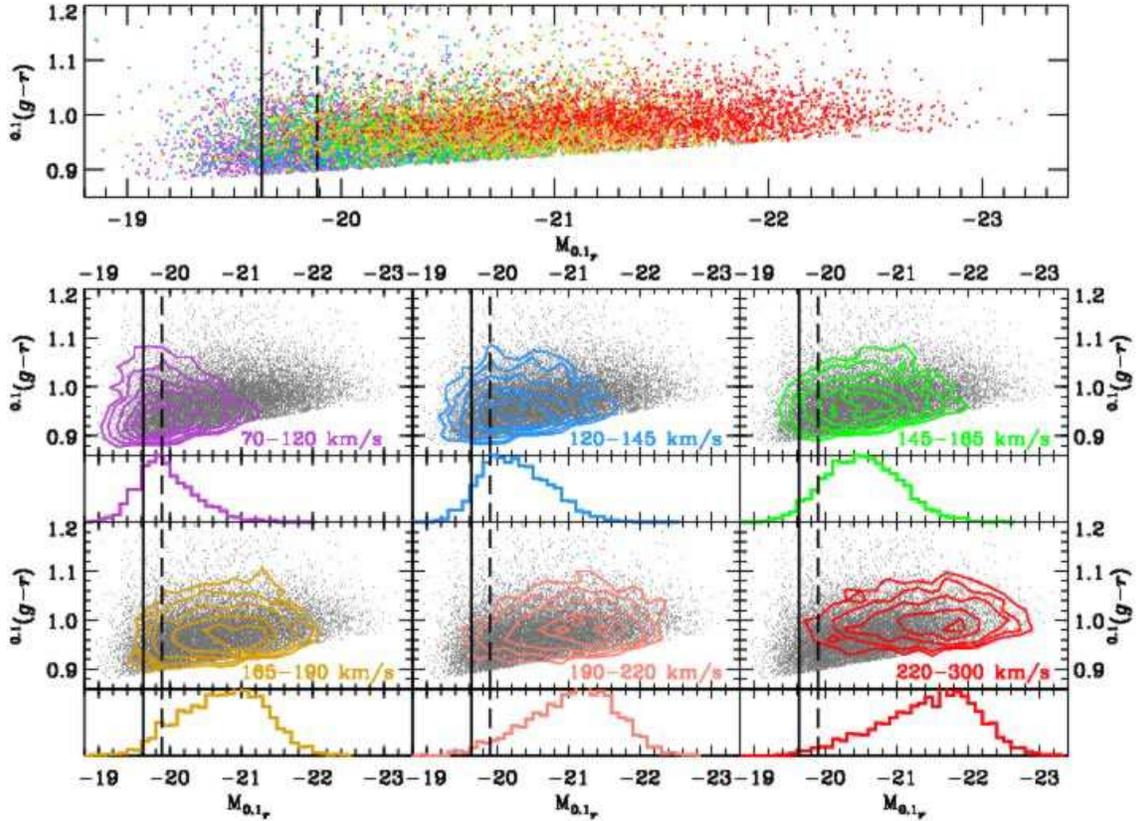}
\caption{Color-magnitude diagrams (CMDs) of quiescent and LINER-like
  red sequence galaxies.  The top panel shows the CMD of all
  (quiescent and LINER-like) galaxies in the sample color-coded by
  aperture-corrected velocity dispersion ($\sigma$): $\sigma =
  70$--120 km s$^{-1}$ in purple, $\sigma = 120$--145 km s$^{-1}$ in
  blue, $\sigma = 145$--165 km s$^{-1}$ in green, $\sigma = 165$--190
  km s$^{-1}$ in gold, $\sigma = 190$--220 km s$^{-1}$ in salmon, and
  $\sigma = 220$--300 km s$^{-1}$ in red.  The lower panels show
  similar CMDs with all galaxies in gray and overplotted colored
  contours for the distribution of the $\sigma$ range indicated.
  Colors correspond to the colors in the top panel.  Below each
  contoured CMD is a histogram of the magnitude distribution of
  galaxies in that $\sigma$ bin.  In all panels, the black dashed line
  shows the 100\% completeness limit, while the solid black line shows
  the magnitude at which the sample is 50\% incomplete.  The
  histograms show that the highest $\sigma$ bins are effectively
  complete while the lower $\sigma$ bins are seriously incomplete.  A
  correction for this effect is described in \S\ref{lowsig_corr}.
  Higher $\sigma$ galaxies are more luminous and redder than lower
  $\sigma$ galaxies, but there is a large spread in both magnitude and
  color at fixed $\sigma$, typically larger than the separation
  between $\sigma$ bins.  At fixed $\sigma$, there is no clear
  color-magnitude relation.}\label{cmd_sig}
\end{figure*}

This offset in mean stellar population age between the samples is not
unexpected.  The galaxies included in our luminosity-binned sample, as
in the $\sigma$-binned sample used in the main analysis presented
above, have been selected with a fairly stringent color cut, which
excludes the blue side of the red sequence (see Figure
\ref{sample_cuts}).  No such stringent cut has been applied to the
\citet{eis03} galaxies.  That sample therefore contains a sizable
fractin of galaxies which are as much as 0.1 mag bluer in $g-r$ than
our galaxy sample and therefore may include a significant
sub-population of intermediate-age galaxies which are excluded from
our sample. 

Also, as discussed earlier, the \citet{eis03} data likely include a
modest fraction of galaxies with low-level star formation because
galaxies have not been excluded on the basis of emission lines.
Including these galaxies would bias the sample toward younger mean
ages.  It is true that the star-forming galaxies also have stronger
Balmer emission infill, which we have argued is likely to be
under-corrected.  However, the effect on the blue Balmer lines is
small both in terms of the total amount of infill and the amount of
correction applied, thus the inaccuracies in accounting for the infill
in H$\delta_F$ and H$\gamma_F$ should not be large.

This is not true for H$\beta$, where the emission infill is
substantial and therefore the effect of the under-correction could be
significant.  Indeed, H$\beta$ behaves differently in the
\citet{eis03} stacked spectra than in our luminosity-binned spectra;
H$\beta$ remains roughly constant over a range of galaxy absolute
magnitude in the \citet{eis03} data whereas H$\beta$ strengths decline
for brighter galaxies in our sample.  The contant values of H$\beta$
are also qualitatively different from the H$\delta_F$ and H$\gamma_F$
indices in both the \citet{eis03} data and in our data, where Balmer
absorption declines with increasing galaxy luminosity everywhere.  It
seems likely that the H$\beta$ values measured in the \citet{eis03}
data suffer from an improper infill correction which has dramatic
effects on H$\beta$ but less so on H$\delta_F$ and H$\gamma_F$, and
none at all on the metal lines.

The metal lines in fact appear very similar between the two different
data sets.  There is some indication that metal absorption strengths
are slightly lower in the \citet{eis03} data, which would be
consistent with the supposition that those data include a larger
fraction of intermediate-age galaxies than do our data.  The slopes of
the relations between line strengths and galaxy magnitude are in good
agreement between the two samples.  

We therefore conclude that the slight differences in the ages and
abundance patterns reported by \citet{sch07} as compared to the
results presented in this paper are due to differing trends in the
H$\beta$ line strengths that are used to compute the fiducial galaxy
ages (see \S\ref{modeling}), rather than genuine differences in
abundance patterns.  These differing trends in measured H$\beta$ line
strengths are due to differences in sample selection coupled with less
accurate emission infill corrections applied to the \citet{eis03} data
in \citet{sch07}.  In order for this to explain the observed flat
relationship between H$\beta$ absorption strength and galaxy
luminosity, the fainter galaxies would have to be more affected by
under-corrected emission infill (in other words, the fainter galaxies
would have to have a larger fraction of galaxies with emission having
non-LINER-like line ratios).  It seems reasonable that the fainter
\citet{eis03} galaxies, which are observed to have much stronger
H$\delta_F$ absorption than the brighter galaxies, are more likely to
contain residual star formation and weak H\textsc{ii} region emission
than the brighter galaxies.

It should be noted that the differences in abundances shown in
Figure \ref{eis_abuns} between the two data sets are never more than
0.05 dex.  Although different sample selection and inaccurate emission
infill corrections pose problems for precision stellar population
determinations, there does not seem to be a catastrophic discrepancy
created by moderate differences in the input data.

\end{appendix}

\clearpage

\begin{deluxetable}{lclclccc}
\tabletypesize{\scriptsize}
\tablenum{1}
\tablecaption{
Stacked Spectra: Parameters
}\label{bin_table}
\tablewidth{0pt}
\tablehead{
\colhead{$\sigma$} &
\colhead{Fractional} &
\colhead{LINER} &
\colhead{\#} &
\colhead{$S/N_{\rm{med}}$} &
\colhead{$\langle\sigma\rangle$} &
\colhead{$\langle$EW[O\textsc{ii}]$\rangle$} \\
\colhead{(km s$^{-1}$)} &
\colhead{Incompleteness} &
\colhead{emission} &
\colhead{} &
\colhead{({\AA}$^{-1}$)} &
\colhead{(km s$^{-1}$)} &
\colhead{({\AA})}
}
\startdata
\phantom070--120  &51\%   &quiescent       &339      &298.0      &103.2   &-0.02 \\
         &       &weak       &137      &229.1      &105.9   &3.75  \\
         &       &strong     &317      &280.2      &101.7   &9.10  \\
120--145 &28\%   &quiescent       &374      &366.9      &133.4   &0.12  \\
         &       &weak       &248      &344.7      &133.8   &3.63  \\  
         &       &strong     &388      &379.5      &133.5   &8.40  \\
145--165 &19\%   &quiescent       &327      &389.5      &155.2   &-0.03 \\
         &       &weak       &286      &408.8      &155.8   &3.44  \\
         &       &strong     &369      &420.3      &155.2   &8.90  \\
165--190 &11\%   &quiescent       &307      &405.7      &176.5   &-0.23 \\
         &       &weak       &380      &521.5      &177.4   &3.38  \\
         &       &strong     &394      &485.1      &176.7   &8.64  \\
190--220  &4\%   &quiescent       &317      &469.7      &204.1   &-0.11 \\
          &      &weak       &374      &561.9      &203.9   &3.25  \\
          &      &strong     &347      &506.7      &203.2   &9.06  \\
220--300  &...   &quiescent       &284      &526.4      &245.8   &-0.27 \\
          &      &weak       &289      &570.9      &247.1   &3.14  \\
          &      &strong     &326      &550.6      &243.6   &9.67  \\
\enddata
\end{deluxetable}

\begin{deluxetable}{lrrrrrrrrrrrr}
\tabletypesize{\scriptsize}
\tablenum{2}
\tablecaption{
Indices in Stacked Spectra: Balmer Indices\tablenotemark{*}
}\label{indices_table1}
\tablewidth{0pt}
\tablehead{
\colhead{} &
\colhead{} &
\multicolumn{3}{c}{H$\beta$} &
\colhead{} &
\multicolumn{3}{c}{H$\gamma_F$} &
\colhead{} &
\multicolumn{3}{c}{H$\delta_F$} \\
\colhead{Name} &
\colhead{} &
\multicolumn{3}{c}{({\AA})} &
\colhead{} &
\multicolumn{3}{c}{({\AA})} &
\colhead{} &
\multicolumn{3}{c}{({\AA})} \\
\cline{1-1} \cline{3-5} \cline{7-9} \cline{11-13} \\
\colhead{Infill Correction:\tablenotemark{**}} &
\colhead{} &
\colhead{Ho97} &
\colhead{Yan06} &
\colhead{none} &
\colhead{} &
\colhead{Ho97} &
\colhead{Yan06} &
\colhead{none} &
\colhead{} &
\colhead{Ho97} &
\colhead{Yan06} &
\colhead{none}
}
\startdata
\phantom070--120 / quiescent	&&1.88  &1.88 &1.88 &&-0.90 &-0.90 &-0.90 &&0.58 &0.58 &0.58\\
\phantom070--120 / weak	&&1.95  &1.90 &1.69 &&-0.45 &-0.48 &-0.61 &&1.04 &1.02 &0.95\\
\phantom070--120 / strong	&&2.15  &2.03 &1.47 &&-0.03 &-0.12 &-0.45 &&1.39 &1.33 &1.14\\
120--145 / quiescent   &&1.79  &1.79 &1.79 &&-1.06 &-1.06 &-1.06 &&0.53 &0.53 &0.53\\
120--145 / weak   &&1.93  &1.89 &1.67 &&-0.68 &-0.72 &-0.85 &&0.93 &0.91 &0.83\\ 
120--145 / strong	&&2.04  &1.93 &1.43 &&-0.47 &-0.55 &-0.85 &&1.03 &0.98 &0.80\\
145--165 / quiescent   &&1.78  &1.78 &1.78 &&-1.24 &-1.24 &-1.24 &&0.49 &0.49 &0.49\\
145--165 / weak   &&1.87  &1.82 &1.62 &&-0.89 &-0.92 &-1.04 &&0.79 &0.77 &0.69\\
145--165 / strong	&&1.98  &1.85 &1.32 &&-0.73 &-0.82 &-1.15 &&0.85 &0.80 &0.61\\
165--190 / quiescent   &&1.73  &1.73 &1.73 &&-1.33 &-1.33 &-1.33 &&0.47 &0.47 &0.47\\
165--190 / weak   &&1.83  &1.78 &1.58 &&-1.07 &-1.11 &-1.23 &&0.70 &0.68 &0.60\\
165--190 / strong	&&1.93  &1.81 &1.29 &&-0.98 &-1.07 &-1.39 &&0.73 &0.68 &0.49\\
190--220 / quiescent   &&1.70  &1.70 &1.70 &&-1.45 &-1.45 &-1.45 &&0.32 &0.32 &0.32\\
190--220 / weak   &&1.73  &1.68 &1.49 &&-1.29 &-1.32 &-1.44 &&0.51 &0.49 &0.42\\
190--220 / strong	&&1.83  &1.70 &1.16 &&-1.19 &-1.28 &-1.63 &&0.63 &0.57 &0.37\\
220--300 / quiescent   &&1.63  &1.63 &1.63 &&-1.61 &-1.61 &-1.61 &&0.22 &0.22 &0.22\\
220--300 / weak   &&1.63  &1.59 &1.40 &&-1.51 &-1.54 &-1.66 &&0.35 &0.33 &0.26\\
220--300 / strong	&&1.72  &1.58 &1.01 &&-1.41 &-1.51 &-1.88 &&0.47 &0.40 &0.18\\
\enddata
\tablenotetext{*}{Index values are as measured in the stacked
  spectra.  They do {\it not} include the corrections for
  incompleteness described in Appendix \ref{lowsig_corr}}
\tablenotetext{**}{Balmer index measurements are shown for three
  different choices of emission infill correction: (1) assuming
  H$\alpha$/H$\beta = 3.27$ as in \citet{ho97}, (2) assuming
  H$\alpha$/H$\beta = 4.1$ as in \citet{yan06}, and (3) with no infill
  correction. See \S\ref{infill_correction} for details.}
\end{deluxetable}

\begin{deluxetable}{lccccccccc}
\tabletypesize{\scriptsize}
\tablenum{3}
\tablecaption{
Indices in Stacked Spectra: D$_n$4000, Fe, and Mg Indices\tablenotemark{*}
}\label{indices_table2}
\tablewidth{0pt}
\tablehead{
\colhead{} &
\colhead{D$_n$4000} &
\colhead{Fe4383} &
\colhead{Fe4531} &
\colhead{Fe5270} &
\colhead{Fe5335} &
\colhead{Fe5406} &
\colhead{Mg$_1$} &
\colhead{Mg$_2$} &
\colhead{Mg {\it b}} 
\\
\colhead{Name} &
\colhead{} &
\colhead{({\AA})} &
\colhead{({\AA})} &
\colhead{({\AA})} &
\colhead{({\AA})} &
\colhead{({\AA})} &
\colhead{(mag)} &
\colhead{(mag)} &
\colhead{({\AA})}
}

\startdata
\phantom070--120 / quiescent       &1.87   &3.92    &2.89    &2.64    &2.32    &1.56    &0.080   &0.205   &3.43\\
\phantom070--120 / weak       &1.81   &3.83    &2.87    &2.50    &2.25    &1.52    &0.075   &0.195   &3.38\\
\phantom070--120 / strong       &1.76   &3.52    &2.83    &2.49    &2.18    &1.46    &0.069   &0.182   &3.13\\
120--145 / quiescent       &1.91   &4.19    &3.06    &2.70    &2.39    &1.62    &0.091   &0.223   &3.76\\
120--145 / weak       &1.85   &3.99    &2.99    &2.70    &2.40    &1.64    &0.085   &0.213   &3.62\\
120--145 / strong       &1.82   &3.92    &2.99    &2.65    &2.29    &1.50    &0.083   &0.206   &3.58\\
145--165 / quiescent       &1.94   &4.28    &3.03    &2.71    &2.46    &1.64    &0.100   &0.237   &3.91\\
145--165 / weak       &1.89   &4.16    &3.04    &2.70    &2.45    &1.63    &0.094   &0.225   &3.76\\
145--165 / strong       &1.87   &4.13    &3.03    &2.69    &2.42    &1.60    &0.093   &0.222   &3.82\\
165--190 / quiescent       &1.96   &4.25    &3.13    &2.74    &2.53    &1.67    &0.105   &0.244   &4.10\\
165--190 / weak       &1.90   &4.25    &3.10    &2.75    &2.47    &1.65    &0.100   &0.237   &4.02\\
165--190 / strong       &1.92   &4.31    &3.07    &2.76    &2.46    &1.66    &0.103   &0.239   &4.05\\
190--220 / quiescent       &1.99   &4.35    &3.16    &2.81    &2.52    &1.73    &0.114   &0.254   &4.16\\
190--220 / weak       &1.99   &4.42    &3.19    &2.79    &2.57    &1.70    &0.112   &0.256   &4.27\\
190--220 / strong       &1.98   &4.40    &3.15    &2.70    &2.50    &1.67    &0.112   &0.252   &4.27\\
220--300 / quiescent       &2.06   &4.48    &3.28    &2.81    &2.65    &1.70    &0.126   &0.276   &4.55\\
220--300 / weak       &2.01   &4.54    &3.23    &2.85    &2.64    &1.77    &0.125   &0.270   &4.45\\
220--300 / strong       &1.99   &4.45    &3.23    &2.84    &2.58    &1.75    &0.126   &0.271   &4.56\\
\enddata
\tablenotetext{*}{Index values are as measured in the stacked
  spectra.  They do {\it not} include the corrections for
  incompleteness described in Appendix \ref{lowsig_corr}}
\end{deluxetable}

\begin{deluxetable}{lrcccccccc}
\tabletypesize{\scriptsize}
\tablenum{4}
\tablecaption{
Indices in Stacked Spectra: CN, CH, C$_2$, Ca, Na, and TiO Indices\tablenotemark{*}
}\label{indices_table3}
\tablewidth{0pt}
\tablehead{
\colhead{} &
\colhead{CN$_1$} &
\colhead{CN$_2$} &
\colhead{G4300} &
\colhead{C$_2$4668} &
\colhead{Ca4227} &
\colhead{Ca4455} &
\colhead{NaD} &
\colhead{TiO$_1$} &
\colhead{TiO$_2$} \\
\colhead{Name} &
\colhead{(mag)} &
\colhead{(mag)} &
\colhead{({\AA})} &
\colhead{({\AA})} &
\colhead{({\AA})} &
\colhead{({\AA})} &
\colhead{({\AA})} &
\colhead{(mag)} &
\colhead{(mag)} 
}
\startdata
\phantom070--120 / quiescent     &0.0263  &0.0507  &4.83    &5.40    &0.95    &1.12    &2.61    &0.0248  &0.0612\\
\phantom070--120 / weak     &0.0139  &0.0398  &4.56    &5.06    &0.83    &1.02    &2.64    &0.0246  &0.0604\\
\phantom070--120 / strong     &-0.0051 &0.0207  &4.19    &4.63    &0.86    &0.97    &2.77    &0.0246  &0.0573\\
120--145 / quiescent     &0.0386  &0.0659  &4.99    &5.70    &0.99    &1.16    &3.08    &0.0271  &0.0643\\
120--145 / weak     &0.0261  &0.0528  &4.74    &5.62    &0.93    &1.07    &2.98    &0.0234  &0.0627\\
120--145 / strong     &0.0184  &0.0451  &4.59    &5.26    &0.91    &1.07    &3.24    &0.0238  &0.0614\\
145--165 / quiescent     &0.0468  &0.0742  &5.05    &6.02    &1.00    &1.21    &3.18    &0.0280  &0.0677\\
145--165 / weak     &0.0377  &0.0639  &4.82    &5.91    &0.94    &1.16    &3.41    &0.0241  &0.0655\\
145--165 / strong     &0.0357  &0.0626  &4.85    &5.68    &0.93    &1.10    &3.40    &0.0249  &0.0650\\
165--190 / quiescent     &0.0559  &0.0859  &5.17    &6.13    &0.97    &1.21    &3.18    &0.0254  &0.0689\\
165--190 / weak     &0.0487  &0.0769  &4.96    &6.22    &0.96    &1.18    &3.48    &0.0256  &0.0675\\
165--190 / strong     &0.0478  &0.0763  &4.97    &6.06    &0.95    &1.14    &3.50    &0.0265  &0.0689\\
190--220 / quiescent     &0.0694  &0.1025  &5.22    &6.55    &1.02    &1.23    &3.62    &0.0286  &0.0713\\
190--220 / weak     &0.0640  &0.0938  &5.11    &6.63    &1.02    &1.21    &3.83    &0.0280  &0.0711\\
190--220 / strong     &0.0629  &0.0929  &5.14    &6.43    &1.01    &1.21    &3.96    &0.0272  &0.0712\\
220--300 / quiescent     &0.0896  &0.1233  &5.32    &6.91    &1.08    &1.32    &4.16    &0.0321  &0.0762\\
220--300 / weak     &0.0842  &0.1168  &5.20    &6.99    &1.03    &1.28    &4.47    &0.0305  &0.0754\\
220--300 / strong     &0.0823  &0.1134  &5.20    &6.87    &1.03    &1.26    &4.41    &0.0290  &0.0744\\
\enddata
\tablenotetext{*}{Index values are as measured in the stacked
  spectra.  They do {\it not} include the corrections for
  incompleteness described in Appendix \ref{lowsig_corr}}
\end{deluxetable}

\begin{deluxetable}{rccc}
\tabletypesize{\scriptsize}
\tablenum{5}
\tablecaption{Direction of Line Strength Change in SSP Models}\label{stellarpop_indices}
\tablewidth{0pt}
\tablehead{
\colhead{} &
\colhead{Balmer lines} &
\colhead{D$_n$4000, G4300} &
\colhead{Metal lines}
}
\startdata
Age $\Uparrow$      &$\Downarrow$    &$\Uparrow$     &$\Uparrow$  \\
\verb|[|Fe/H\verb|]| $\Uparrow$   &$\Downarrow$    &$\Uparrow$     &$\Uparrow$  \\
\enddata
\end{deluxetable}

\begin{deluxetable}{lccccc}
\tabletypesize{\scriptsize}
\tablenum{6}
\tablecaption{Incompleteness Corrections}\label{incomplete_table}
\tablewidth{0pt}
\tablehead{
\colhead{} &
\colhead{$\sigma$ bin} &
\colhead{$\sigma$ bin} &
\colhead{$\sigma$ bin} &
\colhead{$\sigma$ bin} &
\colhead{$\sigma$ bin} \\
\colhead{Index} &
\colhead{70--120 km s$^{-1}$} &
\colhead{120--145 km s$^{-1}$} & 
\colhead{145--165 km s$^{-1}$} & 
\colhead{165--190 km s$^{-1}$} & 
\colhead{190--220 km s$^{-1}$}  
}
\startdata
D$_n$4000     &\phn0.0255   &\phn0.0089   &\phn0.0041   &\phn0.0013   &\phn0.0001 \\
H$\delta_F$   &\phn0.0240   &\phn0.0030   &   -0.0020   &   -0.0032   &   -0.0020 \\
CN$_1$        &\phn0.0032   &\phn0.0014   &\phn0.0009   &\phn0.0004   &\phn0.0001 \\ 
CN$_2$        &\phn0.0038   &\phn0.0016   &\phn0.0009   &\phn0.0004   &\phn0.0001 \\
Ca4227        &\phn0.0058   &\phn0.0059   &\phn0.0050   &\phn0.0035   &\phn0.0015 \\
G4300         &\phn0.0292   &\phn0.0112   &\phn0.0057   &\phn0.0023   &\phn0.0005 \\
H$\gamma_F$   &   -0.0698   &   -0.0255   &   -0.0123   &   -0.0045   &   -0.0006 \\
Fe4383        &\phn0.0765   &\phn0.0456   &\phn0.0324   &\phn0.0195   &\phn0.0074 \\
Ca4455        &\phn0.0069   &\phn0.0071   &\phn0.0061   &\phn0.0042   &\phn0.0018 \\
Fe4531        &\phn0.0298   &\phn0.0153   &\phn0.0100   &\phn0.0055   &\phn0.0019 \\
C$_2$4668     &\phn0.2019   &\phn0.1160   &\phn0.0807   &\phn0.0478   &\phn0.0178 \\
H$\beta$      &\phn0.0104   &\phn0.0103   &\phn0.0088   &\phn0.0060   &\phn0.0025 \\
Mg$_1$        &\phn0.0019   &\phn0.0008   &\phn0.0004   &\phn0.0002   &\phn0.0001 \\
Mg$_2$        &\phn0.0054   &\phn0.0023   &\phn0.0013   &\phn0.0006   &\phn0.0002 \\
Mg {\it b}    &\phn0.0873   &\phn0.0328   &\phn0.0164   &\phn0.0064   &\phn0.0011 \\
Fe5270        &\phn0.0474   &\phn0.0303   &\phn0.0222   &\phn0.0137   &\phn0.0053 \\
Fe5335        &\phn0.0476   &\phn0.0286   &\phn0.0203   &\phn0.0123   &\phn0.0047 \\
Fe5406        &\phn0.0382   &\phn0.0187   &\phn0.0118   &\phn0.0064   &\phn0.0021 \\
NaD           &\phn0.1200   &\phn0.0563   &\phn0.0345   &\phn0.0180   &\phn0.0058 \\
TiO$_1$       &   -0.0017   &   -0.0006   &   -0.0002   &   -0.0001   &\phn0.0000 \\
TiO$_2$       &\phn0.0006   &\phn0.0004   &\phn0.0003   &\phn0.0002   &\phn0.0001 \\
\enddata
\tablecomments{Values given are the scaled corrections for
  incompleteness, as described in \S\ref{lowsig_corr}.  These are {\it
  subtracted} from the measured index strengths given in Table 2.  The
  same correction is applied to quiescent, weak, and strong LINER-like
  emission composite spectra.  All incompleteness corrections are
  given in the units of the corresponding index (CN$_1$,
  CN$_2$, Mg$_1$, Mg$_2$, TiO$_1$, and TiO$_2$ are in magnitudes,
  D$_n$4000 is dimensionless, and all others are in {\AA}).}
\end{deluxetable}

\begin{deluxetable}{lcc}
\tabletypesize{\scriptsize}
\tablenum{7}
\tablecaption{Abundance Trends with $\sigma$}\label{abun_table}
\tablewidth{0pt}
\tablehead{
\multicolumn{3}{c}{$Abundance = a + b \log \sigma$} \\
\cline{1-3} \\
\colhead{$Abundance$} &
\colhead{$b$} &
\colhead{$a$}
}
\startdata
\verb|[|Fe/H\verb|]|      &$0.451 \pm 0.034$ &$-1.14 \pm 0.08$\\
\verb|[|Mg/Fe\verb|]|     &$0.362 \pm 0.043$ &$-0.61 \pm 0.10$\\
\verb|[|C/Fe\verb|]|      &$0.555 \pm 0.027$ &$-1.06 \pm 0.06$\\
\verb|[|N/Fe\verb|]|      &$0.640 \pm 0.035$ &$-1.34 \pm 0.08$\\
\verb|[|Ca/Fe\verb|]|     &$0.130 \pm 0.039$ &$-0.29 \pm 0.09$\\
\\
\verb|[|Z/H\verb|]| (\verb|[|O/Fe\verb|]| = 0.0)       &$0.786 \pm 0.050$ &$-1.80\pm0.11$\\
\verb|[|Z/H\verb|]| (\verb|[|O/Fe\verb|]| = 0.3)       &$0.705 \pm 0.051$ &$-1.47\pm0.11$\\
\verb|[|Z/H\verb|]| (\verb|[|O/Fe\verb|]| = 0.5)       &$0.656 \pm 0.052$ &$-1.23\pm0.12$\\
\verb|[|Z/H\verb|]| (\verb|[|O/Fe\verb|]| = \verb|[|Mg/Fe\verb|]|)   &$0.916 \pm 0.054$ &$-2.00\pm0.12$\\
\verb|[|Z/H\verb|]| (\verb|[|O/Fe\verb|]| = \verb|[|Mg/Fe\verb|]| + 0.3) &$0.895 \pm 0.055$ &$-1.77\pm0.12$\\
\enddata
\end{deluxetable}

\begin{deluxetable}{llcccccc}
\tabletypesize{\scriptsize}
\tablenum{8}
\tablecaption{Trends with $\sigma$: Comparison with Other Work}\label{abun_comp_table}
\tablewidth{0pt}
\tablehead{
\multicolumn{2}{c}{$\log Age = a + b \log \sigma$} &
\colhead{This Work} &
\colhead{Trager et al.} &
\colhead{Thomas et al.} &
\colhead{Nelan et al.} &
\colhead{Smith et al.} &
\colhead{Bernardi et al.} \\
\multicolumn{2}{c}{$Abundance = a + b \log \sigma$} &
\colhead{} &
\colhead{(2000b)} &
\colhead{(2005)} &
\colhead{(2005)} &
\colhead{(2007)} &
\colhead{(2003b)} 
}
\startdata
Age                  &$b$   &$0.35\pm0.03$\tablenotemark{b}
&$0.6\pm0.2$\tablenotemark{c}   &$0.24$ $(0.32)$\tablenotemark{d}
&$0.59\pm0.13$ &$0.52\pm0.06$  &$1.15$ \\
\verb|[|Fe/H\verb|]| &$b$   &$0.45\pm0.03$ &$0.48\pm0.12$\tablenotemark{c} &... &... &... &...\\
\verb|[|Z/H\verb|]|  &$b$   &$0.79\pm0.05$\tablenotemark{e}
&$0.76\pm0.13$\tablenotemark{c} &$0.55$ $(0.57)$\tablenotemark{d}
&$0.53\pm0.08$ &$0.34\pm0.04$ &$0.38$ \\
\verb|[|$\alpha$/Fe\verb|]|\tablenotemark{a} &$b$ &$0.36\pm0.04$
&$0.33\pm0.01$ &$0.28$ $(0.28)$\tablenotemark{d} &$0.31\pm0.06$ &$0.23\pm0.04$ &$0.32$ \\
\\
\cline{1-8} \\
SSP Models\tablenotemark{f}  &  &S06  &W94  &TMB03  &TMB03  &TMB03 &TMB03\\
\enddata
\tablenotetext{a}{Measured using Mg {\it b} in all works except Nelan
  et al.(2005) which uses Mg {\it b} and CN$_1$ as $\alpha$ sensitive
  indices.  Corresponds to [Mg/Fe] in this work.}
\tablenotetext{b}{From fit to H$\beta$ age, quiescent galaxies only.}
\tablenotetext{c}{Trager et al.(2000b) posit the existence of an
  ``age-metallicity hyperplane'' such that galaxies display a range of
  ages and [Z/H] (or [Fe/H]) at fixed $\sigma$, with an {\it
    anticorrelation} between age and [Z/H]; they therefore give [Z/H]
  as a function of $\sigma$ and age.  The age relation
given here is from the Nelan et al.(2005) fit to the Trager et
al. data, while the [Z/H] and [Fe/H] values include only the $\sigma$
portion of the effect on [Z/H] and [Fe/H] reported in Trager et
al.(2005b), ignoring the effect of age.}
\tablenotetext{d}{Thomas et al.(2005) report two sets of relations for
galaxies in low and high density environments.  Values for both
environments are show as: high density (low density).}
\tablenotetext{e}{Calculated with [O/Fe] = 0.0, as described in
  \S\ref{abuns}. For other choices of [O/Fe], see Table 5.}
\tablenotetext{f}{S06: \citet{sch07}, W94: \citet{wor94}, TMB03: \citet{tho03}}
\end{deluxetable}

\clearpage

\thispagestyle{empty}

\clearpage

\clearpage

\clearpage

\clearpage

\begin{figure*}[b]
\epsscale{0.6}
\plotone{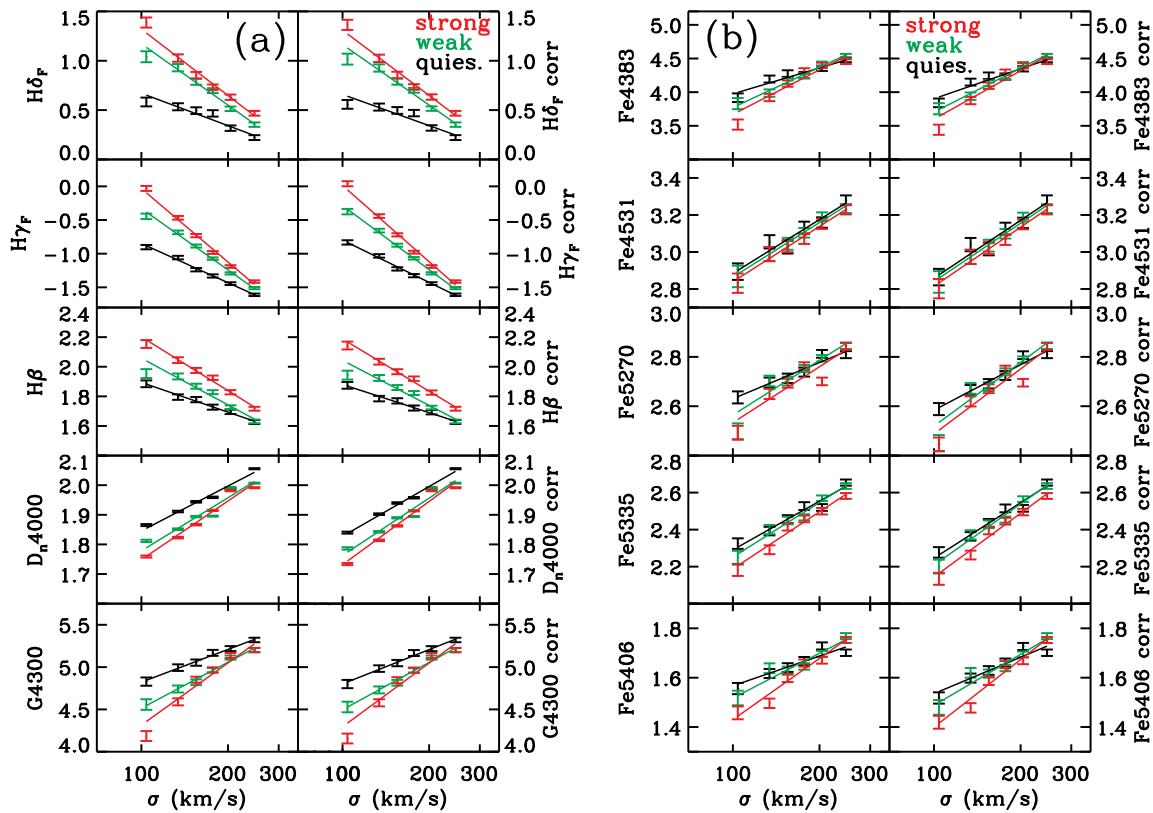}
\caption{Lick index measurements in the stacked spectra as a function
  of the mean $\sigma$ and emission line properties of the galaxies in
  each stacked spectrum.  Quiescent galaxies are shown in black, weak
  LINER-like galaxies in green, and strong LINER-like galaxies in red.
  Error bars show the 1$\sigma$ line strength measurement errors.
  Solid lines show linear least-squares fits of index strength onto
  $\sigma$, independently for each bin in emission line properties.
  (a) Indices strongly sensitive to age.  (b) Indices dominated by
  iron absorption lines.  In each panel, the left column shows raw
  measured indices, while the right column shows the indices corrected
  for incompleteness in the lower $\sigma$ bins (see
  \S\ref{lowsig_corr}).  The Balmer line strengths decrease with
  increasing $\sigma$ while all other line strengths increase with
  increasing $\sigma$.  This is consistent with higher $\sigma$
  galaxies being older and/or more metal rich than lower $\sigma$
  galaxies.  Galaxies with LINER-like emission have stronger Balmer
  lines and weaker metal lines, consistent with LINER-like galaxies
  having younger age and/or lower metallicities than quiescent
  galaxies.  (c) and (d) Indices strongly affected by non-solar
  abundance ratios.  The same trends are visible as in panel (b), that
  higher $\sigma$ galaxies have stronger absorption lines than lower
  $\sigma$ galaxies while LINER-like galaxies have weaker absorption
  lines than quiescent galaxies (with the exception of Na D, where
  LINER-like galaxies have stronger absorption than quiescent
  galaxies; see \S\ref{linestrength_trends} for details).
  }\label{indices}
\end{figure*}

\clearpage

\begin{figure*}[b]
\epsscale{0.6}
\figurenum{\ref{indices}}
\plotone{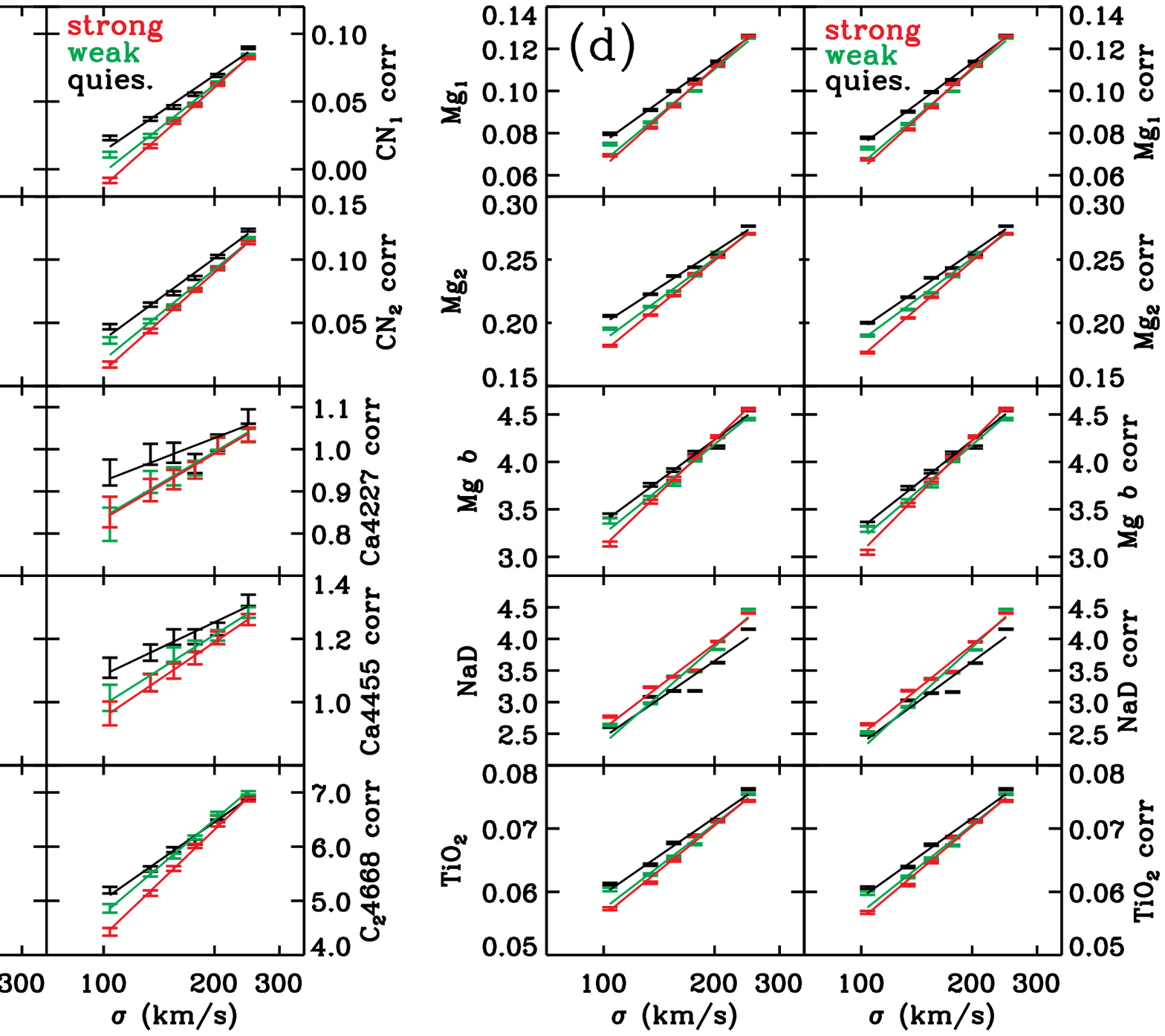}
\caption{cont. }
\end{figure*}

\clearpage

\begin{figure*}[b]
\epsscale{0.6}
\plotone{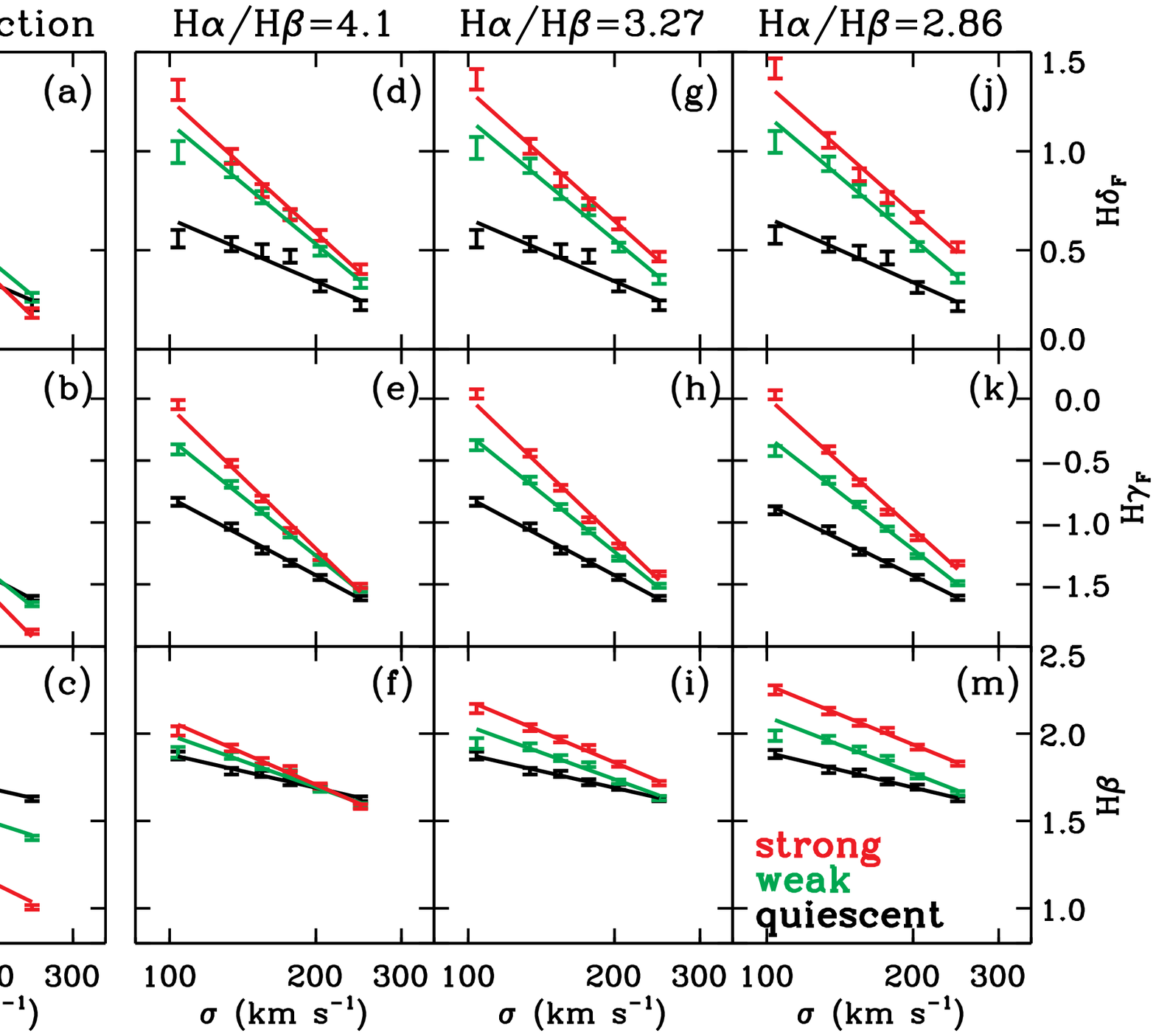}
\caption{The effect of emission infill and various corrections for
  emission infill on Balmer absorption line strengths.  As before,
  quiescent galaxies are shown in black, weak LINER-like galaxies in
  green, and strong LINER-like galaxies in red; error bars show
  measurement errors for the stacked spectra, and solid lines are
  linear least squares fits of line strength onto $\sigma$. (a--c)
  Balmer absorption line measurements with no infill correction
  applied.  Note that in (c) H$\beta$ is substantially filled in by
  emission in the LINER-like galaxies while in (a) H$\delta_F$ is only
  marginally affected.  Panels (d--m) show the effects of making a
  range of reasonable emission infill corrections.  (d--f) show the
  Balmer lines when the Balmer decrement measured by \citet{yan06} is
  used.  H$\beta$ line strengths vary little between quiescent and
  LINER-like galaxies, while H$\delta_F$ line strengths vary
  substantially.  (g--i) show the Balmer lines using the Balmer
  decrement measured in red sequence LINERs of \citet{ho97}.  These
  two choices of the Balmer decrement likely bracket the true mean
  value of the Balmer decrement in these systems.  (j--m) show the
  Balmer lines using a ``maximal'' Balmer decrement, with
  H$\alpha$/H$\beta$ characteristic of stellar-ionized H\textsc{ii}
  regions and no reddening.  The trend of stronger Balmer absorption
  in LINER-like galaxies exists for all reasonable values of the
  Balmer decrement, and exists even in the uncorrected H$\delta_F$,
  where emission infill is minimal.  }\label{balmer_comp}
\end{figure*}

\clearpage

\begin{figure*}[b]
\epsscale{0.30}
\plotone{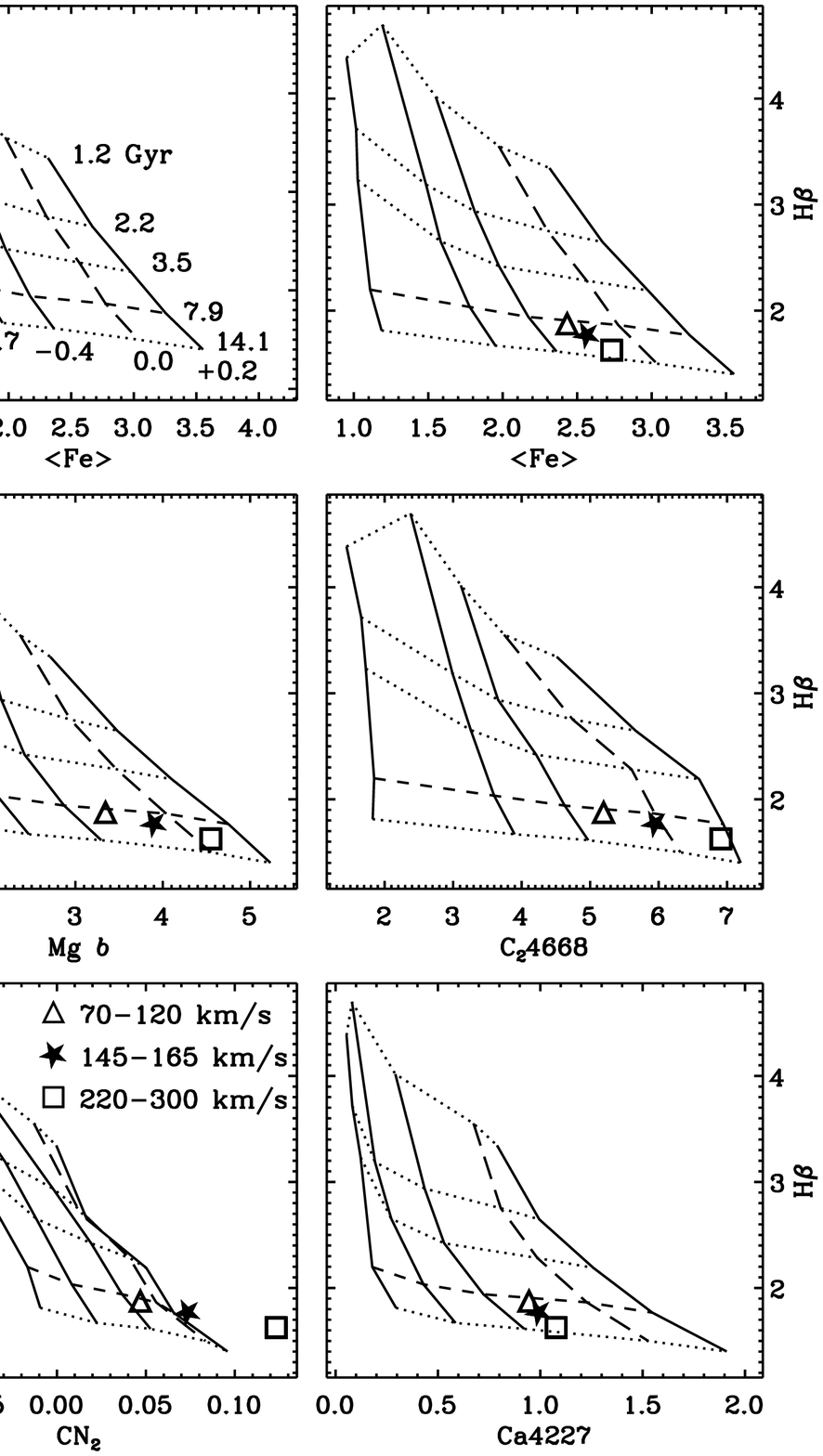}
\caption{Comparison of Lick index measurements for 3 stacked
  spectra with a simple stellar population model from \citet{sch07}.
  The triangle, star, and square, show measurements for the $70 <
  \sigma < 120$ km s$^{-1}$, $145 < \sigma < 165$ km s$^{-1}$, and
  $220 < \sigma < 300$ km s$^{-1}$ quiescent galaxy bins,
  respectively.  Solid lines connect models of the same [Fe/H], from
  left to right: $-1.3$, $-0.7$, $-0.4$, 0.0, and +0.2.  Dotted lines
  connect models of the same age, with ages from top to bottom: 1.2,
  2.2, 3.5, 7.9, and 14.1 Gyr, as labelled in the top left plot.  For
  reference, the [Fe/H] = 0.0 and $age$ = 7.9 Gyr lines are shown as
  the long-dash and short-dash lines, respectively.  The model has
  abundance ratios [Mg/Fe] = [Na/Fe] = [Si/Fe] = [Ti/Fe] = 0.120,
  [C/Fe] = 0.100, [N/Fe] = $-0.002$, and [Ca/Fe] = 0.010, calculated
  with a solar isochrone and with [O/Fe] = [Cr/Fe] = 0.0.  This is the
  best-fitting model for the $70 < \sigma < 120$ km s$^{-1}$ data
  (triangle), as determined by {\bf EZ\_Ages} \citep{gra07}.  Note
  that the data points for the $70 < \sigma < 120$ km s$^{-1}$ stacked
  spectrum fall at the same values of $age$ and [Fe/H] in each panel,
  as expected since this model gives consistent parameters for all
  indices.  The data points from the $145 < \sigma < 165$ km s$^{-1}$
  (star) and $220 < \sigma < 300$ km s$^{-1}$ (square) stacked spectra
  do {\it not} fall at the same values in each panel; higher values of
  [Mg/Fe], [C/Fe], and [N/Fe] are needed to make the age and [Fe/H]
  measured in Mg {\it b}, C$_2$4668, and CN$_2$ match the other
  panels, suggesting that the higher $\sigma$ galaxies are enhanced in
  [Mg/Fe], [C/Fe], and [N/Fe] over the $70 < \sigma < 120$ km s$^{-1}$
  galaxies.  }\label{grid1_a}
\end{figure*}

\clearpage

\begin{figure*}[b]
\epsscale{0.3}
\plotone{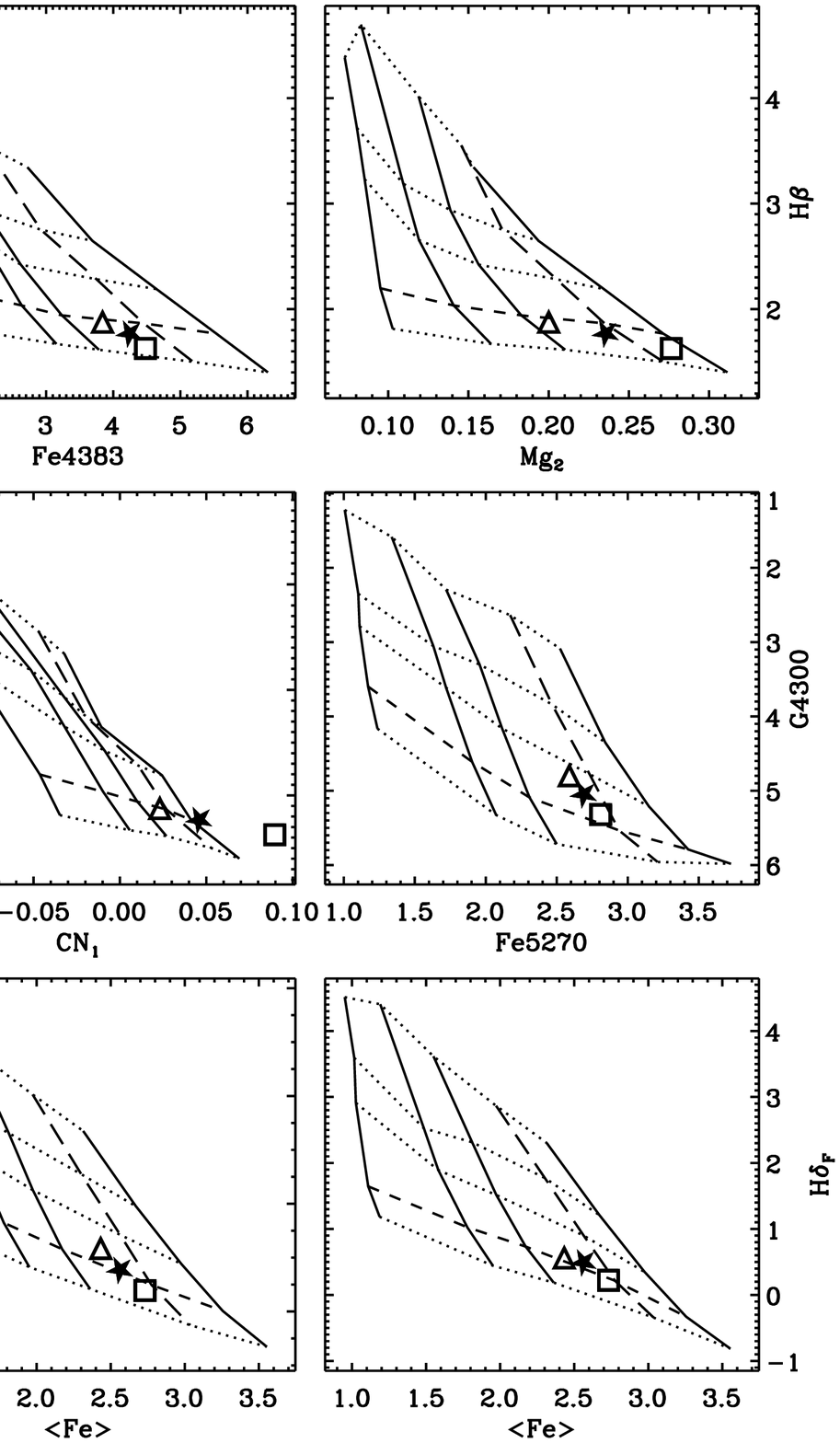}
\caption{Index-index plots for indices {\it not} used by {\bf
  EZ\_Ages} \citep{gra07} in the abundance fitting process.  All lines
  and symbols are as in Figure \ref{grid1_a}.  The $age$ and [Fe/H]
  values for the $70 < \sigma < 120$ km s$^{-1}$ galaxies (triangle)
  are in good agreement with those seen in Figure \ref{grid1_a}, with
  the exception of G4300 and H$\gamma_F$.  Also, the higher $\sigma$
  galaxies have stronger Mg$_2$ and CN$_1$ than predicted by this
  model, suggesting they have higher [Mg/Fe], [C/Fe], and [N/Fe] than
  the $70 < \sigma < 120$ km s$^{-1}$ galaxies, as was also implied by
  Figure \ref{grid1_a}.}\label{grid1_b}
\end{figure*}

\clearpage

\begin{figure*}[b]
\epsscale{0.3}
\plotone{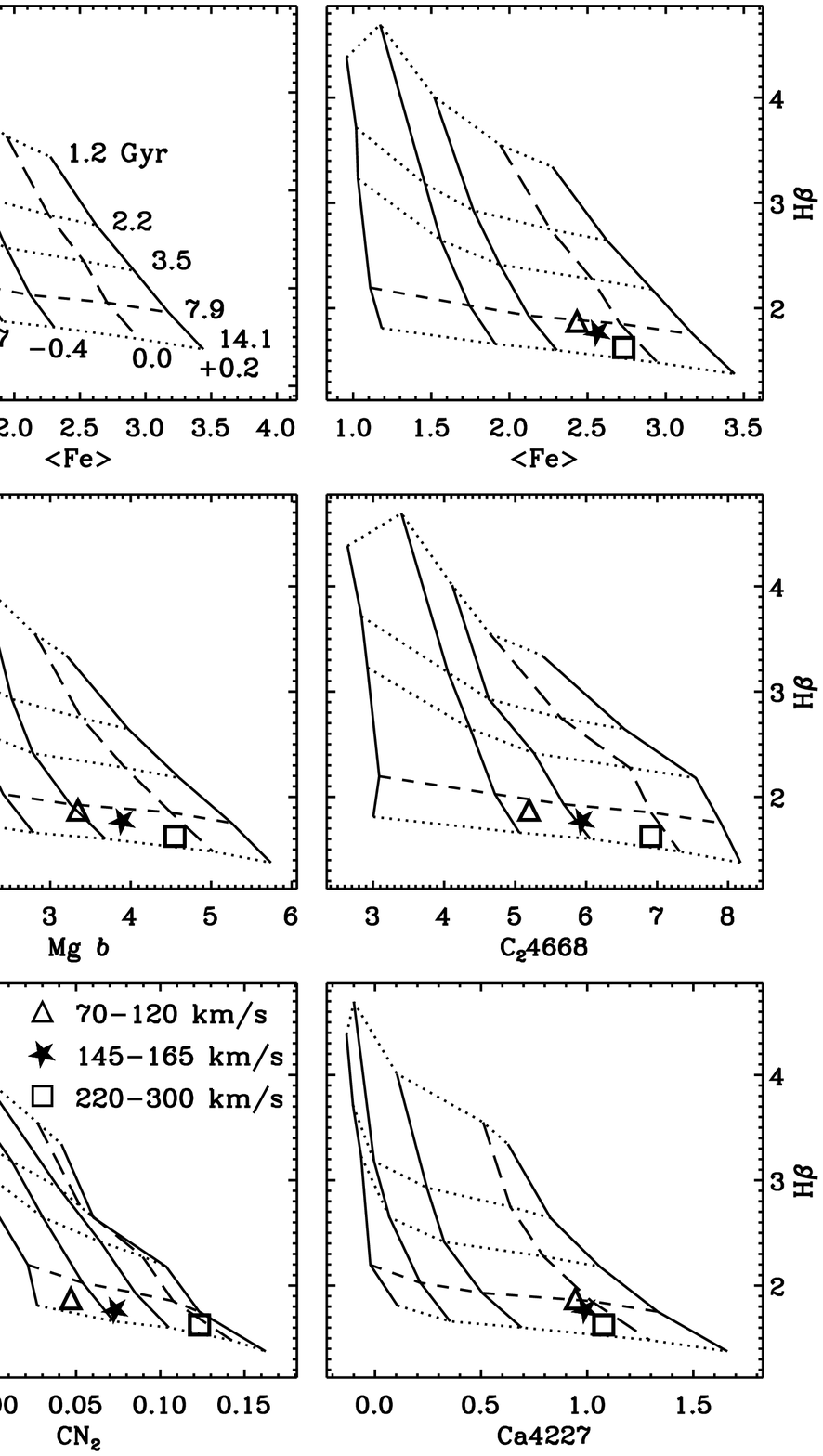}
\caption{Same as Figure \ref{grid1_a} except that grids are computed
  using the best-fit abundances for the quiescent $220 < \sigma < 300$
  km s$^{-1}$ data, as determined by {\bf EZ\_Ages}.  This model has
  [Mg/Fe] = [Na/Fe] = [Si/Fe] = [Ti/Fe] = 0.260, [C/Fe] = 0.270,
  [N/Fe] = $0.220$, and [Ca/Fe] = 0.040, calculated with a solar
  isochrone and with [O/Fe] = [Cr/Fe] = 0.0.  The triangle, star, and
  square show the data for the $70 < \sigma < 120$ km s$^{-1}$, $145 <
  \sigma < 165$ km s$^{-1}$, and $220 < \sigma < 300$ km s$^{-1}$
  quiescent stacked spectra, respectively.  Here, note that the data
  point for the $220 < \sigma < 300$ km s$^{-1}$ stacked spectrum
  falls in the same position on the grid in each index-index plot,
  indicating that this is the correct abundance pattern for this data.
  The lower $\sigma$ galaxies, by contrast, are clearly less enriched
  in Mg, CH, and CN than this model as shown by their location in the
  plots of H$\beta$ against Mg $b$, C$_2$4668, and CN$_2$.
  }\label{grid6_a}
\end{figure*}

\clearpage

\begin{figure*}[b]
\epsscale{0.3}
\plotone{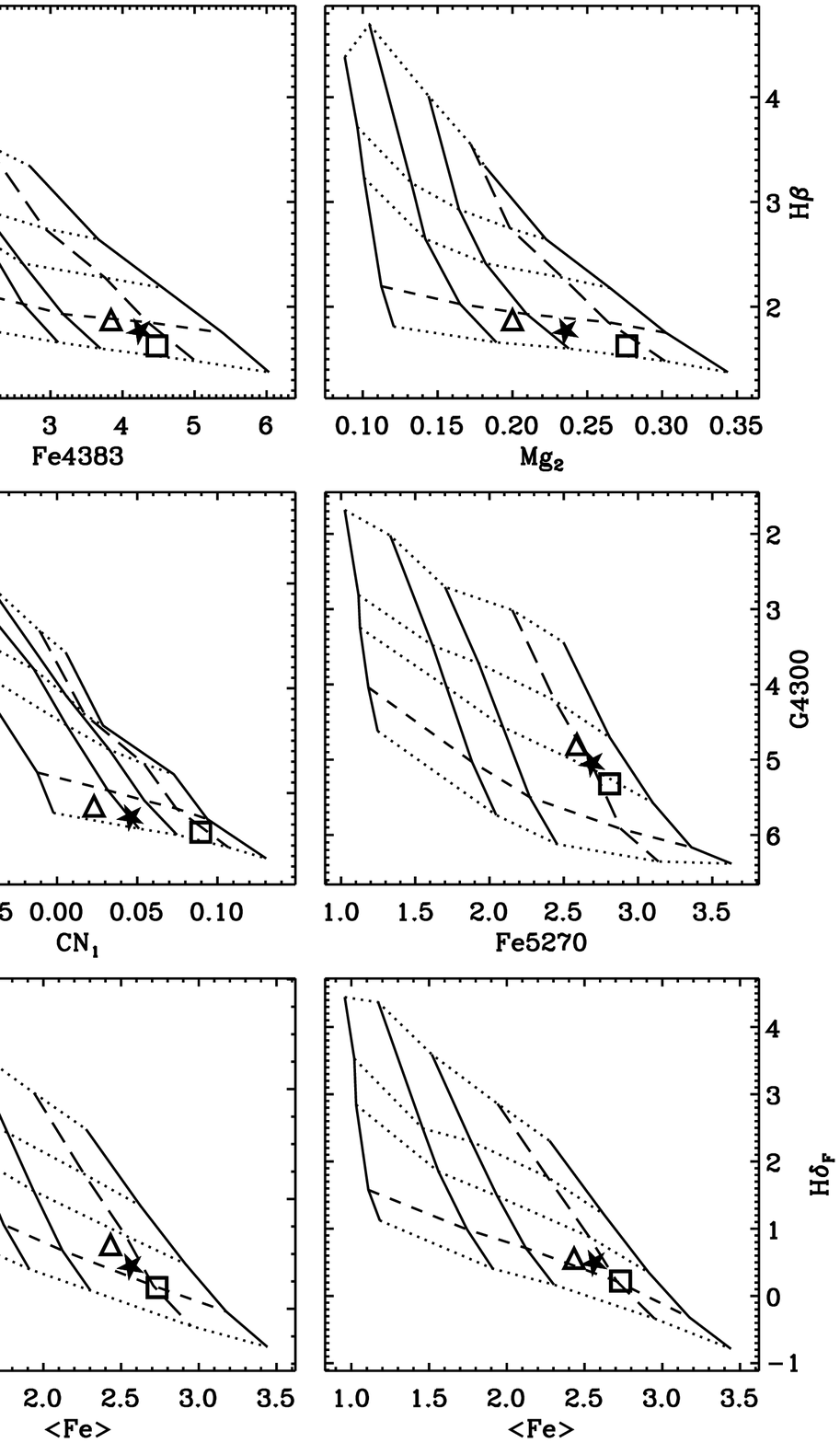}
\caption{Same as Figure \ref{grid1_b} except that grids are computed
  using the best-fit abundances for the quiescent $220 < \sigma < 300$
  km s$^{-1}$ data, showing indices not used in the fitting process.
  The plots of H$\beta$ against Fe4383, Mg$_2$, and CN$_1$ give ages
  and [Fe/H] values for the $220 < \sigma < 300$ km s$^{-1}$ data
  which are consistent with those in Figure \ref{grid6_a}, indicating
  that this abundance pattern reproduces even indices which are not
  used in the fitting process.  The G4300 index is inconsistent with
  other indices, as discussed in the text.  The ages measured from
  H$\gamma_F$ and H$\delta_F$ are slightly younger than the ages
  measured from H$\beta$.}\label{grid6_b}
\end{figure*}

\clearpage

\begin{figure*}[b]
\epsscale{0.6}
\plotone{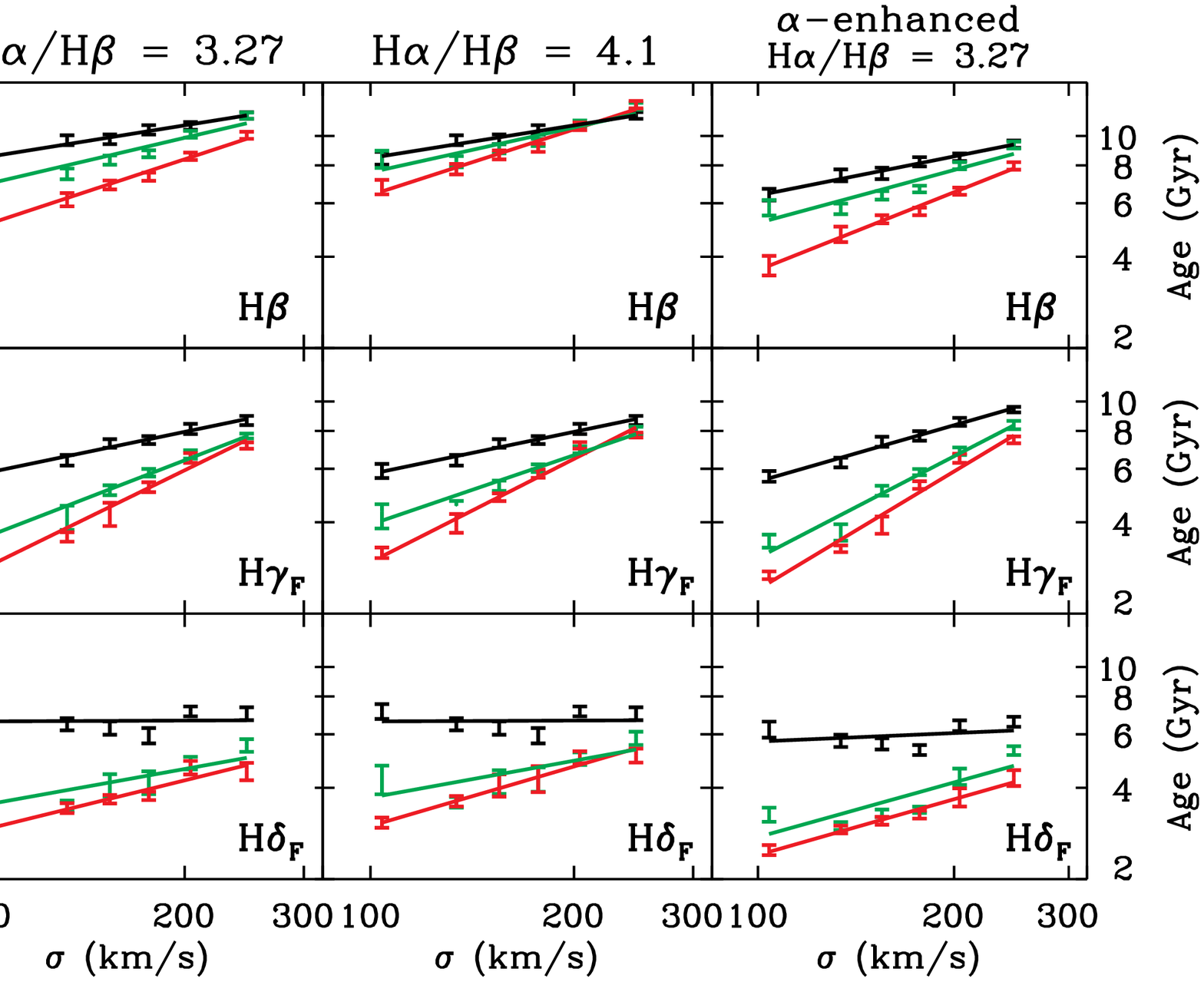}
\caption{SSP ages as a function of $\sigma$ and emission line
  properties.  As in previous plots, quiescent, weak, and strong
  LINER-like galaxies are shown in black, green, and red,
  respectively.  Error bars show the uncertainty in the ages due to
  measurement errors in the Balmer lines and $\langle$Fe$\rangle$
  index only.  Solid lines show linear least squares fits of age onto
  $\sigma$ for each data set.  Ages are determined separately for the
  three Balmer lines H$\beta$, H$\gamma_F$, and H$\delta_F$.  The left
  and center columns show the ages measured with the \citet[our
  preferred correction]{ho97} and \citet{yan06} Balmer decrements,
  respectively, to illustrate the effect of different infill
  corrections.  Both are determined by {\bf EZ\_Ages} \citep{gra07}
  using a solar-scale isochrone.  The right column shows the effect on
  the age measurements when an $\alpha$-enhanced isochrone and [O/Fe]
  = +0.3 are adopted.  Higher $\sigma$ galaxies have older ages than
  lower $\sigma$ galaxies, and LINER-like galaxies are younger than
  quiescent galaxies at fixed $\sigma$, independent of the choice of
  Balmer decrement, stellar population model isochrone, or Balmer line
  used to make the measurement.  However age differences are more
  pronounced using higher-order Balmer lines, as expected if the
  stellar populations have a {\it distribution} of ages.  A smaller
  Balmer decrement (i.e., larger infill correction) leads to larger
  differences between LINER-like and quiescent ages.  The use of an
  $\alpha$-enhanced isochrone produces systematically younger ages in
  H$\beta$, but qualitatively similar trends with $\sigma$ and
  emission line properties.  }\label{ages_plot}
\end{figure*}

\clearpage

\begin{figure*}[b]
\epsscale{0.7}
\plotone{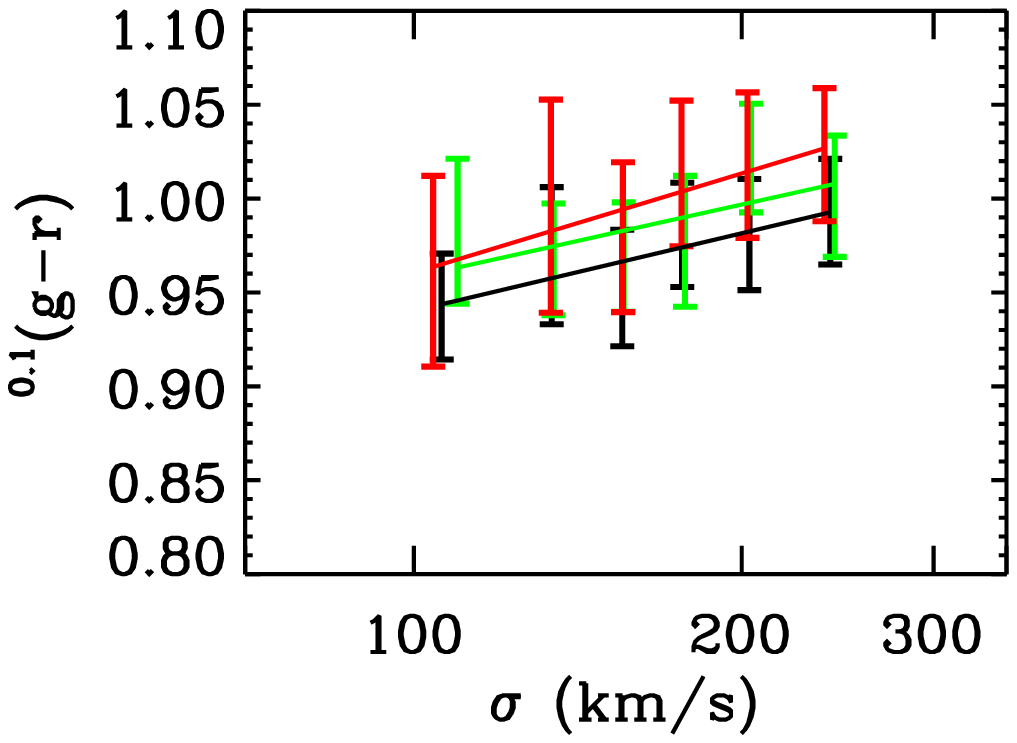}
\caption{Median $^{0.1}(g-r)$ colors as a function of mean $\sigma$ for
  the galaxies that went into each stacked spectrum.  The quiescent
  galaxies are shown in black, the weak LINER-like galaxies in green,
  and the strong LINER-like galaxies in red.  Solid lines show a
  linear least-squares fit of color onto log $\sigma$.  The error bars
  show the 1$\sigma$ spread in the color values for each bin of
  galaxies.  Although the spectral data show that the LINER-like
  galaxies are typically {\it younger} than their quiescent
  counterparts, the LINER-like galaxies on average have slightly {\it
  redder} colors.  This is not a metallicity effect (see Figure
  \ref{abuns_plot}) but is probably an indication that the LINER-like
  galaxies are typically more dust-reddened than the quiescent
  galaxies, consistent with the larger amounts of interstellar NaD
  absorption seen in the LINER-like galaxies in Figure \ref{indices}.
  See \S\ref{ages} for details.}\label{g_r_sig}
\end{figure*}

\clearpage

\begin{figure*}[b]
\epsscale{0.7}
\plotone{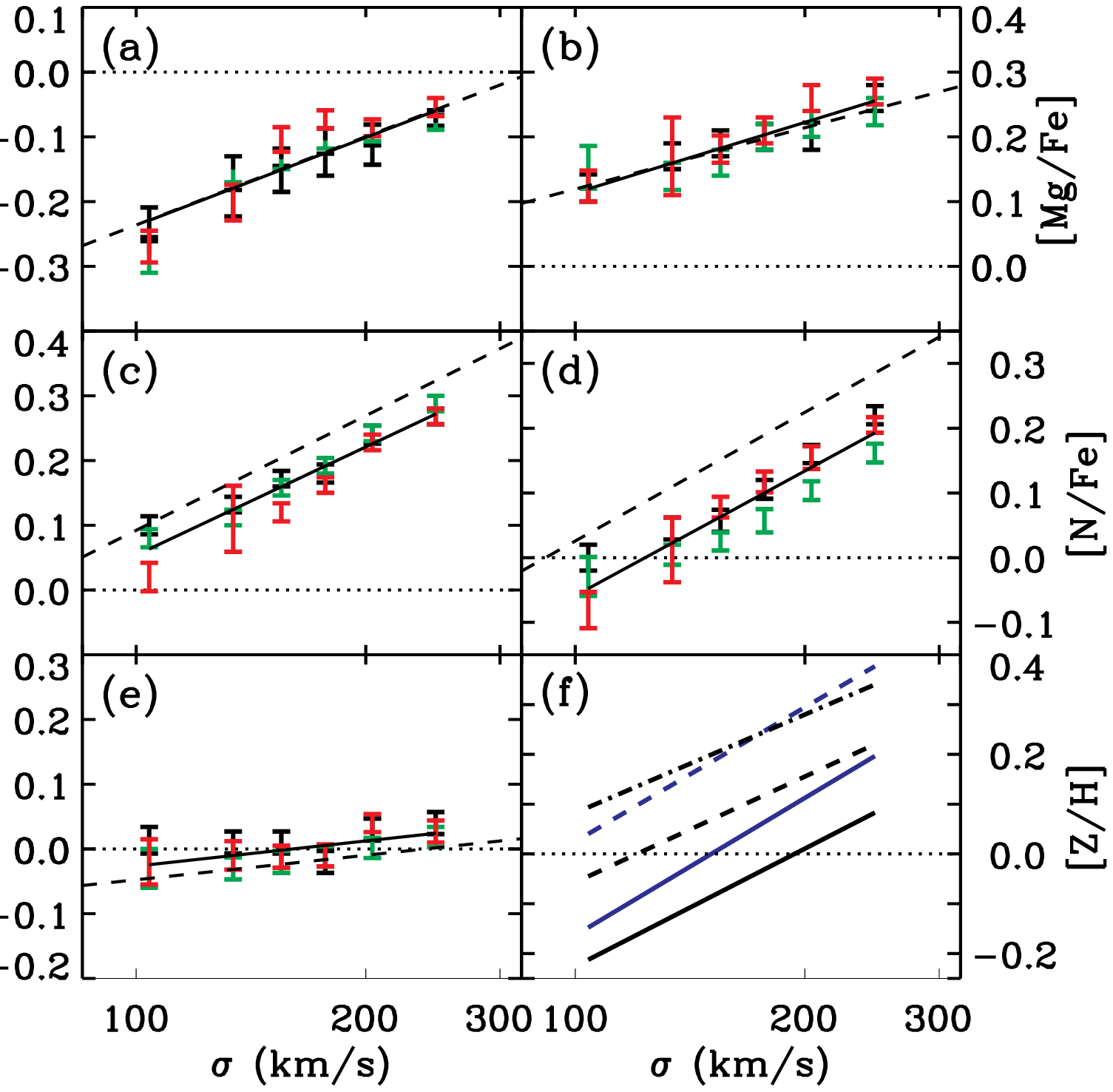}
\caption{(a--e) [Fe/H] and abundance ratios for quiescent and
  LINER-like galaxies as a function of $\sigma$.  Quiescent,
  weak, and strong LINER-like galaxies are shown in black, green, and
  red, respectively.  Error bars show the uncertainties in each
  abundance due to measurement errors in the Lick
  indices as described in \S\ref{abuns}.  The dotted line in each
  panel shows solar abundance.  There are no significant variations in
  abundances or abundance ratios between quiescent and LINER-like
  galaxies.  The solid line show a linear least squares fit of
  abundance onto $\sigma$ for quiescent and LINER-like galaxies
  simulatenously.  Abundances are computed using a solar-scale
  isochrone and assuming [O/Fe] = 0.0, with Na, Si, and Ti set to
  track Mg, and [Cr/Fe] = 0.0.  The abundances computed using the
  $\alpha$-enhanced isochrone are not significantly different.  The
  dashed line shows the effect of computing abundances with [O/Fe] =
  +0.3.  All abundances increase with $\sigma$, [Fe/H], [C/Fe], and
  [N/Fe] most strongly. All galaxies are Mg-enhanced.  (f) Linear
  least squares fits of [Z/H] as a function of $\sigma$. The black
  solid, dashed, and dash-dot lines show [Z/H] computed assuming
  [O/Fe] = 0.0, [O/Fe] = +0.3, and [O/Fe] = +0.5, respectively.  The
  blue solid and dashed lines show [Z/H] computed assuming [O/Fe] =
  [Mg/Fe], and [O/Fe] = [Mg/Fe] + 0.3, respectively.  Total
  metallicity is dominated by [O/Fe] which cannot be measured and is
  thus highly uncertain.  }\label{abuns_plot}
\end{figure*}

\clearpage

\begin{figure*}[b]
%\figurenum{A1}
\epsscale{0.6}
\plotone{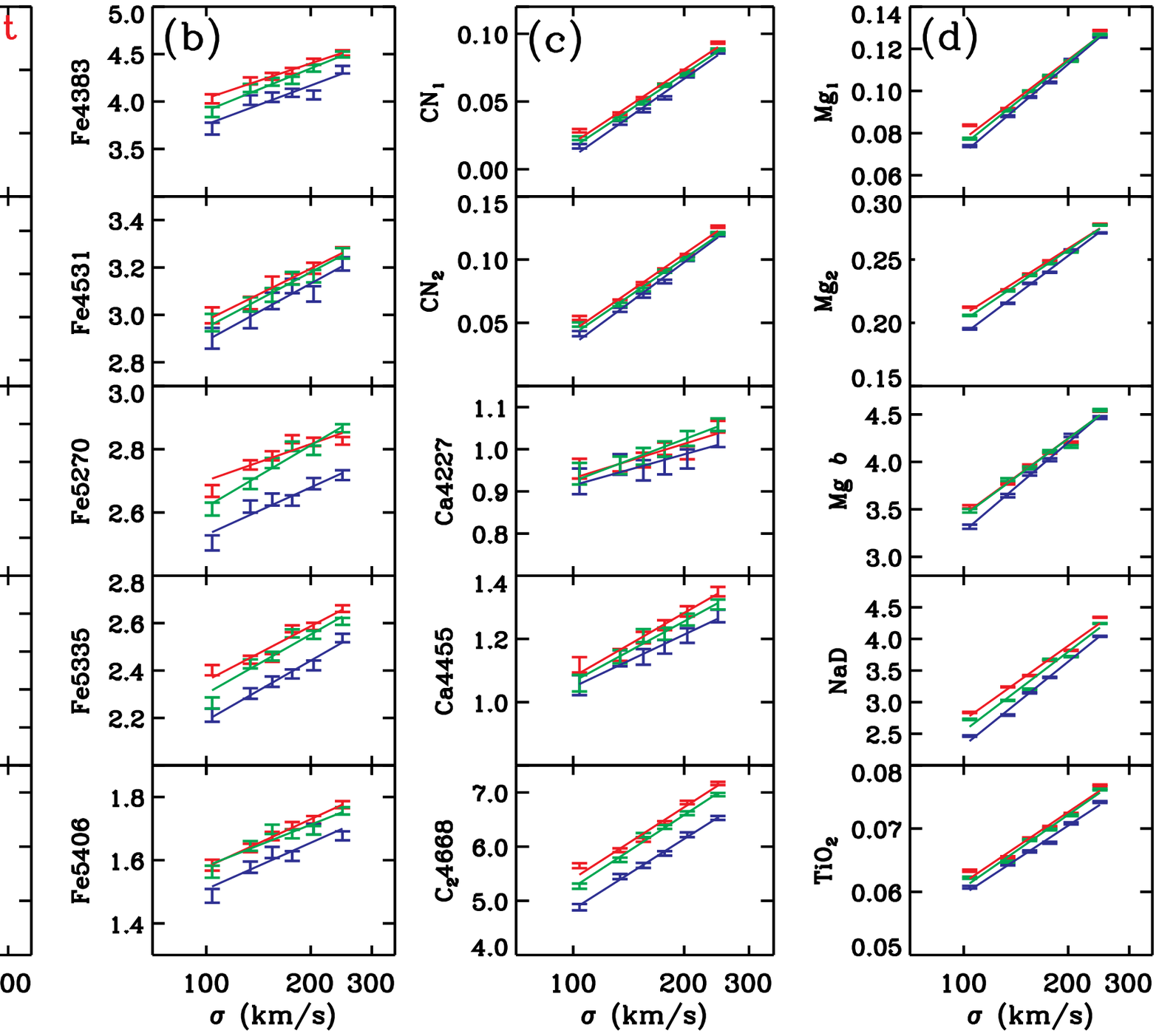}
\caption{Lick index line strengths in stacked spectra of quiescent red
  sequence galaxies, as a function of $\sigma$ and magnitude.  Each
  bin in $\sigma$ has been divided into faint, mid-magnitude, and
  bright sub-bins, with equal numbers of galaxies in each sub-bin.
  The faintest galaxies at each $\sigma$ are shown in blue, the
  mid-magnitude galaxies are shown in green, and the brightest
  galaxies are shown in red.  Error bars show the 1$\sigma$ index
  measurement errors for each stacked spectrum.  Solid lines show
  linear least-squares fits of index strength onto $\sigma$,
  independently for each bin in magnitude.  As in Figure
  \ref{indices}, panel (a) shows strongly age-sensitive indices, panel
  (b) shows indices dominated by iron absorption lines, and panels (c)
  and (d) show indices that are sensitive to non-solar abundance
  ratios.  Faint galaxies appear to have similar ages but lower metal
  abundances when compared to mid-magnitude and bright galaxies at the
  same $\sigma$.  }\label{siglum_indices}
\end{figure*}

\clearpage

\begin{figure*}[b]
%\figurenum{B1}
\epsscale{0.6}
\plotone{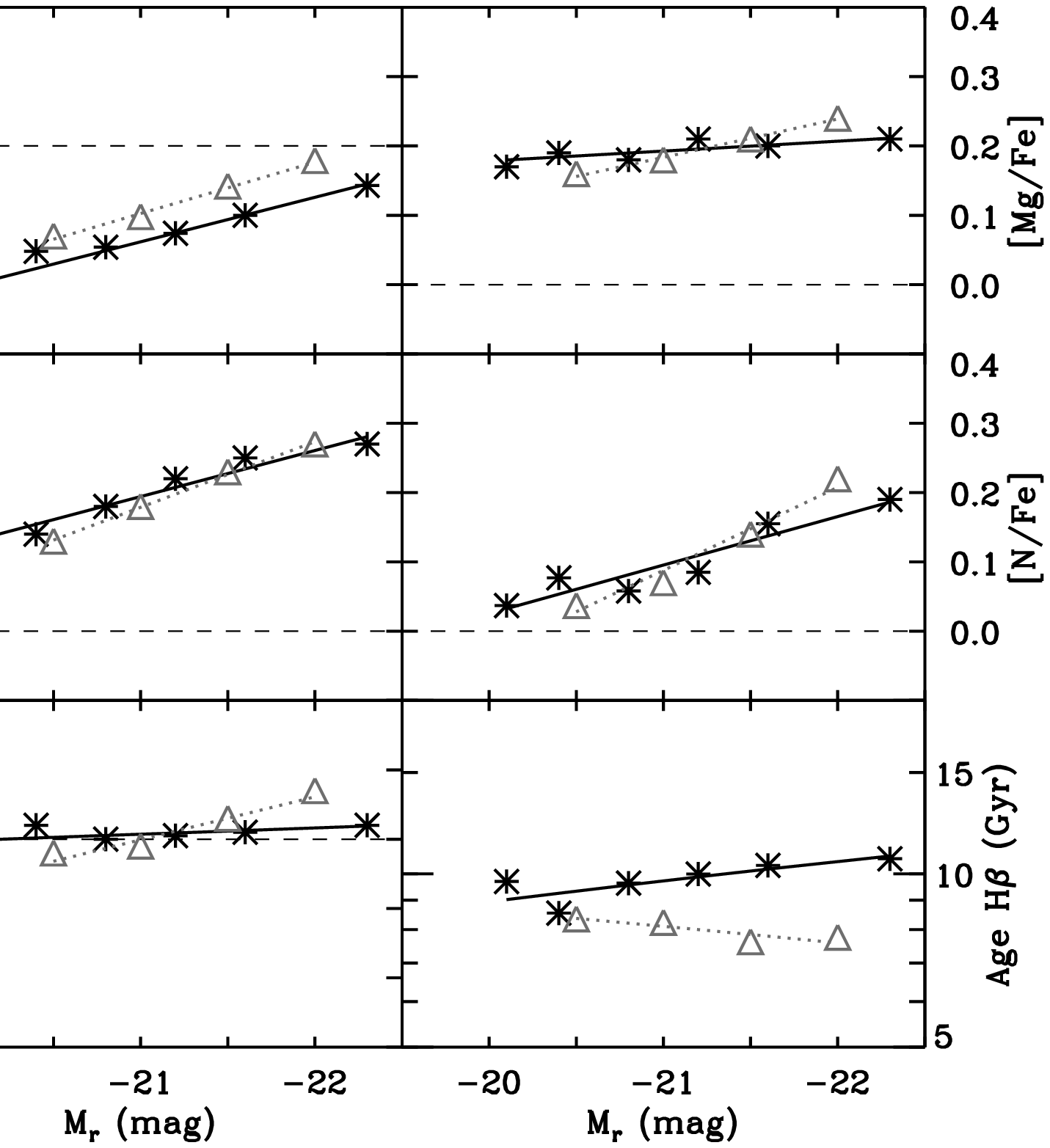}
\caption{[Fe/H], abundance ratios, and ages of stacked spectra, as a
  function of galaxy absolute magnitude.  Black stars are the
  luminosity-binned data from appendix \ref{eis_data}.  Only the
  stacked spectra for quiescent galaxies are shown.  The solid black
  line shows a linear least-squares fit of the stellar population
  parameter onto magnitude for the luminosity-binned data.  Grey
  triangles show the results from \citet{sch07} for the \citet{eis03}
  stacked spectra.  The dotted grey line shows a linear least-squares
  fit for the \citet{eis03} stacked spectra.  Solar abundance and
  abundance patterns are shown for reference as the dashed line in
  each panel.  The \citet{eis03} data (triangles) show younger ages,
  slightly higher [Fe/H], and slightly steeper abundance relations
  versus magnitude than do the luminosity-binned spectra from our
  sample (stars).  }\label{eis_abuns}
\end{figure*}

\clearpage

\begin{figure*}[b]
%\figurenum{B2}
\epsscale{0.6}
\plotone{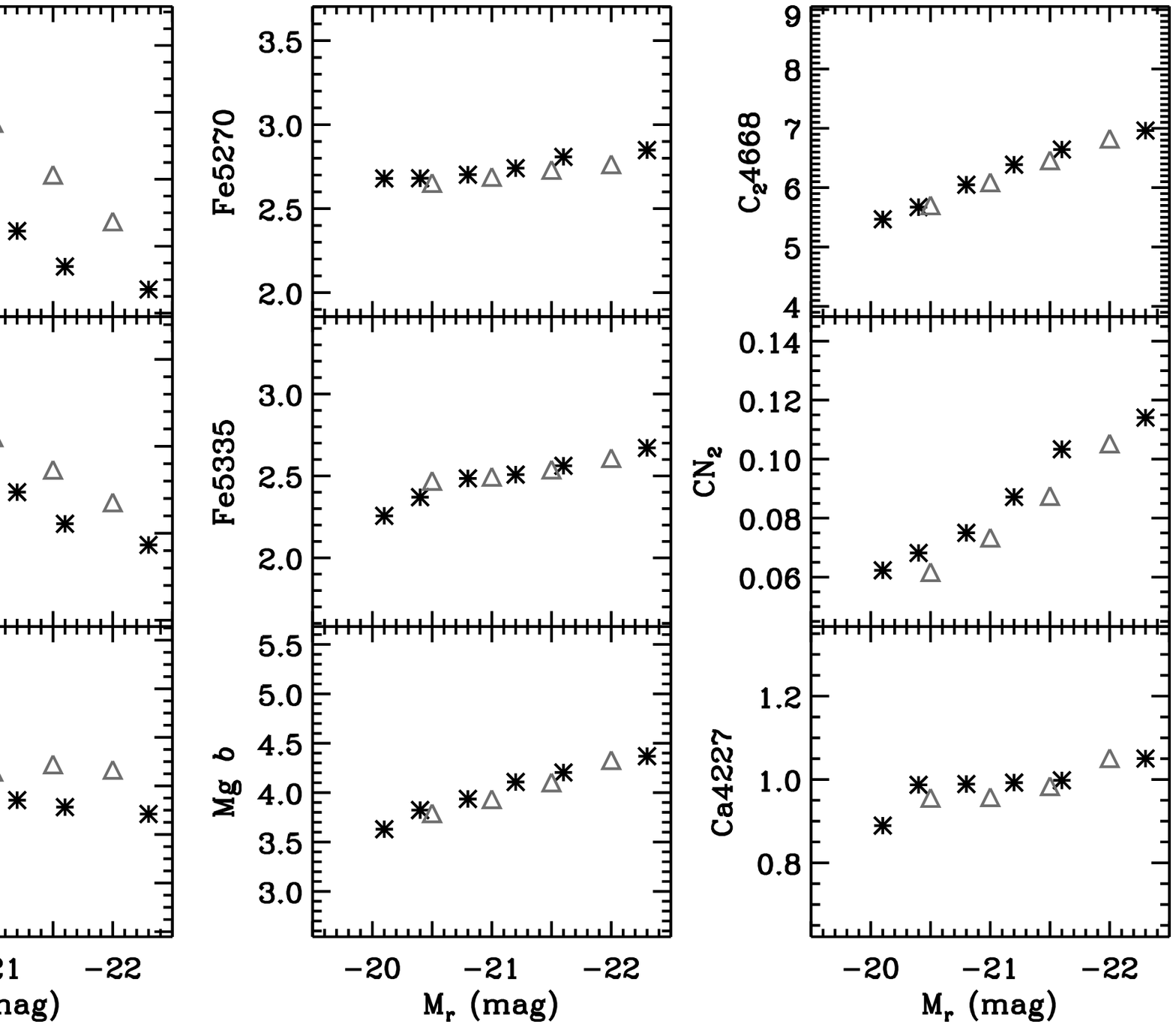}
\caption{Selected Lick index line strengths as a function of galaxy
  magnitude.  Black stars are the luminosity-binned data from appendix
  \ref{eis_data}.  Only the stacked spectra for quiescent galaxies are
  shown. Grey triangles show indices measured in the \citet{eis03}
  stacked spectra.  In both samples, Balmer lines weaken as galaxy
  luminosity increases with the exception of H$\beta$, which behaves
  similarly to H$\delta_F$ and H$\gamma_F$ in our data but is constant
  in the \citet{eis03} data.  H$\delta_F$ and H$\gamma_F$ are stronger
  in the \citet{eis03} data than in ours, consistent with younger mean
  ages.  In contrast to the Balmer lines, the other Lick indices are
  very similar between the two data sets.  The metal lines may be
  slightly weaker in the \citet{eis03} data than in ours, again
  consistent with younger ages.  The consistency of the metal
  absorption line strengths between the two samples implies that the
  differences in the measured abundance trends seen in Figure
  \ref{eis_abuns} are caused by the difference in measured Balmer line
  strengths, rather than intrinsic differences in metallicity.
  }\label{eis_indices}
\end{figure*}

\end{document}